\documentstyle[12pt,epsf,epsfig,here]{article}
\renewcommand{\theequation}{\arabic{section}.\arabic{equation}}
\input epsf

\def\beq{\begin{equation}}
\def\eeq{\end{equation}}
\def\bea{\begin{eqnarray}}
\def\eea{\end{eqnarray}}
\def\Journal#1#2#3#4{{#1} {\bf #2}, #3 (#4)}

\def\NPB{{\em Nucl. Phys.} B}
\def\PLB{{\em Phys. Lett.}  B}
\def\PRL{\em Phys. Rev. Lett.} 
\def\PRD{{\em Phys. Rev.} D}
\def\ZPC{{\em Z. Phys.} C}
\def\PTP{{\em Prog. Theor. Phys.}}
\def\JHEP{\em J. High Energy Phys.}

\def\gappeq{\mathrel{\rlap {\raise.5ex\hbox{$>$}}
{\lower.5ex\hbox{$\sim$}}}}
\def\lappeq{\mathrel{\rlap{\raise.5ex\hbox{$<$}}
{\lower.5ex\hbox{$\sim$}}}}

\def\gsim{{~\raise.15em\hbox{$>$}\kern-.85em
          \lower.35em\hbox{$\sim$}~}}
\def\lsim{{~\raise.15em\hbox{$<$}\kern-.85em
          \lower.35em\hbox{$\sim$}~}}

\def\R{{\cal R}}
\def\L{{\cal L}}
\def\B{{\cal B}}
\def\wi{{\cal W}}
\def\la#1{{\lambda_{#1}}}
\def\pf{{\rm Pf}}
\def\det{{\rm Det}}
\def\tr{{\rm Tr}}

\def\gws{SU(3)\times SU(2)\times U(1)}

\def\Qb{\bar Q}
\def\Fb{\bar \Phi}
\def\F{\Phi}

\def\fourpisq{{1\over 16\pi^2}}
\def\fivecube{SU(5)_1\times SU(5)_2\times SU(5)_W}

\def\one{{\bf 1}}

\def\five{{\bf 5}}
\def\fivebar{{\bf \bar 5}}
\def\identity{1 \hspace{-.085cm}{\rm l}}

\begin{document}
\begin{titlepage}
\begin{center}
January 1998  \hfill    CERN-TH/97-380\\
              \hfill    DFPD-98/TH/06\\
               \hfill    hep-ph/9801271
\vskip .2in
{\large \bf 

Theories with Gauge-Mediated Supersymmetry Breaking}
\vskip .3in

\vskip .3in
G.F. Giudice\footnote{On leave
of absence from INFN, Sez. di Padova, Italy.} 
and R. Rattazzi

\vskip .2in

{\em Theory Division, CERN\\
     CH-1211 Geneva 23, Switzerland}

\end{center}
\vskip .2in
\begin{abstract}
\medskip

Theories with gauge-mediated supersymmetry breaking provide an interesting
alternative to the scenario in which the soft terms of the
low-energy fields are induced by gravity. These theories allow for
a natural suppression of flavour violations in the supersymmetric
sector and have very distinctive phenomenological features. Here we
review their basic structure, their experimental implications, and the
attempts to embed them into models in which all mass scales are dynamically
generated from a single fundamental scale.

\end{abstract}

\vskip 3cm
\begin{center}
{\it Submitted to Physics Reports}
\end{center}

\end{titlepage}

\tableofcontents
\newpage
\baselineskip20pt

\section{Introduction}
\setcounter{equation}{0}

The naturalness (or hierarchy) problem~\cite{wilson79,hooft80} is considered
to be the most serious theoretical argument against the validity of the
Standard Model (SM) of elementary particle interactions beyond the TeV
energy scale. In this respect, it can be viewed as the ultimate
motivation for pushing the experimental research to higher energies.
The naturalness problem arises from the difficulty, in field theory, in
keeping fundamental scalar particles much lighter than the highest energy
within the range of validity of the theory, $\Lambda_{UV}$. 
This difficulty is a consequence
of the lack of a symmetry prohibiting a scalar mass term or, in more
technical terms, of the presence of quadratic divergences in the quantum
corrections to scalar masses. 

The SM Lagrangian contains a single dimensionful parameter, in the mass term
for the Higgs field, which determines the size of the electroweak scale.
A reasonable criterion for ``naturalness"
limits the validity cut-off scale 
$\Lambda_{UV}$ to be at most a loop factor larger than
the mass scale of fundamental scalars. We are then led to the conclusion that
the SM can be valid only up to the TeV scale, if 
the naturalness criterion is satisfied. 
New physics should appear at this scale, modifying the high-energy
behaviour and divorcing the Higgs mass parameter from its ultraviolet
sensitivity.

An elegant solution to the naturalness problem is provided by 
supersymmetry~\cite{golfand71,volkov73,wess74}. Since supersymmetry relates
bosons to fermions, a scalar mass term is never, in a supersymmetric 
theory,  generated
by quantum corrections, if the corresponding fermion mass is forbidden
by a chiral symmetry. Moreover, a supersymmetric theory is free from
quadratic divergences~\cite{wess74a,iliopoulos74,ferrara75,grisaru79}.

In the real world, supersymmetry must be broken. However, if supersymmetry
provides a solution to the naturalness problem, it must then be an
approximate symmetry of the theory above the TeV scale. This is possible
when supersymmetry is broken only 
softly~\cite{girardello82,dimopoulos81b,sakai81}, {\it i.e.} by terms that do
not introduce quadratic divergences. These terms always have dimensionful
couplings, and the naturalness criterion implies that the corresponding
mass scale cannot exceed the TeV region. The soft
terms provide gauge-invariant
masses to all supersymmetric partners of the known SM particles.
These masses give a precise physical
meaning to the SM ultraviolet cut-off $\Lambda_{UV}$.

As the soft terms determine the mass spectrum of the new particles,
the mechanism of supersymmetry breaking is the
key element for understanding the low-energy aspects of supersymmetric
theories. However,
the mechanism of communicating the original supersymmetry
breaking to the ordinary particle supermultiplets plays an equally or even more
important r\^ole. As an analogy, one can think of the case of the SM,
in which the Higgs vacuum expectation value (VEV) determines the scale
of electroweak breaking, but the detailed mass spectrum of bosons and
fermions is dictated by the coupling constants of the forces that
communicate the information of electroweak breaking, {\it i.e.} gauge and
Yukawa interactions, respectively.

A major difference between the case of supersymmetry breaking and this SM
analogy is that the supertrace theorem~\cite{ferrara79} 
essentially rules out the
possibility of constructing simple models in which supersymmetry breaking
is communicated to ordinary supermultiplets by tree-level renormalizable
couplings. Indeed, in a globally supersymmetric theory with a gauge group
free from gravitational anomalies~\cite{alvarez84}, the sum of the particle
tree-level squared masses, weighted by the corresponding number of degrees
of freedom, is equal in the bosonic and fermionic sectors~\cite{ferrara79}:
\beq
{\rm STr} {\cal M}^2= \sum_J (-1)^{2J}(2J+1){\cal M}^2_J =0~.
\eeq
Here ${\cal M}_J$ denotes the tree-level mass of a particle with spin $J$.
Rather generically, this theorem implies, in cases of tree-level 
communication, the existence of a supersymmetric particle lighter than
its ordinary partner.

As a consequence of this difficulty, the paradigm for constructing 
realistic supersymmetric theories is to assume that the sector
responsible for supersymmetry breaking (the analogue of the Higgs sector)
has no renormalizable tree-level couplings with the ``observable sector", which
contains the ordinary particles and their supersymmetric partners. 
Moreover, the effective theory describing the observable sector, obtained
by integrating out the heavy particles in the supersymmetry-breaking sector,
should have a non-vanishing supertrace.

It is not too hard to satisfy these conditions. The supertrace theorem, 
after all, follows from the properties of renormalizability that force
the kinetic terms to have the minimal form. Let us consider the effective
theory describing the ordinary supermultiplets, the goldstino field, and
possibly other associated light fields, but with the heavy fields of the 
supersymmetry-breaking sector integrated out. If this theory has 
non-canonical (and therefore non-renormalizable) kinetic terms for matter
and gauge multiplets, 
involving interactions with the goldstino superfield,
we generally induce scalar and gaugino masses, which
break supersymmetry but violate the supertrace theorem. Therefore, 
to understand the question of supersymmetry-breaking
communication is to identify the interactions that generate the
non-renormalizable effective Lagrangian.

One possibility is to consider a theory that is altogether non-renormalizable,
and such that the supertrace over the whole spectrum is non-vanishing.
The best-motivated example is given by gravity. Indeed
the most general supergravity Lagrangian,
in the presence of supersymmetry breaking, leads 
to~\cite{chamseddine82,barbieri82a,nilles83,cremmer83,hall83,soni83} 
an effective theory for the low-energy modes containing the
desired soft terms. This is the scenario most commonly considered in
phenomenological applications (for reviews, see {\it e.g.}
refs.~\cite{nilles84,haber85,barbieri88a}), 
and it is certainly a very attractive one
as, for the first time, gravity ventures to play an active r\^ole in
electroweak physics.

Another possibility is that the relevant dynamics at the microscopic
level is described
by a renormalizable Lagrangian, 
and, at tree level, the theory has a vanishing supertrace and no mass
splittings inside the observable supermultiplets. However
the low-energy modes are described by an effective Lagrangian, which
has non-renormalizable kinetic terms (and non-vanishing supertrace) at
the quantum level, induced by known gauge interactions.
This is the case of theories with gauge-mediated supersymmetry
breaking, which we are going to review here.

The fundamental difference between the two approaches is related to
the problem of flavour. In the limit of vanishing Yukawa couplings,
the SM Lagrangian is invariant under a global $U(3)^5$ symmetry, with 
each $U(3)$ acting on the generation indices of the five irreducible
fermionic representations of the gauge group $(q_L, u_R^c,d_R^c,\ell_L,
e_R^c)_i$. This symmetry, called flavour (or family) symmetry, follows
from the property that gauge interactions do not distinguish between
the three generations of quarks and leptons. We ignore the dynamical
origin of the Yukawa couplings or, ultimately, of the flavour-symmetry
breaking, but let us just define $\Lambda_F$ to be the relevant energy scale
of the corresponding new physics.
Above $\Lambda_F$ lie some unknown dynamics
responsible for flavour breaking. Below $\Lambda_F$ these dynamics are frozen,
leaving their scars on the flavour-breaking structure of Yukawa couplings.

A possible realization of the flavour dynamics is given by models where
a subgroup $G_F$ of $U(3)^5$ is a fundamental symmetry (local or global)
which is broken spontaneously by the vacuum expectation values of a set
of scalar fields $\phi_F=\{\phi_a\}$, the flavons. In these scenarios the
hierarchical structure of the fermion spectrum can either be obtained
via multiple stages of $G_F$ breaking or by assuming that the flavon VEVs 
are somewhat smaller (one or two orders of magnitude) than the scale $M_F$ 
which sets their coupling to the quarks and leptons. This possibility
corresponds to the Froggatt-Nielsen mechanism \cite{froggatt79} and
the scale $M_F$ could correspond to either the mass of some heavy states
or simply to the Planck scale \cite{leurer93}. It should be clear that,
for the purpose of our discussion, 
the flavour scale $\Lambda_F$ can be indifferently identified with 
either $\langle \phi_F\rangle$ or $M_F$. Consider for instance the 
Froggatt-Nielsen case.
By integrating out the flavour sector, the effective Yukawa couplings are 
functions of $\phi_a/M_F$, whose form is restricteded by the $G_F$ 
selection rules. In a similar way, if soft terms are already present at 
$\Lambda_F$, the effective squark and slepton
masses and $A$-terms are functions of the flavons as well. 
Now these soft term matrices will necessarily contain new sources of flavour
violation, in addition to  those given by the Yukawa matrices.

In the gravity-mediated approach, the soft terms are generated at the
Planck scale, and therefore necessarily at a scale larger than or equal to
$\Lambda_F$. There is then no obvious reason why the supersymmetry-breaking
masses for squarks and sleptons should be flavour-invariant. Even if
at tree level, for
some accidental reason, they are flavour-symmetric, 
loop corrections from
the flavour-violating sector will still distort their structure. Even 
contributions from ordinary grand unified theories (GUTs) can lead to
significant flavour-breaking effects in the soft 
terms~\cite{hall86,barbieri94,barbieri95}.

These flavour-breaking contributions to the soft terms are very 
dangerous~\cite{dimopoulos81b,ellis82,barbieri82}.
The mismatch between the diagonalization matrices for quarks and squarks
(and analogously for leptons and sleptons) leads to flavour-violating
gaugino vertices, and eventually to large contributions to flavour-changing
neutral-current (FCNC) processes. Studies of the
${\bar K}^0$--$K^0$ mass difference, $\mu \to e \gamma$
and similar processes set very stringent
bounds on the relative splittings among different generations of squarks
and sleptons~\cite{hagelin94,gabbiani96}.

Of course this does not mean that gravity-mediated scenarios cannot give a
realistic theory. In models with a spontaneously broken flavour group
$G_F$, like the Froggatt-Nielsen models, the selection rules of $G_F$ can lead
to approximate universality or to approximate alignment between
particle and sparticle 
masses~\cite{dine93a,nir93,leurer94,pouliot93,barbieri97}.
A similar result may also be obtained by a dynamical 
mechanism~\cite{dimopoulos95}. Indeed it may also be that at the level 
of quantum gravity  soft terms are flavour-invariant. At any rate,
it seems unavoidable~\cite{barbieri94,barbieri95} 
that in these scenarios, flavour violations should
be, at best, just at the edge of present bounds and should soon be visible.

On the other hand, in gauge-mediated theories, soft terms are generated at the
messenger scale $M$, which is {\it a priori} unrelated to $\Lambda_F$.
If $M\ll \Lambda_F$, the soft terms feel the breaking of flavour only
through Yukawa interactions. Yukawa couplings are the only relevant
sources of flavour violation, as in the SM. More precisely,
all other sources of flavour violation at the messenger scale correspond
to operators of dimension larger than $4$, suppressed by the suitable powers
of $1/\Lambda_F$. The contribution of these operators to soft masses
is necessarily suppressed by powers of $M/\Lambda_F$.  As a consequence,
the GIM mechanism is fully operative and it can be generalized to a
super-GIM mechanism, involving ordinary particles and their supersymmetric
partners. Since it is reasonable to expect
that $\Lambda_F$ is as large as the GUT or the Planck scales, in gauge-mediated
theories the flavour
problem is naturally decoupled, in contrast to the case of supergravity or,
in a different context, of technicolour theories~\cite{weinberg78,susskind79}.

Moreover, in gauge-mediated theories, it is possible to
describe the essential dynamics without dealing with gravity. This may be
viewed as an aesthetic drawback, as it delays the complete unification
of forces. However, from a more technical point of view, it is
an advantage, because the model can be solved using only field-theoretical
tools, without facing our present difficulties in treating quantum gravity.
This is particularly interesting in view of the recent developments in
the understanding of non-perturbative aspects of supersymmetric theories
(for reviews, see {\it e.g.} refs.~\cite{intriligator96,shifman97,peskin97}).

Finally, as will be illustrated in this review,
gauge-mediated theories are quite predictive in the supersymmetric
mass spectrum and have distinctive phenomenological
features. Future collider experiments can put these predictions fully to test.

This review is organized as follows. In sect.~\ref{structure} we describe
the general structure of models with gauge-mediated supersymmetry breaking.
Their main phenomenological features are discussed in sect.~\ref{phenomenology}.
In sect.~\ref{tools} we briefly outline some theoretical tools needed
to study non-perturbative aspects of supersymmetry breaking. These techniques
are used in sect.~\ref{models}, where we discuss the present status of
models with dynamical supersymmetry breaking and gauge mediation.
Finally, in sect.~\ref{muprob} we consider mechanisms for generating the
Higgs mixing parameters $\mu$ and $B\mu$, in the context of gauge mediation.

\section{The Structure of Models with Gauge-Mediated Supersymmetry Breaking} 
\label{structure}
\setcounter{equation}{0}

In this section we will explain the main features of models with
gauge-mediated supersymmetry breaking and describe the structure of
the emerging soft-breaking terms. 

\subsection{The Building Blocks of Gauge Mediation}
\label{bblocks}

The first ingredient of these models is
an {\it observable sector}, which contains the usual
quarks, leptons, and two Higgs doublets, together with their supersymmetric
partners. 
Then the theory contains a sector responsible for supersymmetry breaking.
We will refer to it as the
{\it secluded sector}, 
to distinguish it from the hidden sector of
theories where supersymmetry breaking is mediated by gravity.
For the moment we will leave the secluded sector
unspecified, since it still lacks a standard description. For our
purposes, all we need to know is that the goldstino field overlaps with
a chiral superfield $X$, which acquires a VEV along the scalar and auxiliary
components
\beq
\langle X\rangle = M+\theta^2 F~.
\label{xvev}
\eeq
As will be discussed in detail in the following, the parameters
$M$ and $\sqrt{F}$, which are the fundamental mass scales in the theory,
can vary from several tens of TeV to almost
the GUT scale. We will start by considering
the simplest case, in which 
$X$ coincides with the goldstino superfield. However in sect.~\ref{variations} 
we will also
discuss secluded sectors in which the goldstino is a linear combination
of different fields.

Finally
the theory has a {\it messenger sector}, formed by some
new superfields that
transform under the gauge group as a real non-trivial
representation and couple at tree level with the goldstino superfield $X$.
This coupling generates a supersymmetric mass of order $M$ for the messenger
fields and mass-squared splittings inside the messenger supermultiplets 
of order $F$. This sector is also unknown and it is the main source of
model dependence. It is fairly reasonable to expect that the secluded and
messenger sectors have a common origin, and models in which these two sectors 
are unified will be discussed in sect.~\ref{models}. 

The simplest messenger sector is described by
$N_f$ flavours of chiral superfields
$\Phi_i$ and ${\bar \Phi}_i$ ($i=1,...,N_f$) transforming as the 
representation ${\bf r}+
{\bf {\overline r}}$ under the gauge group. 
In order to preserve gauge coupling-constant unification,
one usually requires that the messengers form complete GUT multiplets.
If this is the case, the presence of messenger fields at an intermediate
scale does not modify the value of $M_{GUT}$, but the inverse
gauge coupling strength at the unification scale $\alpha^{-1}_{GUT}$
receives an extra contribution 
\beq
\delta \alpha^{-1}_{GUT}
=-\frac{N}{2\pi}\ln \frac{M_{GUT}}{M}~,
\label{dagut}
\eeq
\beq
N=\sum_{i=1}^{N_f}n_i~.
\label{enne}
\eeq
Here $n_i$ is twice the Dynkin index of the gauge representation
${\bf r}$ with flavour index $i$, 
{\it e.g.} $n=1$ or 3 for an $SU(5)$ {\bf 5} or {\bf 10},
respectively. We will refer to $N$ as the {\it messenger index}, a quantity
that will play an important r\^ole in the phenomenology of gauge-mediated
theories. From eq.~(\ref{dagut}) we infer that perturbativity
of gauge interactions up to the scale $M_{GUT}$ implies
\beq
N\lappeq 150/\ln \frac{M_{GUT}}{M}~.
\label{uppern}
\eeq
If $M$ is as low as 100 TeV, then $N$ can be at most equal to 5. However,
this upper bound on $N$ is relaxed for larger values of $M$. For instance,
for $M=10^{10}$ GeV, eq.~(\ref{uppern}) shows that $N$ as large as 10 is
allowed.

It should also be noticed that, in the minimal $SU(5)$ model, the presence
of messenger states at scales of about 100 TeV is inconsistent with
proton-decay limits and with 
$b$--$\tau$ unification, unless large GUT threshold corrections are
added~\cite{carone96,bagger97}. Anyhow, these
constraints critically depend on the GUT model considered and, after all,
minimal $SU(5)$ is not a fully consistent model.

In the case under consideration the interaction between the
chiral messenger superfields $\Phi$ and ${\bar \Phi}$ and the goldstino
superfield $X$ is given by the superpotential term
\beq
W=\lambda_{ij} {\bar \Phi}_i X \Phi_j ~.
\label{phix}
\eeq
After replacing in eq.~(\ref{phix}) the $X$ VEV, see eq.~(\ref{xvev}), we
find that the spinor components of $\Phi$ and ${\bar \Phi}$ form 
Dirac fermions with masses $\lambda M$, while the scalar components have
a squared-mass matrix
\beq
\pmatrix{\Phi^\dagger & {\bar \Phi}}
\pmatrix{(\lambda M)^\dagger (\lambda M) & (\lambda F)^\dagger \cr 
(\lambda F) & (\lambda M)(\lambda M)^\dagger}
\pmatrix{\Phi \cr {\bar \Phi}^\dagger}~.
\eeq
Here we have dropped flavour indices and, with a standard abuse 
of notation, we have denoted the superfields and their scalar components
by the same symbols.
If there is a single field $X$, then the matrices $\lambda M$ and
$\lambda F$ can be simultaneously made 
diagonal and real, and the scalar messenger mass eigenvectors
are $(\Phi +{\bar \Phi}^\dagger )/\sqrt{2}$ and
$({\bar \Phi} - \Phi^\dagger )/\sqrt{2}$, with
squared-mass eigenvalues $(\lambda M)^2\pm (\lambda F)$. It is now
convenient to absorb the coupling constant $\lambda$ in the 
definition of $M$ and $F$, $\lambda_{ii}M\to M_i$, $\lambda_{ii}F\to F_i$.
In the following we will implicitly assume this redefinition.

\subsection{Soft Terms in the Observable Sector}
\label{soft}

\begin{figure}
\hspace{3cm}
\epsfig{figure=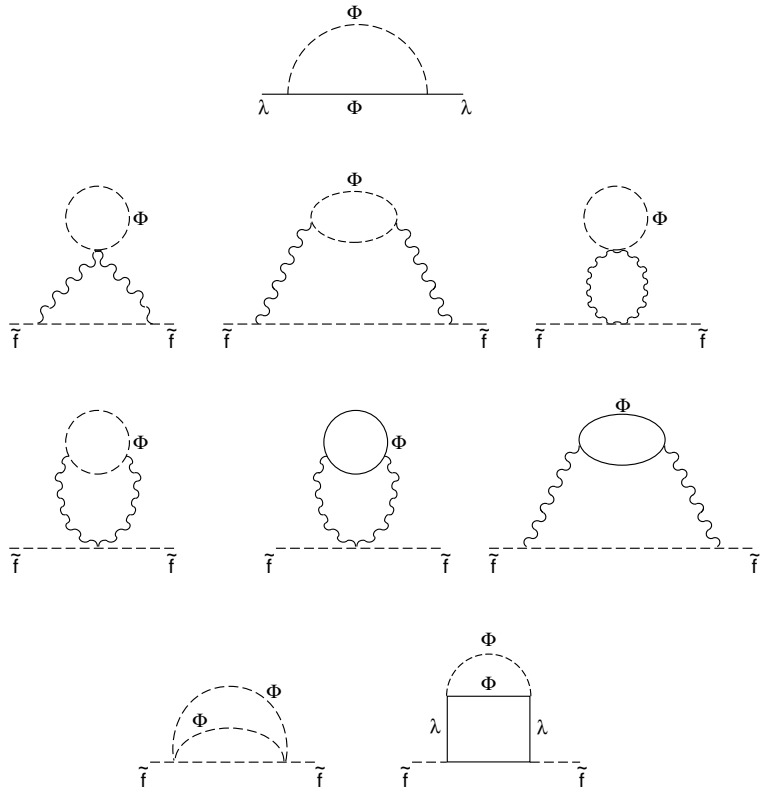,width=10cm}
\caption[]{\it
Feynman diagrams contributing to supersymmetry-breaking
gaugino ($\lambda$) and sfermion ($\tilde f$) masses. The scalar and fermionic
components of the messenger fields $\Phi$ are denoted by dashed and solid
lines, respectively; ordinary gauge bosons are denoted by wavy lines.
\label{gmfig1}}
\end{figure}

The mass
scale $\sqrt{F}$ is the measure of supersymmetry breaking in the messenger
sector. However,
we are of course
mainly interested in the amount of supersymmetry breaking in the
observable sector. 
Ordinary particle supermultiplets are degenerate at the tree level, since
they do not directly couple to $X$, but splittings arise at the quantum
level because of gauge interactions between observable and messenger
fields. While vector bosons and matter fermion masses are protected by
gauge invariance, gauginos, squarks, and sleptons can acquire masses
consistently with the gauge symmetry, once supersymmetry is broken.
Gaugino masses are generated at one loop, but squark and slepton masses
can only arise at two loops, since the exchange of both gauge and 
messenger particles is necessary. The corresponding Feynman diagrams are
drawn in fig.~\ref{gmfig1}. 
They were first evaluated in ref.~\cite{alvarez82}, 
where it
was shown that they lead to an acceptable particle spectrum with positive
scalar squared masses.

Higher-loop calculations in theories with many particles are quite involved.
Therefore it is often 
useful to employ a simple and systematic method~\cite{giudice97}
to extract the soft terms for observable fields in theories in which
supersymmetry breaking is communicated by renormalizable and perturbative
(but not necessarily gauge) interactions. Let us briefly discuss the
method.

The positivity of the messenger squared masses requires $F<M^2$. 
Moreover, as we will see in the following, in most
realistic cases it is appropriate to assume $F\ll M^2$. With
this assumption, in
the effective field theory below the messenger scale
$M$, supersymmetry breaking can be treated as a small effect.
This allows us to use a manifestly
supersymmetric formalism to keep track of supersymmetry-breaking 
effects as well.
First we define the effective theory, valid below
the mass scale $M$, by
integrating out the heavy messenger
fields. We are interested only in
renormalizable terms, which are not suppressed by powers of $M$.
Due to the non-renormalization of the superpotential,
all the relevant $M$ dependence of the low-energy effective theory
is contained in the gauge and matter wave-function 
renormalizations $S$ and
$Z_Q$. In the presence of a single mass scale $M$, this dependence is
logarithmic, and it can be calculated by solving the 
renormalization-group (RG) equations in the exact supersymmetric theory.
At the end, the mass parameter $M$ can be analytically continued in
superspace into a chiral superfield $X$. Holomorphy dictates the
correct analytic continuation $M\to X$ to be performed in $S$.
On the other hand,
the only substitution in $Z_Q$ that is consistent with the chiral
reparametrization
$X\to e^{i\phi}X$ is given by $M\to \sqrt{XX^\dagger}$. 
By replacing $X$ with its background value, given in eq.~(\ref{xvev}),
we can derive all the relevant supersymmetry-breaking effects.
We obtain that
the supersymmetry-breaking
gaugino masses, squark and slepton masses, and coefficients of the 
trilinear $A$-type terms, defined by
\beq
{\cal L}_{\rm soft}=-\frac{1}{2} \left( {\tilde M}_\lambda \lambda_g 
\lambda_g +{\rm h.c.}\right)
-m_{\tilde Q}^2 Q^\dagger Q -\left(
\sum_iA_iQ_i\partial_{Q_i}W(Q)+{\rm h.c.}\right) ~,
\label{lsodef}
\eeq
have the general
expressions~\cite{giudice97}
\beq
{\tilde M}_\lambda (t)=\left. -\frac{1}{2}\frac{\partial \ln S(X,t)}
{\partial \ln X}\right|_{X=M}~\frac{F}{M}~,
\label{gauginoc}
\eeq
\beq
m_{\tilde Q}^2(t)=\left. -\frac{\partial^2 \ln Z_Q(X,X^\dagger , t)}
{\partial \ln X ~\partial \ln X^\dagger}\right|_{X=M}
~\frac{FF^\dagger}{MM^\dagger}~,
\label{squakc}
\eeq
\beq
A_i(t)=\left. \frac{\partial \ln Z_{Q_i}(X,X^\dagger,t)}
{\partial \ln X}\right|_{X=M}~\frac{F}{M}~.
\label{aac}
\eeq
Here $t=\ln M^2/Q^2$ and $Q$ is the low-energy scale at which the soft terms
are defined.
The gauge and chiral wave-function renormalizations
$S$ and $Z_Q$ are obtained by integrating the well-known RG differential
equations in the supersymmetric limit. 

Explicit formulae for the soft terms in gauge mediation will be given in
the next section. Here we just want to remark that, by inserting the
one-loop approximations for $S$ and $Z_Q$ in eqs.~(\ref{gauginoc})--(\ref{aac}),
we can directly obtain both the leading-loop messenger contribution to
the soft terms and the resummation of the leading logarithms in the
evolution
from the messenger scale $M$ to the low-energy scale $Q$.
It is also interesting to notice that the gauge-mediated
two-loop sfermion masses are obtained by integrating the
one-loop RG equation. This is because, at the leading order, $Z_Q$ is a
function of $\alpha_s \log XX^\dagger$. The two derivatives in 
eq.~(\ref{squakc})
are responsible for a leading contribution to soft masses of order $\alpha_s^2$.
In the ordinary Feynman-diagram approach, soft scalar masses arise
from finite two-loop graphs. Following the method of ref.~\cite{giudice97},
they can be reconstructed
from the behaviour of the wave-function renormalization far away from threshold,
derived from the RG equation. This method is computationally much simpler
than the explicit evaluation of the relevant Feynman diagrams.

\subsection{Physical Mass Spectrum}
\label{spectrum}

We present here the complete formulae, in the leading-log approximation,
for the soft terms generated by
chiral messengers communicating via gauge
interactions. The formulae in the next-to-leading order approximation
are given in the appendix.
These results can be derived either by directly evaluating the
relevant Feynman diagrams or by applying the method described in the
previous section.
The supersymmetry-breaking gaugino masses are
\beq
{\tilde M}_{\lambda_r} (t)=k_r\frac{\alpha_r(t)}{4\pi}~\Lambda_G~~~~~
(r=1,2,3)~,
\label{gaugau}
\eeq
\beq
\Lambda_G=\sum_{i=1}^{N_f} n_i \frac{F_i}{M_i}~\left[ 
1+{\cal O}(F_i^2/M_i^4)\right] ~,
\label{lambdagp}
\eeq
where $k_1=5/3$, $k_2=k_3=1$, and the gauge coupling constants are normalized
such that $k_r \alpha_r$ ($r=1,2,3$) are all equal at the GUT scale.
In the simple case in which there is a single $X$ superfield, and 
the ratio 
$F_i/M_i$ is therefore independent of the flavour index $i$, 
eq.~(\ref{lambdagp}) becomes
\beq
\Lambda_G=N \frac{F}{M}~\left[ 
1+{\cal O}(F^2/M^4)\right] ~,
\label{lambdag}
\eeq
where the messenger index $N$ is defined in eq.~(\ref{enne}).
Next-to-leading corrections in $\alpha_s$ can affect the gluino mass prediction
by a significant amount, especially 
at small values of $N$ and $M$~\cite{arkani98,picariello98}.

Neglecting Yukawa-coupling effects, the supersymmetry-breaking scalar
masses at the scale $Q$ ($t=\ln M^2/Q^2$) are
\beq
m_{\tilde f}^2(t)=2\sum_{r=1}^3C_r^{\tilde f}k_r~ \frac{\alpha_r^2(0
)}{(4\pi)^2}~\left[ \Lambda_S^2+h_r\Lambda_G^2
\right] ~,
\label{scalmass}
\eeq
\beq
h_r=\frac{k_r}{b_r}\left[ 1-\frac{\alpha_r^2(t)}{\alpha_r^2(0)}\right]~,
\eeq
\beq
\alpha_r (t)=\alpha_r (0)\left[ 1+\frac{\alpha_r (0)}{4\pi}b_r t\right]^{-1}~.
\eeq
Here $\alpha_r (0)$ are the gauge coupling constants at the messenger scale
$M$ and, in the case of a single $X$ superfield,
\beq
\Lambda_S^2= N\frac{F^2}{M^2}
~\left[ 1+{\cal O}(F^2/M^4)\right]~.
\label{lambdas}
\eeq
In the general case of non-universal values of $F_i$ and $M_i$ 
({\it i.e.} with several superfields $X$ overlapping with the goldstino),
eq.~(\ref{lambdas}) is no longer valid, and the ratio
$\Lambda_G^2/\Lambda_S^2$ is not just given by the messenger index $N$.
However, in this case,
we can treat $\Lambda_S$ and $\Lambda_G$ as free parameters, instead of
$N$ and $F/M$.
In eq.~(\ref{scalmass})
$C_r^{\tilde f}$ is the quadratic Casimir of the $\tilde f$ particle,
$C=\frac{N^2-1}{2N}$ for the $N$-dimensional representation of $SU(N)$, and
$C=Y^2=(Q-T_3)^2$ for the $U(1)$ factor. Here $b_r$ are the $\beta$-function
coefficients
\beq
b_3=-3,~~~b_2=1,~~~b_1=11~.
\eeq
The physical scalar squared 
mass is obtained by adding to eq.~(\ref{scalmass}) the
$D$-term contribution $M_Z^2\cos 2\beta(T_3^{\tilde f}-Q^{\tilde 
f}\sin^2\theta_W)$.

\begin{figure}
\hspace{3cm}
\epsfig{figure=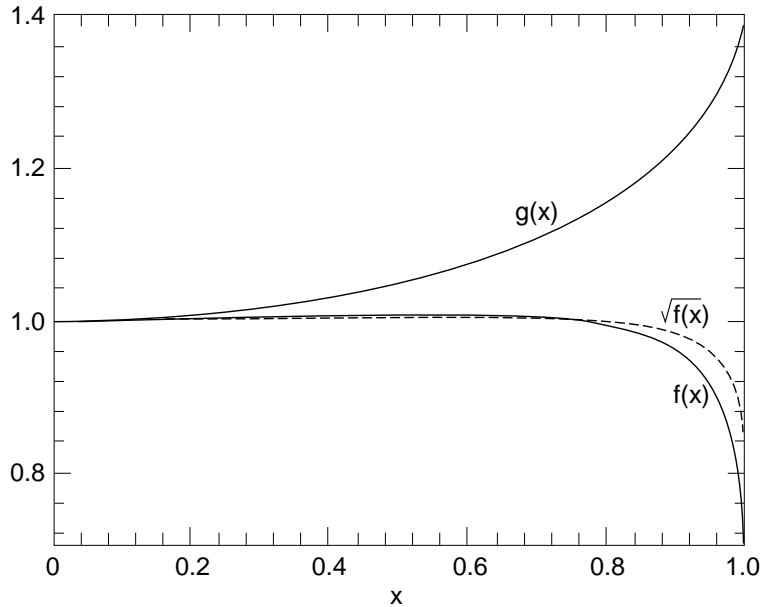,width=10cm}
\caption[]{\it
The functions 
$ f(x)$, $ g(x)$, and $\sqrt{f(x)}$.
\label{gmfig2}}
\end{figure}

In deriving these mass formulae, we have assumed
$F\ll M^2$. Although this approximation is in most cases justified since,
for $F>M^2$, a scalar messenger particle has negative squared mass,
it is not appropriate whenever
$F\simeq M^2$. Notice that in this case 
the method
of ref.~\cite{giudice97}, discussed in sect.~\ref{soft}, fails because
all higher covariant-derivative operators 
contribute to supersymmetry-breaking terms at the same order.
An explicit Feynman diagram calculation is now necessary. This has been
performed in refs.~\cite{dimopoulos96c,martin97}, with the result 
\beq
\Lambda_G= \sum_{i=1}^{N_f}n_i \frac{F_i}{M_i} g(F_i/M_i^2)~,
\label{lgnlo}
\eeq
\bea
g(x)&=&\frac{1}{x^2}\left[ (1+x)\ln (1+x)\right]  + (x\to -x)\nonumber \\
&=& 1+\frac{x^2}{6}+\frac{x^4}{15}+\frac{x^6}{28}+{\cal O} (x^8) ~,
\eea
\beq
\Lambda_S^2= \sum_{i=1}^{N_f}n_i \frac{F_i^2}{M_i^2} f(F_i/M_i^2) ~,
\label{lsnlo}
\eeq
\bea
f(x)&=&\frac{1+x}{x^2}\left[ \ln (1+x) -2{\rm Li}_2\left( \frac{x}{1+x}\right)
+\frac{1}{2}{\rm Li}_2\left( \frac{2x}{1+x}\right) \right]
+ (x\to -x) 
\nonumber \\
&=& 1+\frac{x^2}{36}-\frac{11}{450}x^4-\frac{319}{11760}x^6+{\cal O} (x^8) 
~.
\eea
The functions $g(x)$ and $f(x)$, which represent the corrections with
respect to the $F\ll M^2$ case, are shown in fig.~\ref{gmfig2};
$g(x)$ is always larger
than 1, reaching the maximum value $g(1)=2\ln 2=1.4$; $f(x)$ is
approximately equal to 1, within 1{\%}, for $x<0.8$, 
and reaches the minimum value
$f(1)=(2\ln 2) (1+\ln 2)-\pi^2/6=0.7$.

The leading contributions to $\Lambda_{G,S}$, in an expansion in $F/M^2$,
are universal for the different GUT components
of messengers. This is true because these leading contributions are
proportional to 
the ratio $F/M$, which  is a universal quantity, independent of
the coupling constant $\lambda$
between the messenger superfields and
the goldstino superfield $X$, see eq.~(\ref{phix}). 
On the other hand, the argument
of the functions $g$ and $f$ in 
eqs.~(\ref{lgnlo}) and (\ref{lsnlo}) is $F/M^2$, a quantity which is not
equal for messengers with different SM quantum numbers. 
Therefore, depending
on the particular choice of $\lambda$, the inclusion of the correction functions
$g$ and $f$ can be relevant only for part of the messenger multiplet.
This effect gives an uncertainty on the mass prediction which is typically
small for squarks and sleptons, but could be 
up to 40\% for gauginos. Notice that, if the messengers form degenerate
multiplets at the GUT scale, then $F/M^2$ is larger for a messenger
weak doublet than for a messenger colour triplet. Therefore, the enhancement
of the gaugino mass due to the correction function $g$ is 
larger for the $W$-ino than for the gluino. This will result in
an apparent violation of 
gaugino-mass unification.

The scalar masses have been computed neglecting Yukawa-coupling effects.
This is not a good approximation for the
stop system, which is described by
the $2\times
2$ mass matrix 
\beq
m^2_{\tilde t}=\pmatrix{
{ m^2_{{\tilde Q}_L}}+m^2_t+
(\frac{1}{2}-\frac{2}{3}\sin^2
\theta_W)\cos 2\beta M_Z^2
          & m_t( A_t-\mu\cot\beta)\cr
m_t( A_t-\mu\cot\beta)&
  m^2_{{\tilde t}_R}+m^2_t+\frac{2}{3}
\sin^2\theta_W\cos 2\beta M_Z^2  }~.
\label{stop}
\eeq
The third-generation supersymmetry-breaking masses
$m_{{\tilde Q}_L}^2$ and $m_{{\tilde t}_R}^2$ 
at the scale $Q$ ($t=\ln M^2/Q^2$)
are given by
\bea
m_{{\tilde Q}_L}^2(t)=&&2\sum_{r=1}^3k_r~ \frac{\alpha_r^2(0
)}{(4\pi)^2}~\left[ (C_r^{{\tilde Q}_L}-\frac{K_t}{12}a_r)
\Lambda_S^2+h_rC_r^{{\tilde Q}_L}\Lambda_G^2\right]\nonumber \\ &&
-\frac{K_t}{6}(H_1-K_tH_2^2)\Lambda_G^2~,
\eea
\bea
m_{{\tilde t}_R}^2(t)=&&2\sum_{r=1}^3k_r~ \frac{\alpha_r^2(0
)}{(4\pi)^2}~\left[ (C_r^{{\tilde t}_R}-\frac{K_t}{6}a_r)
\Lambda_S^2+h_rC_r^{{\tilde t}_R}\Lambda_G^2\right]\nonumber \\ &&
-\frac{K_t}{3}(H_1-K_tH_2^2)\Lambda_G^2~,
\eea
Here $a_r=2(C_r^{{\tilde Q}_L}+C_r^{{\tilde t}_R}+C_r^{H_2})=(13/9,3,16/3)$,
and
\beq
E=\prod_{r=1}^3\left[ \frac{\alpha_r(0)}{\alpha_r(t)}\right]^{\frac{a_r}{b_r}},
~~~~~F=\int_0^t dt~E~,
\eeq
\bea
&&H_1=\frac{\alpha_Xt_X}{4\pi}H_3\frac{E}{F}\left(\frac{t}{t_X}-1\right)+
\left(\frac{\alpha_Xt_X}{4\pi}\right)^2\left\{ \left[
\frac{E}{F}\left(\frac{t}{t_X}-1\right)+ 
\frac{1}{F}\right]
\sum_{r=1}^3 a_r\frac{\alpha_r(0)}{4\pi}
\right. 
\nonumber \\
&& \left. 
+\sum_{r=1}^3 a_rb_r\frac{\alpha^2_r(0)}{(4\pi)^2}+
\left[ \sum_{r=1}^3 a_r\frac{\alpha_r(0)}{4\pi}\right]^2\right\}~,
\eea
\beq
H_2=\frac{\alpha_Xt_X}{4\pi}\left[
\frac{E}{F}\left(\frac{t}{t_X}-1\right)+\frac{1}{F}-\frac{1}{t_X}+
\sum_{r=1}^3 a_r\frac{\alpha_r(0)}{4\pi}\right] ~,
\eeq
\beq
H_3=\sum_{r=1}^3\frac{a_r}{b_r}k_r\frac{\alpha_r(0)-\alpha_r(t)}{4\pi}~,
\eeq
\beq
\alpha_X=\frac{k_sb_r-k_rb_s}{b_r\alpha_s^{-1}(0)-b_s\alpha_r^{-1}(0)}~~~~
\forall r\neq s~~~~r,s=1,2,3~,
\eeq
\beq
t_X=\frac{4\pi\left[ k_r\alpha_s^{-1}(0)-k_s\alpha_r^{-1}(0)\right]}{
k_sb_r-k_rb_s}~~~~
\forall r\neq s~~~~r,s=1,2,3~,
\eeq
\beq
K_t=\frac{6F}{E}\frac{h_t^2(t)}{(4\pi)^2}~.
\eeq
Here $h_t$ is the running top-quark Yukawa coupling, related to the
top-quark mass by $h_t=(2\sqrt{2}G_F)^{1/2}m_t/\sin\beta$. 
Since $K_t$ is equal to the squared top-Yukawa coupling in units of the
infrared fixed-point value, the condition that $h_t$ does not reach
the Landau pole before the messenger scale $M$ implies $K_t<1$.
The definitions of 
$\alpha_X$ and $t_X$ are independent of the specific indices $r$ and $s$,
because we are assuming gauge-coupling unification. They correspond to
the unification coupling constant and the unification mass scale
($t_X=\ln M^2/M_X^2$) in a fictitious theory with no messengers.
The assumption of gauge-coupling unification allows us to solve
integrals which, in general, cannot be expressed by elementary functions.

There is no
one-loop messenger contribution to the supersymmetry-breaking
trilinear terms. However $A$ terms are 
generated in the leading-log approximation
by the RG evolution proportional to gaugino masses. In particular, the stop
$A$ term appearing in eq.~(\ref{stop}) is given by
\beq
A_t(t)=(H_3-K_tH_2)\Lambda_G~.
\label{at}
\eeq
The $A$ terms corresponding to the other Yukawa couplings
can be obtained from eq.~(\ref{at}) by setting $K_t=0$
and replacing $a_r$ with the coefficients of the corresponding interaction.

The above formulae assume that Yukawa couplings other
than the one for the top quark are negligible.
Therefore, they are not valid for very large $\tan\beta$ because, as
$\tan\beta$ approaches $m_t/m_b$, bottom-quark Yukawa effects become
significant.

The parameter $\mu$, which appears in eq.~(\ref{stop}), is the Higgs mixing
mass, defined by the superpotential interaction $\mu H_1H_2$. 
This mass term breaks the Peccei--Quinn
symmetry, and therefore it cannot be generated by gauge interactions alone.
The origin of the $\mu$ parameter
is still one of the main problematic issues~\cite{dvali96} 
in theories with gauge mediation, as we will discuss 
in sect.~\ref{muprob}. Here we will simply assume that a hard mass
parameter $\mu$ exists at the messenger scale $M$. Its value at the low-energy
scale is then 
given by
\beq
\mu(t)=\mu(0)~(1-K_t)^{1/4}\prod_{r=1}^3\left[ \frac{\alpha_r(0)}{\alpha_r
(t)}\right]^{\frac{a^\mu_r}{2b_r}}~,
\eeq
where $a_r^\mu =2(C_r^{H_1}+C_r^{H_2}) =(1,3,0)$.

Analogously, a supersymmetry-breaking Higgs mixing mass $B\mu$  will
appear in the scalar potential.
The running value of the coefficient $B$ at low energy is
\beq
B(t)=B(0)+(H_4-\frac{K_t}{2}H_2)\Lambda_G+\delta B^{(NLO)}(t)~,
\eeq
\beq
H_4=\sum_{r=1}^3\frac{a_r^\mu}{b_r}k_r\frac{\alpha_r(0)-\alpha_r(t)}{4\pi}~,
\eeq
\beq
\delta B^{(NLO)}(t)=-\frac{\alpha_s^2(t)h_t^2(t)}{8\pi^4}t~\Lambda_G 
+\sum_{r=1}^3a_r^\mu k_r\frac{\alpha_r^2(t)}{(4\pi)^2}\Lambda_G
~.
\label{delb}
\eeq
Here $\delta B^{(NLO)}$ contains terms suppressed by $1/t$ with respect
to the leading radiative contributions. The total contribution from
this class of terms comes from ultraviolet and infrared
theshold effects and some RG
evolution, and it is scheme-independent. The result in eq.~(\ref{delb})
contains only ultraviolet thresholds and running in $\overline{\rm 
DR}$, and the
scheme dependence is cancelled by the effective-potential 
contribution~\cite{rattazzi96,giudice97}.
The term in eq.~(\ref{delb}) becomes important for relatively low values of
$M$, where $t$ is small.

Notice that the low-energy value of $B$
is non-vanishing even if $B(0)=0$. This has motivated~\cite{babu96} 
several phenomenological
studies~\cite{dimopoulos97a,bagger97a,rattazzi96,borzumati97,gabrielli97} 
of the very predictive case $B(0)=0$, in which the low-energy value
of $B$ is determined in terms of the gaugino 
mass. This case is also interesting
because the theory is automatically free from any dangerous new CP-violating 
parameter,
aside from the SM ones, {\it i.e.} the usual Kobayashi--Maskawa phase and
$\Theta_{QCD}$.

The Higgs soft mass parameters $m_{H_{1,2}}^{2}$ are 
given by
\beq
m_{H_1}^{2}(t)=2\sum_{r=1}^3C_r^{H_1}k_r~ \frac{\alpha^2(0
)}{(4\pi)^2}~\left[ \Lambda_S^2+h_r\Lambda_G^2
\right] +\delta m_{H_1}^{(1-loop)2}(t)~,
\eeq
\bea
&&m_{H_2}^{2}(t)=2\sum_{r=1}^3k_r~ \frac{\alpha^2(0
)}{(4\pi)^2}~\left[ (C_r^{H_2}-\frac{K_t}{4}a_r)
\Lambda_S^2+k_rh_rC_r^{H_2}\Lambda_G^2\right] \nonumber \\ &&
-\frac{K_t}{2}(H_1-K_tH_2^2)\Lambda_G^2+
\delta m_{H_2}^{(NLO)2}(t)+\delta m_{H_2}^{(1-loop)2}(t)~.
\label{h2eff}
\eea
Here, as it is customary, we added to the running $\overline{\rm DR}$
masses  the contributions $\delta m_{H_{1,2}}^{(1-loop)2}$
from infrared thresholds. These
effects can be computed using the one-loop effective potential
$V_{1-loop}$ in $\overline{\rm DR}$~\cite{gamberini90,barger94}:
\beq
\delta m_{H_{1,2}}^{(1-loop)2}
=\left. \frac{1}{2H_{1,2}}\frac{\partial V_{1-loop}}{\partial
H_{1,2}} \right|_{H_{1,2}=\langle H_{1,2}\rangle} ~.
\eeq
Finally $\delta m_{H_2}^{(NLO)2}$ contains the constant ({\it i.e.}
not log-enhanced) term in the 
$\alpha_s^2h_t^2$ three-loop correction to the value of $m_{H_2}^2$ at
the messenger scale $M$. This has been calculated in ref.~\cite{giudice97}
in $\overline{\rm DR}$ (see also the appendix)
\beq
\delta 
m_{H_2}^{(NLO)2}(t)=-\frac{\alpha_s^2(t)h_t^2(t)}{8\pi^4}\Lambda_S^2.
\label{hnlo}
\eeq

Although the parameter $\mu$ is not determined by the underlying theory,
we can compute it from the condition of correct electroweak breaking:
\beq
\mu^2=-\frac{M_Z^2}{2}+\frac{1}{\tan^2\beta -1}\left(m_{H_1}^{2}
-\tan^2\beta ~m_{H_2}^{2}\right)~,
\label{ew1}
\eeq
\beq
B=\frac{\sin 2\beta}{2\mu}\left( m_{H_1}^{2}+m_{H_2}^{2}+2\mu^2
\right) ~.
\label{ew2}
\eeq
The second equation determines the low-energy value of $B$ in terms of
$\tan \beta$. If the condition $B(0)=0$ is imposed, 
eq.~(\ref{ew2}) predicts the value
of $\tan\beta$, which turns out to be rather large. 
In eqs.~(\ref{ew1}) and (\ref{ew2}),
the parameters $\mu$ and $B$ are calculated at the scale $Q^2$. For
physical applications, we can choose $Q^2=m_{\tilde t}^2$. This  minimizes
most low-energy threshold effects,
because below the stop mass scale the RG running is negligible.
Therefore
the term in eq.~(\ref{hnlo}) is significant only 
when the
logarithm of $M/{\tilde m}_t$ is not very large. The full next-to-leading
expressions of the Higgs mass parameters can be found in the appendix. 

It is important to stress that, if $\mu$ and $B$ are generated radiatively
through some new interactions, it is very plausible~\cite{dvali96} that
also $m_{H_{1,2}}^2$ receive extra corrections. These corrections then
modify the values of $\mu$ and $B$ extracted from the electroweak-breaking
conditions.

\subsection{Properties of the Mass Spectrum}
\label{properties}

The most important feature of 
the gauge-mediated mass spectrum is of course flavour universality,
which is guaranteed by the symmetry of gauge interactions.
This property is maintained if gravity-mediated contributions
do not reintroduce large flavour violations, {\it e.g.} 
in the $K^0$--$\bar K^0$ system
or in $\mu \to e \gamma$ transitions. We therefore require that
gravity-mediated contributions
do not account
for more than, say, one 
per mille of the soft squared masses.
Since gravity generates soft terms with
typical size $F/M_P$, the flavour criterion gives a rough upper bound
on the messenger mass scale
\beq
M\lappeq \frac{1}{10^{\frac{3}{2}}}
\frac{\alpha}{4\pi}M_P\sim 10^{15}~{\rm GeV}~,
\label{flapro}
\eeq
where $M_P=(8\pi G_N)^{-1/2}=2.4\times 10^{18}$ GeV is the reduced Planck 
mass. 

An attractive feature of the gauge-mediated mass formulae is that
gaugino mass terms are generated at one loop, and (positive) squark 
mass terms are generated at two loops. Since fermion and boson bilinears
have different canonical dimensions, all supersymmetry-breaking
mass parameters have the same scaling property ${\tilde m}\sim (\alpha/\pi)
F/M$. However, as we will see in sect.~\ref{muprob}, it is not automatic
for the $\mu$ and $B$
parameters to satisfy this property.

Another important property of the mass formulae
derived here is that they allow a high degree of predictivity.
The whole supersymmetric spectrum is determined by 
the effective supersymmetry-breaking scale $\Lambda =F/M$, 
the messenger index $N$, the messenger
mass $M$, and $\tan\beta$. The parameter $\mu$
can be 
fitted from the electroweak-breaking condition (up to a phase ambiguity), 
if we assume the absence of any 
non-minimal contribution to $m_{H_{1,2}}^2$.

\begin{figure}
\hspace{3cm}
\epsfig{figure=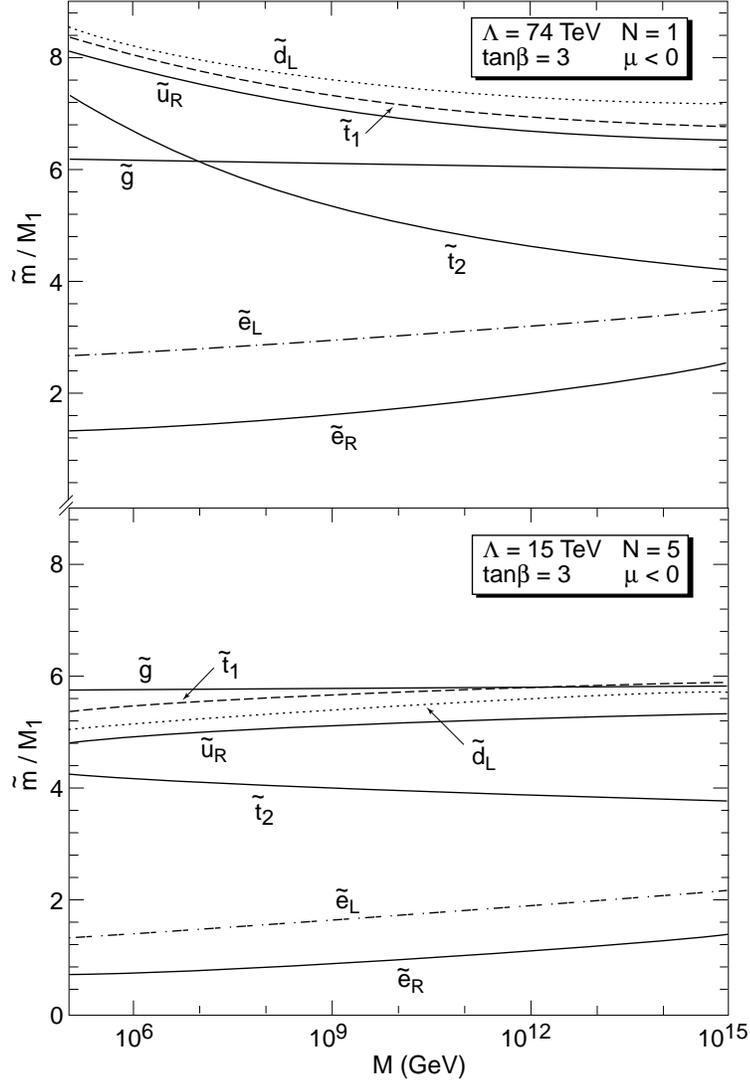,width=10cm}
\caption[]{\it
Different supersymmetric particle masses in units of the $B$-ino mass $M_1$,
as a function of the messenger mass $M$. The choice of parameters is indicated,
and in both cases it corresponds to a $B$-ino mass of 100 GeV.
\label{gmfig3}}
\end{figure}

\begin{figure}
\hspace{3cm}
\epsfig{figure=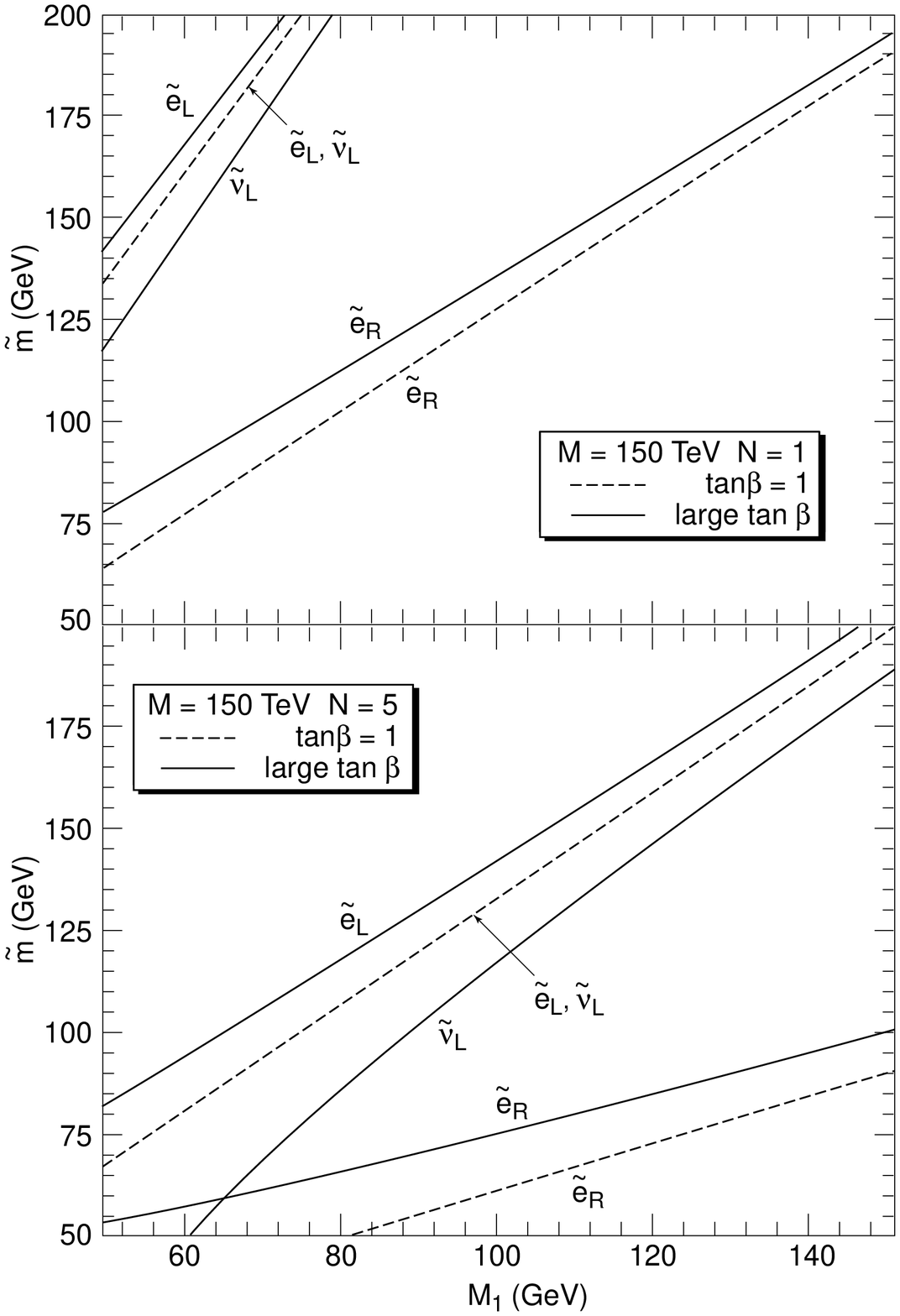,width=10cm}
\caption[]{\it
Slepton masses as a function of the $B$-ino mass $M_1$, for the indicated
choice of parameters. 
Dashed lines correspond to the case of vanishing $D$
terms ($\tan \beta =1$), and the solid lines to the case of maximal
$D$ terms ($\cos 2\beta =-1$).
\label{gmfig4}}
\end{figure}

Comprehensive analyses of the supersymmetric particle masses in
gauge-mediated models have been presented in 
refs.~\cite{dimopoulos97a,bagger97a}.
Examples of the mass spectra are shown in fig.~\ref{gmfig3}. 
We have plotted all sparticle masses in units of the $B$-ino mass
$M_1(t)=5\alpha_1(t)N\Lambda /(12\pi)$ evaluated at the energy scale
$Q=M_1$.
Since $M_1$ is essentially independent of $M$ and $\tan  \beta$, this
choice is equivalent to normalizing the spectrum to $\Lambda$. Moreover,
the mass
ratios plotted in fig.~\ref{gmfig3} (apart from $m_{\tilde t}/M_1$)
are fairly independent of $\Lambda$ and $\tan \beta$, unless
$\Lambda$ (or, equivalently $M_1$) is rather small. In this case,
$D$-term contributions can affect slepton masses, although they are
never very
important for squarks. Figure~\ref{gmfig4} illustrates this regime, showing
how $D$-terms modify the slepton spectrum, and how they can even drive 
$m_{{\tilde \nu}_L}$ lighter than $m_{{\tilde e}_R}$, for large
values of $N$ and small values of $\Lambda$ and $M$. 
This happens, however, only in a very marginal region of parameters, at the
border with the experimental limit on slepton masses.
The stop-mass
eigenvalues are quite sensitive to $\tan \beta$ because of left--right
mixing effects, and the lightest stop mass decreases for smaller
$\tan\beta$. However, the stop squared masses always remain positive, unless
one chooses values of $\tan \beta$ so small that the top-quark Yukawa
Landau pole is reached around the scale $M$.
The case $\mu <0$ gives a larger mixing than the case $\mu >0$, because the
$\mu$ contribution in eq.~(\ref{stop}) 
adds up to the $A_t$ term, which is always
positive, in our notation.

Notice the large 
hierarchy between the strongly interacting and weakly interacting particles.
For small values of $M$, this hierarchy is determined by the ratio
$\alpha_3(0)/\alpha_2(0)$. For larger values of $M$, where the RG running
is important, the value of $N$ plays a crucial r\^ole. If $N$ is small, the
most important effect is the decrease of the ratio
$\alpha_3(0)/\alpha_2(0)$ as $M$ increases, and squarks become lighter and 
closer in mass to sleptons 
(see upper frame of
fig.~\ref{gmfig3}). On the other hand, if $N$ is large, gaugino masses
increase and their effects in the RG evolution dominate the squark and
slepton masses. In this case, squarks become heavier as $M$ grows
(see lower frame of
fig.~\ref{gmfig3}). The ratio between gaugino and scalar masses is
determined by the messenger index $N$. Increasing $N$, scalars become
lighter and the ratio $m_{{\tilde e}_R}/M_1$ can be smaller than 1.

\begin{figure}
\hspace{3cm}
\epsfig{figure=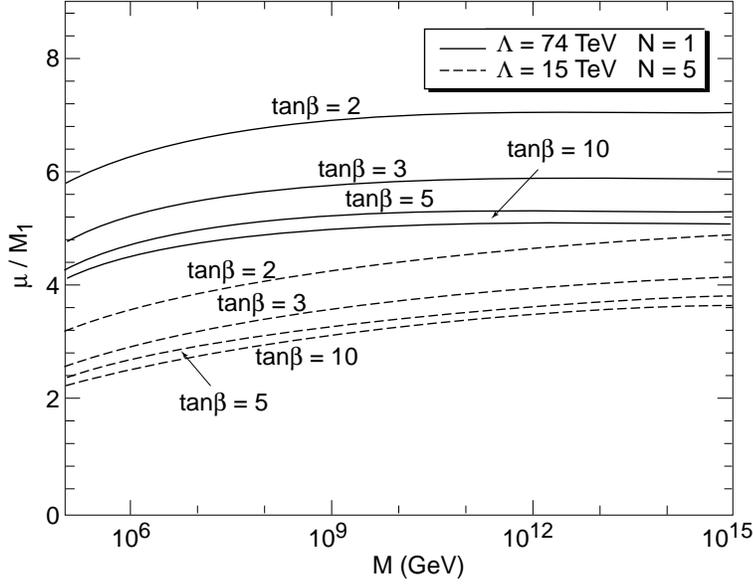,width=10cm}
\caption[]{\it
The ratio between $\mu$ and
the $B$-ino mass $M_1$,
as a function of the messenger mass $M$. The choice of parameters is indicated,
and in both cases it corresponds to a $B$-ino mass of 100 GeV.
\label{gmfig5}}
\end{figure}

Another remarkable success of the gauge-mediated mass spectrum is that 
eqs.~(\ref{ew1}) and (\ref{ew2}) have an acceptable solution~\cite{dine93}, 
and therefore radiative
electroweak-symmetry breaking~\cite{ibanez92} can be achieved.
This happens because the
negative contribution to $m_{H_2}^2$ is proportional to the large stop
mass. 
Indeed for small $M$, eq.~(\ref{h2eff}), with the addition of the
low-energy threshold corrections,  can be approximated by
\beq
m_{H_2}^{2}=m_{{\tilde e}_L}^2
-\frac{3h_t^2}{4\pi^2}m_{\tilde t}^2\left( \ln \frac{M}{
m_{\tilde t}}+\frac{3}{2}\right)~.
\eeq
Therefore, even for moderate values of $M$, the coefficient in front
of the logarithm is large enough 
to drive $m_{H_2}^{2}$ negative and trigger
electroweak-symmetry breaking. 

If the condition of electroweak breaking is imposed, then the value of
$\mu$ is determined by eq.~(\ref{ew1}), with the result
shown in fig.~\ref{gmfig5}.
Rather large values of $\mu$ are required to compensate the stop contribution
in eq.~(\ref{ew1}). 
The ratio $\mu /M_1$ decreases for large $N$, but in this regime $M_1$ is
no longer
the smallest supersymmetric
particle mass in the spectrum. This rather large value of
$\mu$, originating from 
the hierarchy between strongly and weakly interacting particles is at the
basis of fairly
stringent upper bounds on the supersymmetric particle masses obtained
{}from the naturalness criterion. For instance, using the 
criterion~\cite{barbieri88} that no independent parameter 
should be correlated by more than 10\%, it is 
found~\cite{ciafaloni97,bhattacharyya97,agashe97} that 
the right-handed selectron has to be lighter than 100 GeV or less, depending on
the parameters. This limit does not significantly change
for large $N$, since the ratio
$\mu/m_{{\tilde e}_R}$ only slowly increases as $N$ grows.
It was also suggested in ref.~\cite{agashe97} that 
messengers belonging to split GUT multiplets may alleviate the fine-tuning
problem present in the minimal model.

\begin{figure}
\hspace{3cm}
\epsfig{figure=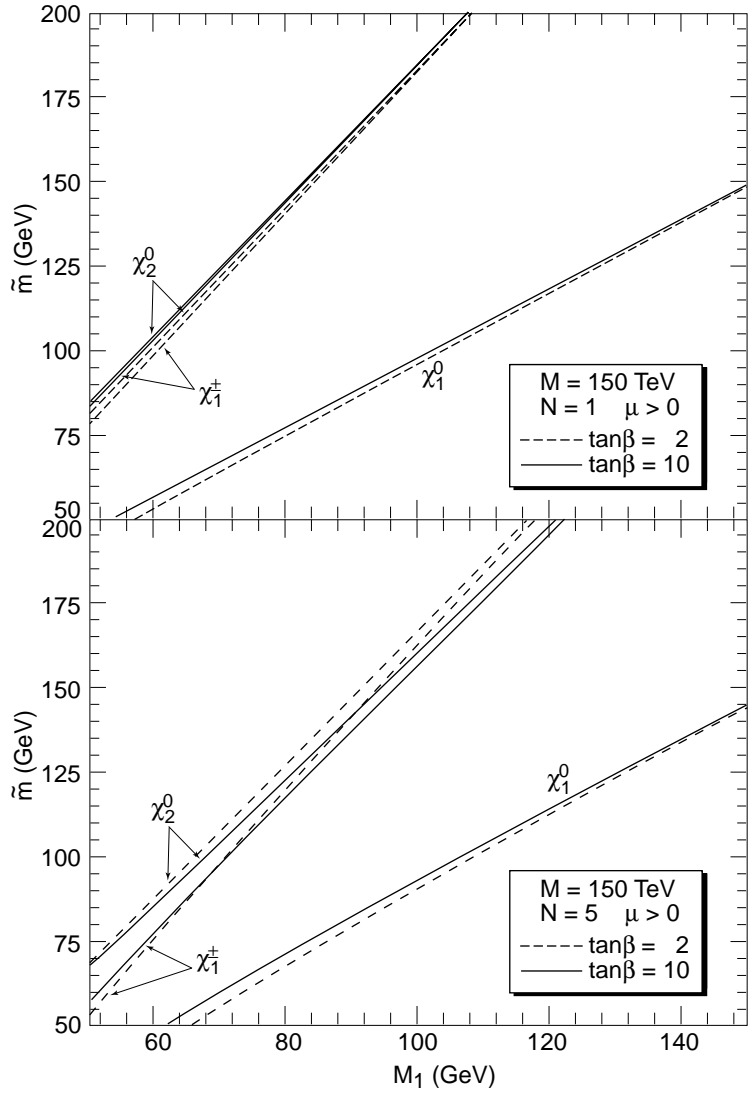,width=10cm}
\caption[]{\it
The masses of the lightest neutralino ($\chi^0_1$) and chargino ($\chi^\pm_1$)
and the next-to-lightest neutralino ($\chi^0_2$) as a function 
of the $B$-ino mass $M_1$, for $\mu >0$ and the indicated
choice of parameters. 
\label{gmfig6}}
\end{figure}

\begin{figure}
\hspace{3cm}
\epsfig{figure=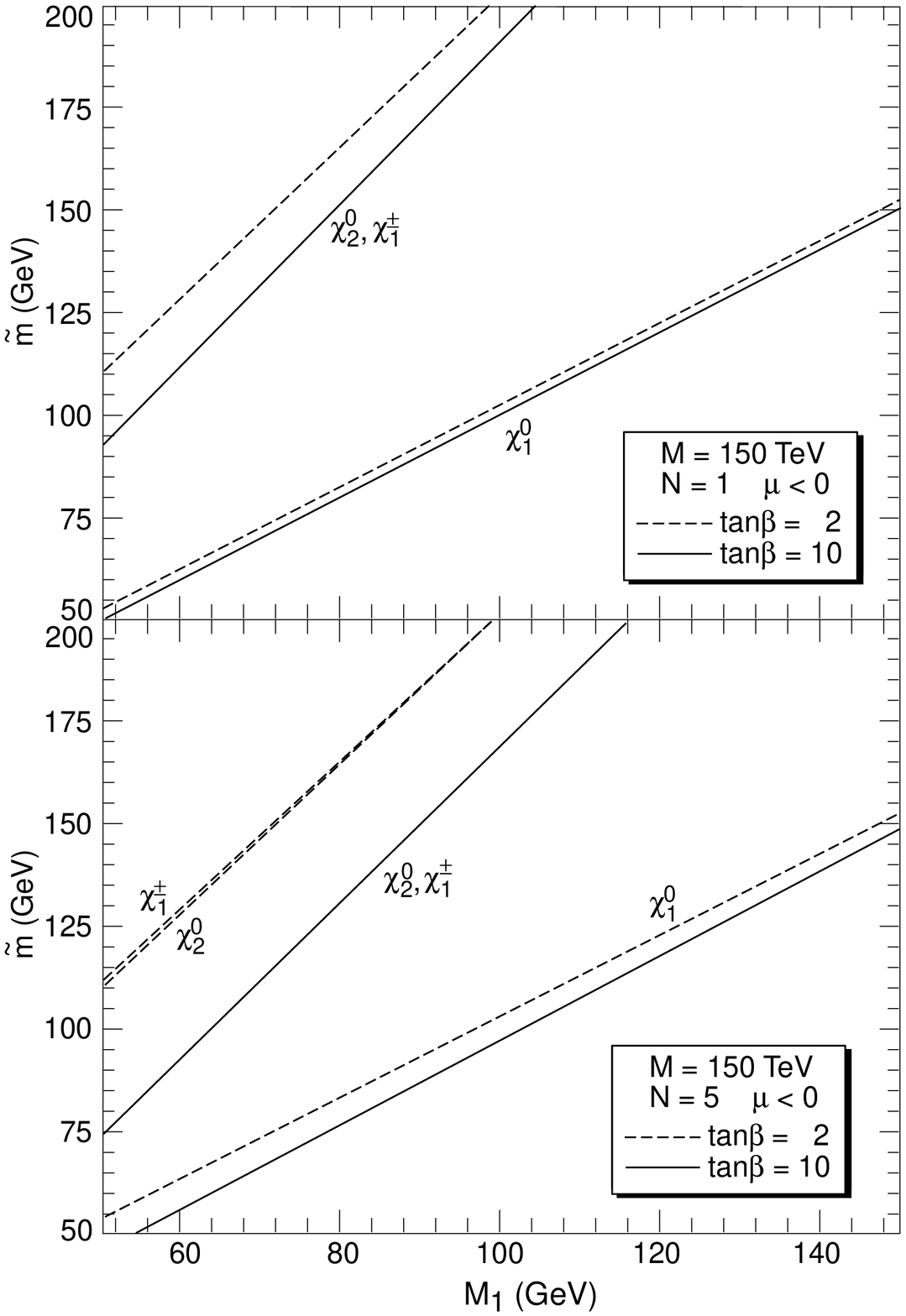,width=10cm}
\caption[]{\it
The masses of the lightest neutralino ($\chi^0_1$) and chargino ($\chi^\pm_1$)
and the next-to-lightest neutralino ($\chi^0_2$) as a function 
of the $B$-ino mass $M_1$, for $\mu <0$ and the indicated
choice of parameters. 
\label{gmfig7}}
\end{figure}

With the value of
$\mu$ extracted from the electroweak-breaking condition, we can now
compute the spectrum of neutralinos and charginos. Since $\mu$ typically
turns out
to be larger than $M_Z$, with a good approximation we can treat electroweak
breaking as a perturbation and obtain~\cite{martin93}
\beq
m_{\chi^0_1}=M_1-\frac{M_Z^2\sin^2\theta_W (M_1+\mu \sin 2\beta )
}{\mu^2-M_1^2}~,
\label{chi01}
\eeq
\beq
m_{\chi^\pm_1}=m_{\chi^0_2}=
M_2-\frac{M_W^2(M_2+\mu \sin 2\beta )}{\mu^2-M_2^2}~,
\eeq
\beq
m_{\chi^\pm_2}=\mu+\frac{M_W^2(\mu +M_2 \sin 2\beta )}{\mu^2-M_2^2}~,
\eeq
\beq
m_{\chi^0_3}=\mu+\frac{M_Z^2(1+\sin 2\beta )(\mu-M_1\cos^2\theta_W
-M_2\sin^2\theta_W )
}{2(\mu -M_1)(\mu -M_2)}~,
\eeq
\beq
m_{\chi^0_4}=-\mu-\frac{M_Z^2(1-\sin 2\beta )(\mu+M_1\cos^2\theta_W
+M_2\sin^2\theta_W )
}{2(\mu +M_1)(\mu +M_2)}~.
\label{chi04}
\eeq
The lightest neutralino is mainly $B$-ino, $\chi_1^\pm$ and $\chi_2^0$
form a degenerate $W$-ino weak
triplet, and the nearly higgsinos $\chi_2^\pm$
and $\chi_{3,4}^0$ have masses roughly equal to $\mu$. Figures~\ref{gmfig6}
and \ref{gmfig7} 
show the light
neutralino and chargino masses in the regime where the approximation in
eqs.~(\ref{chi01})--(\ref{chi04}) is not reliable, namely small $\Lambda$.
Actually, for $N=1$ this approximation is still good in almost all the range
of $\Lambda$, since $\mu$ is still large enough to decouple higgsinos
{}from gauginos. Significant deviations
appear when $N=5$, as shown in figs.~\ref{gmfig6} and \ref{gmfig7},
because both $\mu$ and $M_2$ can both become comparable to $M_Z$.

\begin{figure}
\hspace{3cm}
\epsfig{figure=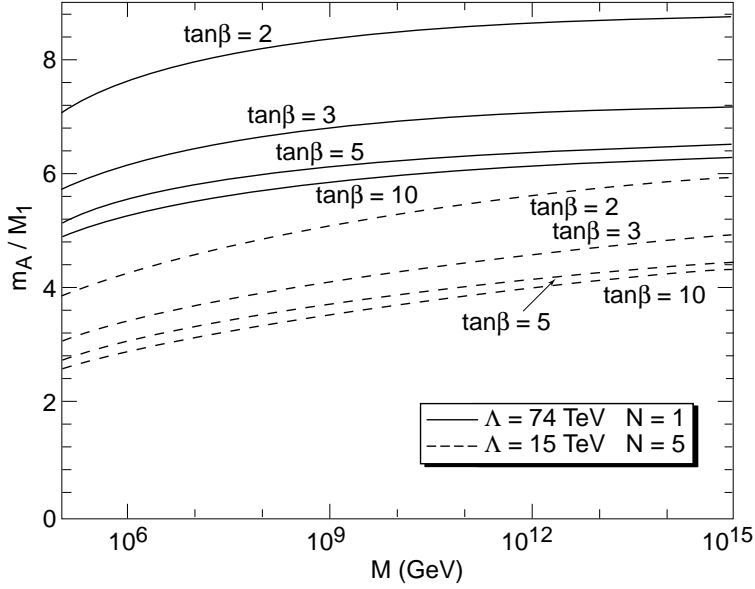,width=10cm}
\caption[]{\it
The ratio between CP-odd Higgs boson mass $m_A$ and
the $B$-ino mass $M_1$,
as a function of the messenger mass $M$. The choice of parameters is indicated,
and in both cases it corresponds to a $B$-ino mass of 100 GeV.
\label{gmfig8}}
\end{figure}

The CP-odd neutral Higgs mass is given by
\beq
m_A^2=\frac{2B\mu}{\sin 2 \beta}\simeq \frac{\mu^2}{\sin^2 \beta}
\eeq
and shown in fig.~\ref{gmfig8}. 
Since $m_A$ is rather large, the theory at low energies is
approximately described by a single Higgs doublet, and
the Higgs phenomenology at 
LEP2 should resemble the SM case with a light Higgs. 
The charged Higgs boson mass is given by
$m_{H^\pm}^2=M_W^2+m_A^2$.

\begin{figure}
\hspace{3cm}
\epsfig{figure=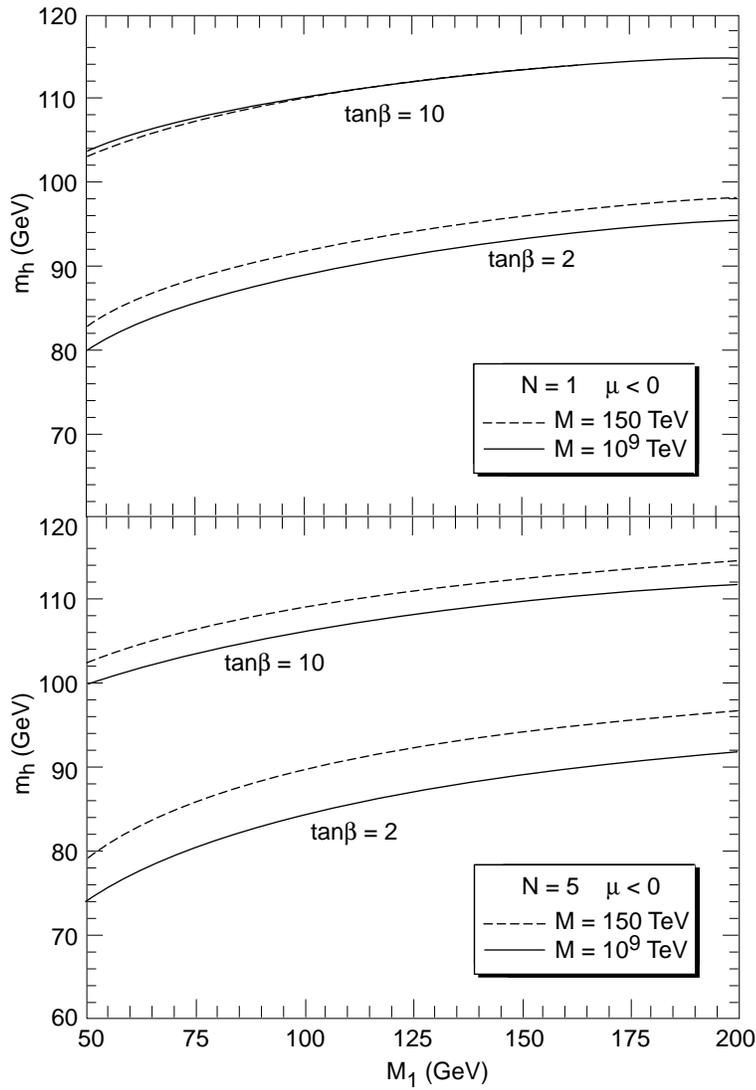,width=10cm}
\caption[]{\it
The mass of the lightest CP-even Higgs boson 
as a function of the $B$-ino mass $M_1$ for the indicated
choice of parameters. 
\label{gmfig9}}
\end{figure}

The lightest CP-even Higgs boson mass $m_h$
receives important radiative corrections proportional to the top quark
Yukawa coupling~\cite{okada91,ellis91,haber91}, and it can be
predicted in terms of the fundamental
parameters $\Lambda$, $M$, $N$, and $\tan 
\beta$~\cite{dimopoulos97a,bagger97a,riotto96}. Including the
leading two-loop effects, in the limit of large $m_A$,
the Higgs boson mass can
be approximated as~\cite{carena95}
\bea
&&m_h^2=M_Z^2\cos^2 2\beta \left( 1-\frac{3\sqrt{2}}{4\pi^2}G_Fm_t^2~t_S
\right) +
\nonumber \\ && 
\frac{3\sqrt{2}}{2\pi^2}G_Fm_t^4\left\{ \frac{{\tilde X}_t}{2}+t_S+
\frac{1}{16\pi^2}\left[ 3\sqrt{2}G_F m_t^2-32\pi \alpha_s(M_t)\right]
({\tilde X}_t +t_S)t_S\right\} ~,
\eea
\beq
t_S=\ln \left( \frac{M_S^2}{M_t^2}\right) ~~~~M_S^2=m_{{\tilde t}_1}
m_{{\tilde t}_2}~,
\eeq
\beq
{\tilde X}_t =\frac{2(A_t-\mu \cot \beta )^2}{M_S^2}\left[ 1
-\frac{(A_t-\mu \cot \beta )^2}{12M_S^2}\right] ~,
\eeq
\beq
m_t=\frac{M_t}{1+\frac{4}{3\pi}\alpha_s(M_t)} ~.
\eeq
Here $m_t$, $M_t$ are the running and $\overline{\rm MS}$ pole top-quark masses,
respectively.
The Higgs boson mass is shown in fig.~\ref{gmfig9}
in the case $\mu<0$, where stop mixing effects are maximized. The largest
values of $m_h$ are obtained at large $\tan\beta$. For large $\tan\beta$
and small $N$, the stop mixing is negligible and the value of $m_h$
shown in fig.~\ref{gmfig9} corresponds to the value obtained in the
usual supersymmetric model with vanishing stop mixing~\cite{carena95}.
As we increase $M$, $A_t$ grows and the typical squark mass decreases;
the two effects roughly compensate each other and the value of $m_h$ is
not modified. At larger $N$, the squarks are lighter and the value of
$m_h$ is smaller. The effect of increasing $M$ is now more important
since $A_t$ receives a large renormalization proportional to the
gaugino mass. In any case,
a considerable fraction of the parameter
space is within the reach of LEP2. If the Higgs boson is not discovered,
the LEP2 search will provide an extremely severe bound on the model.

In conclusion, gauge-mediated models have a very predictive mass spectrum.
If supersymmetry is discovered, it seems very likely that these predictions
can be used to distinguish these models from the generic spectrum of
gravity mediation. In the presence of unification relations, 
gaugino masses are
the same in both scenarios, but  scalar masses are different.
Even if $M$ is close to the GUT scale, the initial condition of
scalar masses is not unified. This is because the gauge bosons are already
split and do not form a complete GUT multiplet. From eq.~(\ref{scalmass}),
we find
\beq
\frac{2}{7}\frac{m_{{\tilde Q}_L}^2(0)}{m_{{\tilde E}_R}^2(0)}=
\frac{3}{8}\frac{m_{{\tilde U}_R}^2(0)}{m_{{\tilde E}_R}^2(0)}=
\frac{3}{7}\frac{m_{{\tilde D}_R}^2(0)}{m_{{\tilde E}_R}^2(0)}=
\frac{2}{3}\frac{m_{{\tilde L}_L}^2(0)}{m_{{\tilde E}_R}^2(0)}=
1~.
\eeq
However, for large $N$ 
and large $M$ the mass spectrum is ``gaugino-dominated" and
the signal of gauge mediation is more difficult to be distinguished. 
Studies
of the comparison
between the mass spectrum of gauge-mediated and gravity-mediated models
can be found in refs.~\cite{dimopoulos97a,bagger97a,carena97a,strumia97}.

\subsection{Variations of the Minimal Model}
\label{variations}

We have seen that the mass formulae obtained in
gauge mediation are very predictive.
It is then important to assess how much these predictions depend on
the assumptions of the minimal model. Moreover, some of the variations we
consider below have important virtues and they could be the natural outcome of
simple fundamental models.

As mentioned before, the prediction for the $\mu$ parameter extracted
{}from the electroweak-breaking conditions is modified by possible
new contributions
to $m_{H_{1,2}}^2(0)$, which arise in theories where all mass parameters are
generated radiatively~\cite{dvali96}, 
see sect.~\ref{muprob}. However,
the parameter $\mu$ is not significantly modified as long as the new
contributions to the Higgs mass parameters are smaller than the usual
gauge-mediated contributions, since the two effects add in quadratures.
The effects of  arbitrary new contributions
to $m_{H_{1,2}}^2(0)$ in the prediction of $\mu$ and of the mass spectrum have
been discussed in ref.~\cite{dimopoulos97a}.

In our study of the minimal gauge-mediated model, we have assumed that 
the goldstino resides in a single chiral superfield, and therefore the
matrices $F$ and $M$ are proportional to each other. If this is not the
case, in the basis in which $M$ is diagonal and real, $F$ is still a
generic matrix in flavour space. Now
supersymmetry-breaking scalar masses can receive a new contribution
{}from an induced 
Fayet--Iliopoulos term~\cite{fayet74}. The one-loop contribution,
proportional to the sfermion ($\tilde f$) and messenger ($\Phi$)
hypercharge $Y$, is~\cite{dimopoulos97}
\beq
\Delta m_{\tilde f}^2=\frac{\alpha_1}{4\pi}Y_{\tilde f}~ {\rm Tr} Y_\Phi ~
 { \Lambda}^ 2_D ~,
\label{1loo}
\eeq
where the trace is taken over the complete GUT messenger representation.
At leading order in $F/M^2$, ${ \Lambda}^2_D$ is
independent of the specific component of the GUT multiplet $\Phi$:
\beq
{\Lambda}^2_D = \frac{1}{2}
\sum_{i,j=1}^{N_f} \frac{|F_{ji}|^2-|F_{ij}|^2}{M_i^2}f_D\left(
\frac{M_j^2}{M_i^2}\right)~,
\label{ld}
\eeq
\beq
f_D(x)=\frac{2}{(1-x)}+\frac{(1+x)}{(1-x)^2}\ln x~.
\label{fx}
\eeq
Equation (\ref{1loo}) vanishes either if the messengers form complete
GUT multiplets (${\rm Tr} Y_\Phi$=0) or if the messenger sector is invariant
under a ``messenger parity", defined in ref.~\cite{dimopoulos97}, 
which guarantees
$\Lambda_D =0$, since one can choose a basis in which $M$ is diagonal
and real and $F$ Hermitian.
This cancellation is welcome, since the one-loop contribution
to scalar masses in eq.~(\ref{1loo}) is proportional to the sfermion 
hypercharge and therefore
is not positive-definite. Were it to dominate over the
ordinary two-loop gauge contribution, it would lead to an
unacceptable mass spectrum. However one can consider theories
with messengers embedded in
a GUT, but with no ``messenger parity". Now,
even if
${\rm Tr} Y_\Phi =0$,
a hypercharge
$D$-term contribution
to scalar masses is generated either at higher order in $F/M^2$
($m_{\tilde f}^2\sim [\alpha_1/\pi ]F^4/M^6$) or at two loops 
($m_{\tilde f}^2\sim [\alpha_1\alpha_3/\pi^2]F^2/M^2$).
Complete expressions
for these contributions can be found in ref.~\cite{dimopoulos97}. 
Numerically, these
effects can be important for slepton masses and for the Higgs mass parameters,
but are insignificant for squark masses.

Another possible variation is described by renormalizable superpotential
couplings between messenger and matter superfields. These couplings 
are usually assumed to vanish, because generically they break flavour
invariance, reintroducing the flavour problem, whose solution was
the primary motivation for gauge mediation. However they
are allowed by gauge invariance and their consequences have been 
analysed in ref.~\cite{dine97}. It has been found~\cite{dvali96,dine97} 
that messenger-matter 
couplings do not induce soft masses at one loop, at leading order
in $F/M^2$, if the goldstino is contained in a single superfield $X$. 
However they contribute to scalar masses at two loops, and 
(in contrast with gauge mediation) give
a non-vanishing $A$-type trilinear interaction at the one-loop level.
Complete formulae for the supersymmetry-breaking terms induced by
messenger-matter couplings can be found in ref.~\cite{giudice97}.

So far we have considered the case in which the messenger
particles belong to chiral superfields. However, it is also possible that
gauge supermultiplets behave as messengers. We are envisaging a situation
in which the supersymmetry-breaking VEV is also responsible for spontaneous
breaking of some gauge symmetry containing the SM 
as an unbroken subgroup. 
The vector bosons corresponding to the broken generators, together with
their supersymmetric partners,
receive masses proportional to $M$. However supersymmetry-breaking
effects proportional to $F$ split the gauge supermultiplets at tree level
and consequently soft terms for observable fields are generated 
by quantum effects. These terms have been computed in ref.~\cite{giudice97}.
The gaugino masses have the same expressions as in eq.~(\ref{gaugau}),
once the messenger index $N$, see eq.~(\ref{enne}) is identified with
\beq
N=nN_f-2(C_G-C_H)~.
\label{ennevec}
\eeq
Here $nN_f$ is the usual chiral messenger contribution.
The second term in eq.~(\ref{ennevec}) describes the new contribution from
gauge messengers and it is defined as follows.
We have assumed that the
scalar component VEV of the goldstino superfield $X$
spontaneously breaks the gauge group $G\to H$, in such a way that the
gauge coupling constant is continuous at the threshold. $C_G$ and $C_H$ are the
quadratic Casimirs of the adjoint representations of $G$ and $H$ (equal to $N$
for an $SU(N)$ group). Equation (\ref{ennevec}) shows that
the pure vector-messenger effect is to
reduce the value of $N$, and also to allow negative values of the total $N$.
Complete formulae for the gauge-messenger contribution to soft terms
can be found in ref.~\cite{giudice97}. It turns out that
scalar squared masses receive new negative contributions, which can
destabilize the ordinary gauge group and which pose a serious constraint
to construct acceptable models. It is also interesting
to notice that gauge messengers lead to one-loop contributions to $A$ terms,
in contrast to the case of chiral messengers.

A conceivable possibility, although not realized in any complete model
built so far, is that not only the secluded sector, but the
messengers themselves feel some strongly interacting gauge force. In this
case, the perturbative calculations presented in sect.~\ref{spectrum} 
are no longer 
valid. Exchange of many new ``gluons" along the messenger lines give
unsuppressed effects. Using naive dimensional analysis to count the
factors of $4\pi$, it has been shown~\cite{luty97,cohen97} 
that the soft terms induced
by strongly interacting messengers have the same form as in the case
of weakly interacting messengers, up to (not calculable)
factors of order 1. Furthermore, the leading-order expressions for
the gaugino masses are not corrected by the new strong interactions, since
they are protected by a ``screening theorem"~\cite{arkani98}.

The authors of ref.~\cite{dvali97} have investigated the
interesting 
possibility of identifying the messengers with the GUT multiplets containing
ordinary Higgs bosons.
Unfortunately, this case requires a considerable amount of fine-tuning 
to maintain the weak scale much smaller than the messenger mass scale.
The case in which the messengers do not form complete GUT multiplets
has been studied in ref.~\cite{martin97}. 
Finally, the papers in refs.~\cite{dobrescu97,mohapatra97,chacko97} 
discuss models in which the soft terms
are generated by new gauge interactions beyond those of the ordinary $\gws$.

\section{Phenomenology of Models with Gauge-Mediated Supersymmetry Breaking}
\label{phenomenology}
\setcounter{equation}{0}

\subsection{The Lightest Supersymmetric Particle: the Gravitino}
\label{gravitino}

As a result of the spontaneous breakdown of supersymmetry, the physical
spectrum contains a massless spin-1/2 fermion, the goldstino.
When the globally supersymmetric
theory is coupled to gravity and promoted to a locally supersymmetric
theory, the goldstino provides the longitudinal modes of the spin-3/2
partner of the graviton, the gravitino. As a result of this super-Higgs
mechanism, the gravitino acquires a supersymmetry-breaking 
mass
which, under the condition of vanishing cosmological constant, is given
by~\cite{volkov73a,deser77}
\beq
m_{3/2}=\frac{F_0}{\sqrt{3}M_P}~.
\eeq
Here $M_P=(8\pi G_N)^{-1/2}=2.4\times 10^{18}$ GeV is the reduced Planck 
mass. We denote by $F_0$ the total contribution of 
the supersymmetry-breaking VEV of the auxiliary fields, normalized in such
a way that the vacuum energy of the globally supersymmetric theory is
$V=F_0^2$. Thus $F_0$ does not coincide with the definition of $F$,
which appears in the sparticle masses through $\Lambda =F/M$. While
$F_0$ is the fundamental scale of supersymmetry breaking, $F$ is the
scale of supersymmetry breaking felt by the messenger particles, {\it i.e.}
the mass splitting inside their supermultiplets. The ratio $k\equiv
F/F_0$ depends
on how supersymmetry breaking is communicated to the messengers. If the
communication occurs via a direct interaction, this ratio is just given
by a coupling constant, like the parameter $\lambda$ in the case
described by eq.~(\ref{phix}).
It can be argued that this coupling should be smaller than 1, by requiring
perturbativity up to the GUT scale~\cite{ambrosanio97}. 
If the communication occurs
radiatively, then $k$ is given by some loop factor, and therefore
it is much smaller than 1. We thus rewrite the gravitino mass as
\beq
m_{3/2}=\frac{F}{k\sqrt{3}M_P}=\frac{1}{k}\left( \frac{\sqrt{F}}{100~
{\rm TeV}}\right)^2 2.4~{\rm eV}~,
\label{gravmass}
\eeq
where the model-dependent coefficient $k$ is such that
$k<1$, and possibly $k\ll 1$.

In gauge-mediated models, the gravitino is the lightest supersymmetric
particle (LSP) for any relevant value of $F$. Indeed, as argued in 
sect.~\ref{properties},
a safe solution to the flavour problem requires that gravity-mediated
contributions to the sparticle spectrum should be much smaller than
gauge-mediated contributions. Since $m_{3/2}$ is exactly the measure of
gravity-mediated effects, it is indeed the solution of the flavour problem
in gauge mediation, see eq.~(\ref{flapro}), 
which implies that the gravitino is the LSP.

If $R$ parity is conserved, all supersymmetric particles follow
decay chains that lead to
gravitinos. In order to compute the decay rate we need to know the 
interaction Lagrangian at lowest order in the gravitino field. Since,
for $\sqrt{F}\ll M_P$, the dominant gravitino
interactions come from its spin-1/2 component, the interaction
Lagrangian can be computed
in the limit of global
supersymmetry. In the presence of spontaneous supersymmetry breaking, the
supercurrent $J_Q^\mu$ satisfies the equation
\beq
\partial_\mu J_Q^\mu = -F_0\gamma^\mu \partial_\mu {\tilde G}~,
\eeq
which is the equivalent of the usual current algebra relation for soft pions. 
This can be viewed
as the goldstino equation of motion. The corresponding
interaction Lagrangian is
\beq
{\cal L}=-\frac{1}{F_0}J_Q^\mu \partial_\mu {\tilde G}~.
\label{golds}
\eeq
This shows how the goldstino interacts with derivative couplings suppressed
by $1/F_0$, which are
typically more important than the gravitational couplings
suppressed by powers of
$1/M_P$. Since we are interested in the goldstino 
couplings to field bilinears, we can replace $J_Q^\mu$ in eq.~(\ref{golds})
by its
expression for free fields and obtain
\beq
{\cal L}=-\frac{k}{F}\left(
{\bar \psi}_L\gamma^\mu \gamma^\nu \partial_\nu \phi -
\frac{i}{4\sqrt{2}}{\bar \lambda}^a\gamma^\mu \sigma^{\nu \rho } F^a_{\nu \rho}
\right) \partial_\mu {\tilde G}
+{\rm h.c.}
\label{golds2}
\eeq
Here $\phi$ and $\psi$ are the scalar and fermionic components of a 
generic chiral supermultiplet and $\lambda^a$ and $F^a_{\mu \nu}$ 
are the Majorana
spinor and gauge field strength belonging to a vector supermultiplet.

The Lagrangian in eq.~(\ref{golds2}) can also
be derived by using the supersymmetric
analogue of the equivalence 
theorem~\cite{fayet77,fayet79a,casalbuoni88,casalbuoni89}. 
This theorem allows the
replacement, in high-energy processes, of an external-state
gravitino field with $\sqrt{2/3}~\partial_\mu {\tilde G}/
m_{3/2}$. If this substitution is done in the
relevant
supergravity Lagrangian, one indeed recovers eq.~(\ref{golds2}). For this
reason, it is perfectly adequate for our purposes
to describe the LSP in terms of the goldstino
properties. The only r\^ole played by gravity is to generate the LSP mass
given in eq.~(\ref{gravmass}).
 
For on-shell particles, by using the equations of motion,
the goldstino Lagrangian in eq.~(\ref{golds2}) can be written as a Yukawa
interaction with chiral fields and a magnetic moment-like interaction with
gauge particles, 
\beq
{\cal L}=\frac{k}{F}\left[ (m_\psi^2-m_\phi^2)
 {\bar \psi}_L \phi +
\frac{M_\lambda}{4\sqrt{2}}{\bar \lambda}^a\sigma^{\nu \rho } F^a_{\nu \rho}
\right] {\tilde G}
+{\rm h.c.}
\label{golds3}
\eeq
Notice that both goldstino interactions are proportional to 
the mass splitting inside the supermultiplet and inversely 
proportional to the scale of supersymmetry
breaking~\cite{fayet77,fayet79a}.

For our purposes the interaction Lagrangians in eqs.~(\ref{golds2}) and
(\ref{golds3}) are sufficient to describe the relevant processes involving
the goldstino. Derivations of the complete effective Lagrangian 
involving multi-goldstino
interactions can be found in 
refs.~\cite{clark96,gherghetta97,brignole97,luty97a,brignole97a,clark97}.

\subsection{The Next-to-Lightest Supersymmetric Particle}
\label{nlsp}

The next-to-lightest supersymmetric particle (NLSP)
plays an important r\^ole in the phenomenology of
gauge mediation. Assuming $R$-parity conservation, we expect that
all supersymmetric particles will promptly decay into cascades leading to the
NLSP, with the NLSP then decaying into the gravitino via $1/F$ 
interactions. Therefore the nature of the NLSP determines the signatures
in collider experiments and some cosmological properties of gauge mediation.
{}From the study in sect.~\ref{properties}, 
we have seen that the NLSP can be, depending
on the parameter choice, the neutralino, the stau, or, in a very restricted
region of parameters, the sneutrino. Let us review these possibilities.

The NLSP neutralino has, in most cases, a dominant $B$-ino component, 
since the ratio $\mu/M_1$ is typically larger than 1. 
An exception occurs
for
large $N$ and small $M$ and $\Lambda$,
as the NLSP neutralino is a non-trivial superposition of
different states. Another exception is given by
models where the physics generating the
parameter $\mu$ also contributes to supersymmetry-breaking
Higgs masses.

{}From eq.~(\ref{golds3}) we find the following
NLSP $\chi_1^0$ decay 
rates~\cite{cabibbo81,ambrosanio96a,dimopoulos96b,bagger97a}:
\beq
\Gamma (\chi^0_1\to \gamma {\tilde G})=\frac{k^2~{\kappa}_\gamma 
m_{\chi_1^0}^5}{16\pi F^2}= k^2~{\kappa}_\gamma
\left( \frac{m_{\chi_1^0}}{100~{\rm GeV}}\right)^5
\left( \frac{100~{\rm TeV}}{\sqrt{F}}\right)^{4}~2\times 10^{-3}~{\rm eV}~,
\label{chidec}
\eeq
\beq
\frac{\Gamma (\chi^0_1\to Z^0 {\tilde G})}{\Gamma (\chi^0_1\to \gamma 
{\tilde G})}
=\frac{{\kappa}_Z}{{\kappa}_\gamma}\left( 1-\frac{M_Z^2}{m_{\chi_1^0}^2}
\right)^4~,
\eeq 
\beq
\frac{\Gamma (\chi^0_1\to h^0 {\tilde G})}{\Gamma (\chi^0_1\to \gamma 
{\tilde G})}
=\frac{{\kappa}_h}{{\kappa}_\gamma}\left( 1-\frac{m_h^2}{m_{\chi_1^0}^2}
\right)^4~,
\eeq 
\beq
{\kappa}_\gamma =|N_{11}\cos \theta_W +N_{12}\sin\theta_W|^2~,
\label{kapgam}
\eeq
\beq
{\kappa}_Z =|N_{11}\sin \theta_W -N_{12}\cos\theta_W|^2+\frac{1}{2}
|N_{13}\cos \beta -N_{14}\sin\beta |^2~,
\eeq
\beq
{\kappa}_h =
|N_{13}\sin \alpha -N_{14}\cos\alpha |^2~.
\eeq
Here $N_{1i}$ are the $\chi^0_1$ components in standard notation
(see, {\it e.g.} ref.~\cite{haber85}) 
and $\tan 2\alpha =\tan 2\beta (m_A^2+M_Z^2)/
(m_A^2-M_Z^2)$. The decay mode into photon and goldstino is very likely
to dominate. Even if the decay modes into the $Z^0$ boson or the neutral
Higgs boson are kinematically allowed, they are quite suppressed by the
$\beta^8$ phase-space factor.
Moreover, if $\chi^0_1$ is mainly $B$-ino, ${{\kappa}_Z}/{{\kappa}_\gamma}=
0.3$ and ${{\kappa}_h}/{{\kappa}_\gamma}$ is negligible. Complete expressions
for the neutralino decay rates into three-body final states can be found in 
ref.~\cite{bagger97a}.

For roughly
\beq
N> \frac{66}{5(13\xi_1 -2)}~,~~~\xi_1\equiv \frac{\alpha_1^2(M_1)}{
\alpha_1^2(M)}=\left[
1+\frac{11}{4\pi}\alpha_1(M_1)\ln \frac{M_1^2}{M^2}\right]^2~,
\label{nlim}
\eeq
the right-handed slepton is lighter than $\chi^0_1$. The transition occurs
at moderate values of $N$ for small $M$ ({\it e.g.} $N=1.7$ for $M=10^5$
GeV), but requires large values of $N$ for large $M$ ({\it e.g.} $N>5$ 
for $M=10^{12}$ GeV). This is the result of the significant renormalization
of $m_{{\tilde E}_R}$ proportional to the gaugino mass, in the regime of large
$N$ and $M$.

Among the three generations of right-handed sleptons, ${\tilde \tau}_R$ is
the lightest because of mixing effects proportional to $m_\tau$ in the
stau mass matrix:
\beq
m^2_{\tilde \tau}=\pmatrix{
{ m^2_{{\tilde L}_L}}+m^2_\tau -
(\frac{1}{2}-\sin^2
\theta_W)\cos 2\beta M_Z^2
          & m_\tau( A_\tau -\mu\tan\beta)\cr
m_\tau ( A_\tau -\mu\tan\beta)&
  m^2_{{\tilde E}_R}+m^2_\tau -
\sin^2\theta_W\cos 2\beta M_Z^2  }~.
\label{stau}
\eeq
As the mixing grows with $\tan\beta$, see eq.~(\ref{stau}), the stau can become
the NLSP for values of $N$ quite smaller than shown in eq.~(\ref{nlim}). In
particular, for extremely large values of $\tan\beta$, the determinant of
the matrix in eq.~(\ref{stau}) can become negative and destabilize
the electromagnetically neutral vacuum~\cite{babu96,rattazzi96}.

The NLSP stau decay rate is
\beq
\Gamma ({\tilde \tau}\to \tau {\tilde G})=\frac{k^2
m_{\tilde \tau}^5}{16\pi F^2}= k^2
\left( \frac{m_{\tilde \tau}}{100~{\rm GeV}}\right)^5
\left( \frac{100~{\rm TeV}}{\sqrt{F}}\right)^{4}~2\times 10^{-3}~{\rm eV}~.
\label{staudec}
\eeq
If the mixing is small, the lightest stau is mainly right-handed. 
In this case, ${\tilde e}_R$ and ${\tilde \mu}_R$ are so close in mass to
${\tilde \tau}$ that their three-body decays into the NLSP are very much
suppressed by phase space. Under these conditions, all three right-handed
sleptons decay directly into the corresponding charged lepton and goldstino.
Particles which, in spite of not being the NLSP, have a dominant two-body
decay into their supersymmetric partner and a ${\tilde G}$ have been
called~\cite{ambrosanio97}
``co-NLSPs".
For larger values of $\tan \beta$, typically $\tan\beta >4$--8 depending
on the parameter choice, the first two generations of sleptons decay
as ${\tilde \ell}_R\to \ell \tau {\tilde \tau}$ and the stau is the
``only" NLSP. Another possibility is that $\chi_1^0$, although not the
NLSP, is nearly mass degenerate with $\tilde \tau$, and therefore
decays dominantly into a photon and a goldstino. In this case, $\tilde \tau$
and $\chi_1^0$ are again ``co-NLSPs".

The possibility that a sneutrino is the NLSP is very marginal. As seen in
sect.~\ref{properties}, 
it requires large values of $N$ and values of $\Lambda$ so low that 
the discovery of supersymmetry should take place quite soon. In this case, the
sneutrino decays into a neutrino and a goldstino.

These various cases correspond to a quite different phenomenology in
high-energy experiments, as we will discuss in sect.~\ref{collider}.

Finally, a completely different option is that the gluino is
a stable LSP. This occurs in models in which the coloured
messengers are heavier than their weak GUT partners, and in which
the mass scale $M$ is close to the GUT scale~\cite{raby97,raby97a}. 
This could be a natural
possibility in theories with Higgs--messengers mixings, as a consequence
of a doublet--triplet splitting mechanism.

\subsection{Signals in Collider Experiments}
\label{collider}

The phenomenology of supersymmetric-particle production in collider
experiments is a well-studied subject (see {\it e.g.} 
refs.~\cite{baer95,drees95,peskin96,tata97}
and references therein).
Here we will not try to give a comprehensive review of these studies, but
merely discuss the main peculiarities of gauge-mediated models with
respect to the ordinary searches for supersymmetry.

Depending on the nature of the NLSP, its decay modes, and its decay rate,
the signals in collider experiments predicted by gauge-mediated models
can be quite different. As we have seen in sect.~\ref{properties}, 
the NLSP can be the
lightest neutralino $\chi^0_1$, the lightest stau $\tilde \tau$ or, in a
very marginal corner of parameter space, the lightest sneutrino. 
Moreover, there is the possibility of having co-NLSPs, {\it i.e.} 
particles other than the NLSP
that have a direct two-body decay into the gravitino. This 
occurs when the mass difference between NLSP and co-NLSP is small enough
to suppress the ordinary supersymmetric decay and when
$F_0$ is adequately low to allow for a sizeable decay rate into gravitinos.
Candidates for co-NLSP are
$\chi^0_1$ (with $\tilde \tau$ NLSP), $\tilde \tau$ (with 
$\chi^0_1$ NLSP), and ${\tilde e}_R$ and ${\tilde \mu}_R$ 
(with $\tilde \tau$ NLSP or co-NLSP). All supersymmetric particles decay
in cascade processes leading to an NLSP or a co-NLSP which, depending on
the case, decays into a photon or a charged lepton and a gravitino. An
NLSP sneutrino decays into neutrino and gravitino and therefore, from
the detector point of view, it behaves like a stable invisible particle.

{}From the NLSP decay rate, see
eqs.~(\ref{chidec}) and (\ref{staudec}), we obtain that the average distance
travelled by an NLSP with mass $m$ and produced with energy $E$ is
\beq
L=\frac{1}{\kappa_\gamma}\left( \frac{100~{\rm GeV}}{m}\right)^5
\left( \frac{\sqrt{F/k}}{100~{\rm TeV}}\right)^4\sqrt{\frac{E^2}{m^2}-1}
~\times 10^{-2}~{\rm cm}~.
\eeq
Here $\kappa_\gamma$ is given in eq.~(\ref{kapgam}) for $\chi^0_1$ and it is
equal to 1 for the stau.
Mainly depending on the unknown value of $\sqrt{F/k}$, the NLSP can either decay
within microscopic distances or decay well outside the solar system. Therefore
{}from the collider experiments' point of view, there are two relevant regimes,
which correspond to different search strategies.

If $\sqrt{F/k}$ is large (roughly larger than $10^6$ GeV), 
the NLSP decays outside the detector and therefore behaves like a 
stable particle. If $\chi_1^0$ is the NLSP, the collider signatures closely
resemble those of the ordinary supersymmetric scenarios with a stable
neutralino. The only handle to distinguish a gauge-mediated origin is given
by the properties of the mass spectrum, described in sect.~\ref{properties}. 
On the other hand, for a $\tilde \tau$ NLSP, the signature is quite novel,
with a stable charged massive particle going through the detector,
leaving an anomalous ionization track.

For small $\sqrt{F/k}$ (typically $\sqrt{F/k}\lappeq 10^6$ GeV), the NLSP
promptly decays and the experimental signature is given by events
with missing transverse energy, imbalance in the final-state momenta and
a pair of photons or charged leptons, possibly accompanied by other
particles. This is a very characteristic signal, which typically allows
better detection efficiency than the usual missing-energy signal of
ordinary supersymmetry. Moreover, in this case, looking for NLSP
pair production, it is possible to extend
the search to a portion of the parameter space not accessible in the
corresponding gravity-mediated scenario, where the LSP is invisible.

The intermediate region between the two regimes is particularly interesting.
In this case the NLSP decay length could be measurable as a vertex 
displacement of the final-state photon (or tau). It is interesting
experimentally, because it allows a better background rejection, but also
theoretically, because a measurement of the decay length
gives direct information on the value of $\sqrt{F/k}$. This is a unique
opportunity, since other measurements are mainly sensitive only to
the mass scale $\Lambda =F/M$, which roughly determines the mass spectrum.
A study of how hadron collider sensitivity to long-lived NLSP can be
maximized can be found in ref.~\cite{chen97}.
There has also been a proposal~\cite{maki97} for 
extending the search to NLSP lifetimes much longer than allowed by
present detectors. The suggestion is to design
dedicated 
collider experiments
where the interaction point is shielded
and the detector measuring the NLSP decay is located in a tunnel some
distance away.

\begin{figure}
\hspace{3cm}
\epsfig{figure=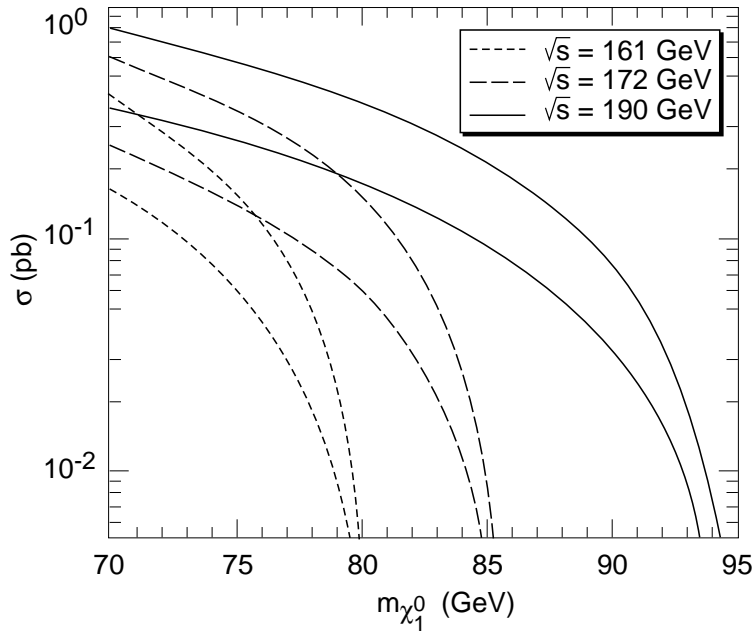,width=10cm}
\caption[]{\it
The minimum and maximum cross section for 
$ e^+e^-\to \chi_1^0 \chi_1^0$
at the indicated centre-of-mass energies, as calculated
in ref.~\cite{ambrosanio97}. 
These bounds hold
only if the dynamics generating $\mu$ and $B$ does not influence the 
supersymmetry-breaking Higgs scalar masses. Here it is also assumed that
$\chi_1^0$ is the NLSP. If this is not the case, the minimum of
the cross section still holds, but cross sections larger than the maximum
value can be obtained. (Courtesy S.~Ambrosanio,
G.~Kribs, and S.~Martin.)
\label{gmfig10}}
\end{figure}

Let us first discuss the phenomenology of gauge mediation at $e^+e^-$
colliders, and more specifically at LEP. If $\chi_1^0$ is the NLSP with
a prompt decay mode $\chi_1^0 \to \gamma \tilde G$, NLSP pair production
leads to events with two photons and missing
energy~\cite{stump96,dimopoulos96,ambrosanio96}.
As we have seen in sect.~\ref{properties}, 
if we assume that the supersymmetry-breaking
Higgs mass parameters are not influenced by the dynamics generating $\mu$,
then
$\chi_1^0$ is mainly $B$-ino and 
its production cross section is therefore determined by the slepton mass. 
In gauge-mediated models, the ratios $m_{{\tilde e}_{L,R}}/m_{\chi_1^0}$ 
are bounded from
above. This guarantees that, for a given $m_{\chi_1^0}$, the cross section
for
$e^+e^-\to \chi_1^0 \chi_1^0$ has a minimum and a maximum, which were
calculated in ref.~\cite{ambrosanio97} and 
are reproduced in fig.~\ref{gmfig10}, for values of $\sqrt{s}$ relevant for
LEP.

The two photons coming from $\chi_1^0$ pair production have a flat energy
distribution in the range $E_{min}<E_\gamma <E_{max}$, with
\beq
E_{max,min}=\frac{1}{4} \left( \sqrt{s} \pm \sqrt{s-4m_{\chi_1^0}^2}\right)~.
\eeq
At LEP, the main source of background comes from 
$e^+e^-\to \gamma \gamma \nu {\bar
\nu}$, which has an invariant missing mass peaked around $M_Z$, although the
distribution has a non-negligible tail due to the contribution from
$t$-channel $W^\pm$ exchange. The background from $e^+e^-\to \gamma \gamma 
\gamma$, with one photon unobserved, contributes only to events with
small missing mass. With some acceptance cuts on the photon energy and angle
and a cut on the invariant missing mass 5 GeV $<M_{inv}<$ 80 GeV, one obtains
a good background rejection, with a signal efficiency between 50 and 
80\%~\cite{ambrosanio96a,ambrosanio97a} (see also 
refs.~\cite{ghosal97,datta98,ghosal98,mukhopadhyaya98}).

All LEP collaborations have searched for diphoton plus missing energy
events. Their combined 95\% CL
limit~\cite{janot,dann97} on the cross section for producing
$\chi_1^0$ with $\chi_1^0 \to \gamma {\tilde G}$ at $\sqrt{s}=172$ GeV
is about $\sigma_{\chi_1^0}<
0.3$ pb for $m_{\chi_1^0}<70$ GeV, and $\sigma_{\chi_1^0}<0.15$ pb 
for 70 GeV $<m_{\chi_1^0}<85$ GeV. From fig.~\ref{gmfig10} we see that, in a
fairly parameter-independent way, neutralino masses below 73 GeV are
excluded, as long as the neutralino decay occurs inside the detector.

Slepton pair production, in the scenario with $\chi_1^0$ as NLSP, leads to
a final state with two leptons, two photons, and missing energy. The two
leptons are expected to be soft at LEP, because of the limited available
phase space. However, if the mass difference between $\tilde \ell$ and
$\chi_1^0$ is large enough for the final-state lepton to be observed,
the process of slepton pair production is essentially free from SM
background and it represents a clean signal for the $\chi_1^0$ NLSP
scenario.

Let us now turn to the case of $\tilde \tau$ NLSP, with moderate $\tan \beta$
such that ${\tilde e}_R$ and ${\tilde \mu}_R$ are almost degenerate with
${\tilde \tau}$ and behave as co-NLSPs.
The production cross sections for $\tilde \mu$ and $\tilde \tau$ are
model-independent (see {\it e.g.} ref.~\cite{giudice96}) 
while the cross section
for $\tilde e$ depends on the neutralino masses and often suffers a
destructive interference between $s$-channel and $t$-channel exchange
in the parameter region relevant for gauge mediation. Therefore one
can expect an excess of $\tilde \mu$ events over $\tilde e$ events.
If all sleptons decay inside the detector, the signature of slepton
pair production is given by two leptons and missing energy. The signal
is analogous to the one in gravity mediation, 
where the slepton decays into
the LSP neutralino and the corresponding lepton.
Above the $W$ threshold, the main SM background comes from $W^+W^-$
production. Since the gravitino is nearly massless, the 
slepton-production kinematics is more similar to the SM background than
in the gravity-mediated case, where the LSP neutralino has a non-zero
mass. The
best handle to identify the signal is then given by the angular distribution
of the charged leptons~\cite{ambrosanio97,dicus97a}. For the background,
the initial- and final-state leptons with same charge are preferentially
in the same direction, while the angular distribution of the signal
is flatter.

For large values of
$\tan \beta$, the mass differences
$m_{\tilde \tau}-m_{\tilde e}$ and $m_{\tilde \tau}-m_{\tilde \mu}$ 
become significant. Now $\tilde e$ and $\tilde \mu$ are no longer
co-NLSPs
and new decay channels are open. 
The three-body decays
${\tilde \ell }\to {\tilde \tau } \nu_\ell \bar \nu_\tau$ ($\ell =e,\mu$)
are always negligible because of the small couplings between charginos
and right-handed sleptons. However, the three-body decays
${\tilde \ell }^-\to {\tilde \tau }^\pm \tau^\mp \ell^-$ 
($\ell =e,\mu$), mediated by
virtual neutralinos, can have significant rates~\cite{ambrosanio97a}. 
Depending on the
value of $\tan \beta$ (which determines $m_{\tilde \tau}-m_{\tilde \ell}$)
and the value of $F_0$ (which determines the gravitino coupling), the 
three-body decay may or may not dominate over the two-body mode
$\tilde \ell \to \ell {\tilde G}$. If it does, slepton production leads
to a striking signal with $\ell^+ \ell^- \tau^\pm \tau^\pm {\tilde \tau}^\mp
{\tilde \tau}^\mp$ or $\ell^+ \ell^- \tau^\pm \tau^\mp {\tilde \tau}^\mp
 {\tilde \tau}^\pm$ in the final state, with possible vertex 
displacements~\cite{ambrosanio97a,cheung97}. 
As the neutralino mass (which determines the $\tilde \ell$ decay widths)
is increased,
the final state with same-sign $\tilde \tau$'s
is suppressed with respect to the case of opposite-sign $\tilde \tau$'s.

The case in which the $\tilde \tau$ NLSP decays outside the detector is also
of great experimental interest, since it leads to observable anomalous
ionization tracks.
LEP experiments have studied stable slepton production, and their
combined unsuccessful searches at $\sqrt{s}=172$ GeV~\cite{janot,cerutti97} 
set a 95\% CL limit of 76 GeV on
the mass of an NLSP $\tilde \tau$ decaying outside the detector.

In the  $\tilde \tau$ NLSP scenario, $\chi_1^0$ pair production, if
kinematically allowed, plays an important r\^ole and can become the
discovery mode. This is because the production cross section for 
$\chi_1^0$, not suppressed by the $\beta^3$ factor, can 
be larger than for $\tilde \ell$, even if $m_{\chi_1^0}> m_{\tilde \ell}$.
The signal from $\chi_1^0$ pair production is given by four leptons and
missing energy, because of the decay chain $\chi_1^0\to \ell {\tilde \ell}
\to \ell^+ \ell^- \tilde G$. In the large $\tan \beta$ region, where
$m_{\tilde \tau}<m_{\tilde e},m_{\tilde \mu}$, the final-state leptons
are predominantly $\tau$'s. Two out of the four leptons in the final
state can be very soft because of the limited phase space available
in the $\chi_1^0$ decay. However, because of the Majorana nature of
$\chi_1^0$, there is equal probability that the two hard leptons have
the same or opposite charge~\cite{dicus97,ambrosanio97a}. This provides
a clean discovery signal.

Let us now consider the search at hadron colliders. 
Because of the 
large mass hierarchy between strongly and weakly interacting particles,
production of (mainly right-handed) sleptons dominates over squark
production and, similarly, chargino--chargino and chargino--neutralino 
productions have larger cross sections than gluino production. 
This is in contrast to ordinary searches for supersymmetry, which have
focused on gluino and squarks as the discovery modes.

If $\chi_1^0$ is the NLSP with a prompt decay, the typical signal is given by
photons and missing energy, accompanied by charged leptons and/or jets
(for a detailed study of the case with a single messenger flavour $N=1$
and $M\simeq \sqrt{F}$, see ref.~\cite{baer97}).
Diphoton events with large missing transverse energy have been searched
for by the CDF~\cite{toback96} and the D0~\cite{abachi96} experiments
at the Tevatron. The measured
${E\llap{/}}_T$ distributions were found to be in agreement with SM
background by both collaborations. D0~\cite{abachi96} has
set a bound on the cross section $\sigma( p\bar p \to \gamma \gamma
E\llap{$/$}_T +X)<185$ fb at 95 \% CL, imposing cuts on the photon
transverse energies, pseudorapidities, and the transverse missing energy of
$E_T^\gamma >12$ GeV, $|\eta^\gamma |<1.1$, $E\llap{$/$}_T >25$ GeV, 
respectively.
This
bound roughly corresponds~\cite{ambrosanio96a} 
to excluding neutralinos lighter than
70 GeV and charginos lighter than 120 GeV, in scenarios with $\chi_1^0$
decaying into a photon and a gravitino.

Actually CDF has detected one controversial event~\cite{park95} 
with two energetic electrons, two
energetic photons, and missing energy that cannot be attributed to the SM.
This event has been 
interpreted~\cite{dimopoulos96,ambrosanio96,dimopoulos96b,ambrosanio96a}
as a possible signal coming from production of
sleptons or
charginos, decaying into unstable neutralinos. This possibility is now 
excluded~\cite{baer97}
in the uni-messenger case $N=1$, because of the absence of large anomalous
rates for jets + $\gamma$'s + ${E\llap{/}}_T$ events. No detailed study
has been presented for $N>1$, although this case appears more promising since,
for a given slepton mass, it predicts heavier charginos and neutralinos, and
therefore a reduction in anomalous events containing hadronic jets.

The phenomenology of long-lived $\tilde \tau$ NLSP at hadron colliders has
been considered in ref.~\cite{feng97}. Long-lived sleptons are penetrating
particles that deposit little energy in the hadron calorimeter, but 
appear in the tracking and muon chambers. If they are highly relativistic,
they are misidentified as muons. At the Tevatron,
in the mass range most relevant for discovery, they are often only
moderately relativistic and therefore have a fast rate of energy loss
through ionization. Their signal is then given by anomalous highly ionizing
tracks.

CDF has searched for highly ionizing
tracks with pseudorapidity $|\eta |<0.6$ and energy loss corresponding to
a relativistic factor $0.4<\beta \gamma <0.85$~\cite{hoffman97}. By computing
slepton production in Drell--Yan processes, the authors of ref.~\cite{feng97}
conclude that these cuts reduce the signal by 25, 44, and 65\% for 
$m_{\tilde \tau}=$ 100, 200, and 300 GeV, and eliminate the background.
The resulting lower limit on $m_{\tilde \tau}$ is about 50 GeV, which is weaker
than the above-mentioned LEP limit of 76 GeV. However, searches at the upgraded 
Tevatron with $\sqrt{s}=2$ TeV and large integrated luminosity
are highly competitive with the future
LEP results, since the limit on $m_{\tilde \tau}$ can be improved to
110 GeV for $\int {\cal L}= 2~{\rm fb}^{-1}$, and 
230 GeV for $\int {\cal L}= 30~{\rm fb}^{-1}$. 
Of 
course, these considerations apply not only to the $\tilde \tau$, but also to
$\tilde e$ and $\tilde \mu$, if they behave as co-NLSPs. 
More relativistic
sleptons can be searched by studying anomalous dimuon events and 
apparent violations of universality in the ratio $\sigma(\mu^+\mu^-)/
\sigma(e^+e^-)$. Because of the significant SM background, these searches
are less powerful and can at most be used as an independent confirmation of
a previously discovered signal.

For large $\tan \beta$, $\tilde e$ and $\tilde \mu$ are no longer co-NLSPs
and have a three-body decay into $\tilde \tau$, as previously discussed. 
Their pair production then
leads to very characteristic multilepton events with little hadronic 
activity. The Tevatron reach for $m_{\tilde e}$ and $m_{\tilde \mu}$
in this mode is
140 GeV for $\int {\cal L}= 2~{\rm fb}^{-1}$ and
230 GeV for $\int {\cal L}= 30~{\rm fb}^{-1}$~\cite{feng97}. This is 
comparable to the discovery reach of the highly ionizing track analysis.

Because of the $\beta^3$ suppression of scalar-production cross sections,
searches for neutralinos and charginos at the Tevatron can often probe a larger
region of parameter space than slepton searches, even in models where 
$\tilde \tau$ is the NLSP. Under the conditions specified in 
sect.~\ref{properties}, the Higgs mixing parameter
$\mu$ is large, and the two dominant processes of gaugino production are
$p\bar p \to W^* \to \chi_1^\pm \chi_2^0$ and
$p\bar p \to \gamma^*, Z^*\to \chi_1^\pm \chi_1^\pm$, where $(\chi_1^\pm ,
\chi_2^0)$ is the $W$-ino $SU(2)$ triplet. These states decay into 
left-handed sleptons, which then decay into right-handed sleptons along
the chains
\beq
\chi_2^0 \to \ell {\tilde \ell}_L \to \ell \ell \ell {\tilde \ell}_R ~,
\eeq
\beq
\chi_2^0 \to \nu {\tilde \nu}_L \to \ell \nu \nu {\tilde \ell}_R ~,
\eeq
\beq
\chi_1^\pm \to \nu {\tilde \ell}_L \to \ell \ell \nu {\tilde \ell}_R ~,
\eeq
\beq
\chi_1^\pm \to \ell {\tilde \nu}_L \to \ell \ell \nu {\tilde \ell}_R ~.
\eeq
The signature is given 
by multilepton production with little associated hadronic 
activity and possibly highly ionizing tracks. Present searches are sensitive
to $W$-ino masses up to about 190~GeV, but in the future they can be
extended to 
310 GeV for $\int {\cal L}= 2~{\rm fb}^{-1}$ and
410 GeV for $\int {\cal L}= 30~{\rm fb}^{-1}$~\cite{feng97}.

In conclusion, although gauge-mediated models are more predictive than
gravity-mediated models in the determination of the new particle spectrum,
they allow for a complicated taxonomy of collider signals, which depend
on the specific parameter choice. Typically, the newly predicted signals
are cleaner than the usual ones and lead to a better background
rejection. This is particularly interesting for the Tevatron because
any luminosity increase will allow for considerable improvements
in the discovery reach, as most of the relevant processes are almost
background-free. On the other hand, searches at $e^+e^-$ colliders are
limited by $\sqrt{s}$. It should also be recalled that the Higgs search
plays a significant r\^ole in probing gauge-mediated models, as
emphasized in sect.~\ref{properties}. Certainly the LHC has the capability of
exhausting the search for supersymmetric particles. However, much more
work on event simulations has to be done, as the study of the experimental
signals for gauge mediation has just begun.

\subsection{Searches in Low-Energy Experiments}
\label{lowenergy}

The search for new physics can be pursued either by directly
producing new particles
in high-energy colliders or by investigating effects caused by
virtual-particle exchange in processes at low energies. The study of FCNC
processes is a typical example of the latter experimental strategy.
The major theoretical success of gauge-mediated models is to predict a nearly
exact flavour invariance of the supersymmetry-breaking terms. 
Therefore it may seem hopeless to look for new effects in FCNC processes.
Exceptions are the cases in which the process is sensitive to new particle
exchange, even for flavour violations coming from ordinary 
Cabibbo--Kobayashi--Maskawa effects. The inclusive $B$-meson decay $B\to
X_s \gamma$ is known to be an example of such a 
case~\cite{bertolini91,barbieri93,oshimo93,okada93,garisto93}.

The contributions to $BR(B\to X_s \gamma )$ from supersymmetric particles in
gauge-mediated models can be discussed in two different regimes. For
moderate $\tan \beta$, all squarks are heavy and 
the dominant effect comes from loop
diagrams involving top quark and charged Higgs boson. They constructively
interfere with the SM diagrams and increase the value of
$BR(B\to X_s \gamma )$. 
Using the present 95\% CL upper limit on $BR(B\to X_s \gamma )$ from 
CLEO~\cite{alam95}, the complete next-to-leading theoretical calculation
leads to the limit $m_{H^\pm}>340$ GeV~\cite{ciuchini97}.
As we have seen in sect.~\ref{properties},
if the
Higgs mass parameters are not affected by the dynamics generating $\mu$,
the electroweak-breaking condition typically 
requires large $\mu$ and therefore a
rather heavy charged Higgs boson (see fig.~\ref{gmfig8} and recall that
$m_{H^\pm}^2=M_W^2+m_A^2$).
In this respect,
although the
bound from $BR(B\to X_s \gamma )$ for moderate $\tan\beta$ 
is important, it is not too
restrictive on the parameter space. On the other hand, 
independently of
any consideration of electroweak-breaking conditions, one can view
the constraint from
$BR(B\to X_s \gamma )$ as a requirement for a large value of $\mu$, in
gauge-mediated models with moderate $\tan\beta$.

In the region of very large $\tan\beta$, the loop diagrams with stop and
chargino exchange become important and, depending on the sign of $\mu$,
they can constructively or destructively interfere with the SM contribution.
In the first case (which occurs for positive $\mu$), one finds
very stringent constraints: {\it e.g.} $\mu >700$ GeV, for $\tan \beta =42$, 
$N=1$, and
$M=1.1~\Lambda$~\cite{deshpande97}. In the second case, when $\mu$ is negative,
there is an approximate cancellation between different contributions and no
useful bound can be derived. On the other hand, 
$BR(B\to X_s \gamma )$ can now be reduced with respect to the SM 
value. This could be 
interesting, since the present CLEO measurement~\cite{alam95}
is about 2~$\sigma$ below the SM prediction, although preliminary 
results from ALEPH~\cite{parodi97} indicate larger rates. 

The case in which $B=0$ at the messenger scale reproduces a situation with
large $\tan \beta$ and negative $\mu$. For small values of the messenger
mass $M$, the contributions from charged Higgs and charginos approximately
cancel each other, and deviations from the SM value of $BR(B\to X_s \gamma )$ are
predicted to be small. However, for larger values of $M$ ($M\gappeq
10^3$--$10^4 \Lambda$), the chargino contribution is more important
and $BR(B\to X_s \gamma )$ can be significantly lowered~\cite{gabrielli97}.

Another potentially interesting source of experimental information for
gauge-mediated models comes from measurements of the muon anomalous
magnetic moment $a_\mu$. 
In the limit of large $\tan\beta$, the supersymmetric contribution is
approximately given by (see refs.~\cite{chattopadhyay96,moroi96,carena97} 
for complete formulae)
\beq
\delta a_\mu \simeq \frac{\alpha}{8\pi \sin^2\theta_W}\frac{m_\mu^2}{
{\tilde m}^2}\tan\beta \simeq 15\times 10^{-10}\left( \frac{100~{\rm 
GeV}}{\tilde m}\right)^2 \tan \beta ~,
\eeq
where $\tilde m$ is the typical mass scale of weakly interacting
supersymmetric particles. From the SM result and the experimental
measurement~\cite{barnett96}, one obtains a bound on new contributions,
$|\delta a_\mu | <200\times 10^{-10}$. This already constrains gauge-mediated
models with light sleptons and very large $\tan \beta$~\cite{carena97}, but the
significance of this constraint may be much more important in the future.
The experimental precision in the measurement of $a_\mu$ is expected to
be improved by the E821 experiment at the Brookhaven National Laboratory to
the level of $4\times 10^{-10}$. If the theoretical prediction
in the SM, unfortunately
affected by hadronic uncertainties, could also be calculated 
at the same level of accuracy, the measurement of $a_\mu$ could compete
with high-energy experiments. 

Although gauge-mediated models have the same source of flavour violation
as the SM, {\it i.e.} the Yukawa couplings, they can have additional sources
of CP violation. In the minimal case with a single $X$ field, the situation
is particularly simple. The phases of
the parameters $M$ and $F$ can be reabsorbed in the definition of the
messenger fields through a global-symmetry and an $R$-symmetry transformation.
However, once the Higgs mixing mass parameters 
$\mu$ and $B$ are introduced, we are left with an irremovable
phase; this, for instance, can be chosen to be arg$(B^*M_3)$, the relative
phase between $B$ and the gluino mass. This phase is constrained by
limits on the electric dipole moments of the electron and the neutron.
The corresponding bounds~\cite{fischler92,grossman97} are 
arg$(B^*M_3)\lappeq 0.1(m_{\tilde \ell}/100~{\rm GeV})^2$ and
arg$(B^*M_3)\lappeq (m_{\tilde q}/{\rm TeV})^2$, respectively. 

If the dynamical mechanism that generates $\mu$ and $B$ correlates
their phases with the phase of the original supersymmetry-breaking
scale $F/M$, then all soft terms are CP-conserving. A simple example
of this possibility is the condition $B(M)=0$~\cite{babu96,dine97}.
In this case, all relative phases (and, in particular, all relative
signs) of the different parameters are
determined. 

Theories with gauge-mediated supersymmetry breaking also offer a possible
scenario for implementing a solution of the strong CP problem through
the Nelson--Barr mechanism~\cite{nelson83,barr84}. 
If the theory has a (spontaneously broken)
CP invariance and the determinant of the mass matrix for all coloured
fermions is real, then $\theta_{QCD}$ vanishes at tree level and it is
computable in perturbation theory. Since, in the supersymmetric limit,
$\theta_{QCD}$ is not renormalized~\cite{ellis82a}, all corrections
must be proportional to supersymmetry-breaking effects. In a general
gravity-mediated scenario, $\theta_{QCD}$ receives dangerously large
corrections from GUT or Planck mass particles that invariably exist in
models attempting to solve the strong CP problem~\cite{dine93a}. 
In gauge-mediated
models, supersymmetry shields $\theta_{QCD}$ from these effects, since
the observable sector feels supersymmetry breaking only at much lower 
scales~\cite{barr97}. The computable corrections to $\theta_{QCD}$ from CP
violation in the Yukawa sector are too small to give a significant effect
to the neutron electric dipole moment~\cite{dugan85}, and the strong
CP problem can be solved.

Finally a more unusual (and more speculative) way of searching for 
low-scale supersymmetry-breaking effects was suggested in 
ref.~\cite{dimopoulos96a}.
In superstring theories we expect the occurrence of gravitationally coupled
massless scalars, called moduli. It is possible that these states acquire
masses only after supersymmetry breaking. If the mass
scale
$\sqrt{F}$ is low, the moduli are very light and their Compton
wavelengths can lie in the range between a micron and a millimeter.
Although moduli may cause cosmological difficulties (see sect.~\ref{cosmology}),
it is interesting that their couplings to matter can
induce new forces, stronger than gravity, in this range of distances.
New experiments have been proposed~\cite{price96,kapitulnik97} to search 
for non-gravitational forces at distances down to 100 or even 10 microns.

\subsection{Gravitino Cosmology}
\label{cosmology}

In models in which supersymmetry breaking is mediated by gravity, the gravitino
mass sets the scale for the soft terms, and therefore it is expected to lie
in the range between 100 GeV and 1 TeV. Its lifetime is dictated by gravity
to be $\tau \sim 10^6~{\rm sec}~({\rm TeV}/m_{3/2})^3$. 
This late decay leads to an enormous entropy production after nucleosynthesis,
unless the 
gravitino number
density has not been diluted after the original
thermalization. If the gravitino
is lighter than few TeV,
a successful nucleosynthesis sets an upper bound of
about $10^{10}$ GeV~\cite{dimopoulos89,ellis92} 
to $T_{max}$, the temperature at which the
ordinary radiation-dominated Universe starts. Here
$T_{max}$ could correspond to
the reheating temperature after an inflationary epoch, or to the temperature
of a significant entropy production. This bound is rather uncomfortable
for many inflation scenarios and necessarily requires some low-temperature
mechanism for baryogenesis.

In gauge-mediated models the gravitino effects on cosmology are quite
different. The gravitino is stable, and therefore an important bound comes
{}from its contribution to the energy density and not from lethal effects of
its decay products. If gravitinos are in thermal equilibrium at early times
and freeze out at the temperature $T_f$, their contribution to the
present energy density is~\cite{pagels82}
\beq
\Omega_{3/2}h^2=\frac{m_{3/2}}{\rm keV}~\left[
\frac{100}{g_* (T_f)}\right] ~.
\label{omegag}
\eeq
Here $h$ is the Hubble constant in units of 100 km sec$^{-1}$ Mpc$^{-1}$ and
$g_* (T_f)$ is the effective number of degrees of freedom at $T_f$, typically
between 100 and 200 in a supersymmetric model. Therefore, if 
$m_{3/2}<$ keV, or equivalently 
\beq
\sqrt{F}<\sqrt{k}~2\times 10^6~{\rm GeV}~,
\label{omeg}
\eeq 
gravitinos do not
lead to overclosure of the Universe and, in contrast with the gravity-mediated
case, there is no need for mechanisms of late entropy production.

The situation is less satisfactory if $m_{3/2}>$ keV, since some means of
gravitino dilution are now necessary, and stringent bounds on $T_{max}$ are
found. At temperatures below $T_{max}$, gravitinos are produced by sparticle
decay and scattering processes. For keV $<m_{3/2}< 100$~keV, or equivalently
$2\times 10^6$ GeV $<\sqrt{F}/\sqrt{k}<2\times 10^7$ GeV, the decays
of thermalized sparticles dominate the gravitino production 
mechanisms~\cite{moroi93}. Then
the bound on the Universe energy density implies $T_{max}<{\tilde m}$, where
${\tilde m}$ is the typical mass of the supersymmetric particles.
For $m_{3/2}> 100$ keV, the scattering processes dominate and the upper
bound on $T_{max}$ is given by~\cite{moroi93}
\beq
T_{max}<10~{\rm TeV}\times h^2
\left( \frac{m_{3/2}}{100~{\rm keV}} \right)
\left( \frac{\rm TeV}{M_3}\right)^2~,
\eeq
where $M_3$ is the gluino mass.
These constraints are quite stringent and they require an inflation with
very low reheating temperature or a mechanism of late entropy production.
Moreover, in this case, we have to rely on a baryogenesis scenario occurring
at temperatures not much higher than the weak scale.

Although the gravitino is stable, a danger for ordinary nucleosynthesis
predictions comes from the NLSP decay. From 
eqs.~(\ref{chidec}) and (\ref{staudec}), we find that the NLSP lifetime is
roughly 
\beq
\tau_{NLSP}=\frac{1}{k^2}\left( \frac{100~{\rm GeV}}{m_{NLSP}}\right)^5
\left( \frac{\sqrt{F}}{100~{\rm TeV}}\right)^4
~3\times 10^{-13}~{\rm sec}~.
\eeq
The damage that the NLSP decay can make depends on the final-state particles.
Decays into photons are safe as long as $\tau_{NLSP}<10^7$ sec. This condition
is satisfied as soon as eq.~(\ref{flapro}) is verified. However, much
stronger constraints arise for hadronic decays of the 
NLSP~\cite{dimopoulos89,ellis92,dimopoulos97b}. 
These require that the messenger mass should be lower than values 
between $10^{12}$--$10^{14}$~GeV, depending on the parameter choice and
the nature of the NLSP~\cite{gherghetta99}. When combined with the
requirement of no gravitino overabundance, this bound also implies that
the reheating temperature after inflation must be less than 
$10^7$~GeV~\cite{gherghetta99}.
Notice that these limits
can be evaded in theories with $R$-parity breaking,
in which the NLSP lifetime is decoupled from the supersymmetry-breaking
scale.

Another potential cosmological difficulty is the ``moduli 
problem"~\cite{carlos93,banks94}.
This problem, also present in gravity-mediated theories, is actually
exacerbated in models with low-energy supersymmetry breaking. It arises
because in string theory there are flat directions in field space, called
moduli, which 
may acquire masses only
after supersymmetry breaking. They correspond to excitations that are only
gravitationally coupled and have masses of the order of $m_{3/2}$. As the
Universe cools down, the moduli oscillate around their minima, storing
an enormous amount of energy density. In gravity-mediated scenarios, they
decay at the same time as gravitinos. If their mass is below
about 10 TeV, their decay products erase the nucleosynthesis predictions. 
Even if they
are heavier than 10 TeV, they are cosmologically worrysome since the entropy
production at the time of their decay dilutes any pre-existing baryon
asymmetry. One again has to invoke a low-temperature baryogenesis scenario.

In gauge-mediated theories with low $F$, 
the moduli are practically stable, at least
when cosmological times are concerned. The contribution to the present
energy density of the Universe from the kinetic energy stored in their
oscillations is
\beq
\Omega h^2 = 4\times 10^{13} \left( \frac{m_{3/2}}{\rm 
keV}\right)^{\frac{1}{2}}~.
\eeq
A very efficient way of entropy production should be found to dilute this
unacceptably large energy density. In order to avoid the regeneration of
moduli, the entropy production should occur at temperatures below the
supersymmetry-breaking scale $\sqrt{F}$. 
Strong constraints on the moduli masses in gauge mediation also come from
the limits on the observed fluxes of gamma 
rays~\cite{kawasaki97,hashiba97,asaka98}

Finally, we mention that limits on light gravitinos have been obtained
{}from various astrophysical considerations of star cooling. These limits only
apply to models with other ultralight particles or for values of $\sqrt{F}$
much below the range of interest to gauge mediation.

\subsection{Dark Matter}
\label{dark}

Theories with gravity-mediated supersymmetry breaking have the appealing
feature that the lightest neutralino is a viable dark matter candidate, as
long as $R$ parity is (almost) exactly conserved. In gauge-mediated theories
the neutralino decays with a lifetime shorter than the age of the Universe,
and therefore cannot constitue the galactic
halos. On the other hand, gravitinos could be the dark matter and give the
dominant contribution to the present energy density if $m_{3/2}$
is about a keV, see eq.~(\ref{omegag}), or equivalently, if $\sqrt{F}$
saturates the bound in eq.~(\ref{omeg}). Gravitinos
would behave as ``warm" dark 
matter, since they just become non-relativistic at the time at which the
growing horizon encompasses a perturbation of a typical galactic size.
Gravitinos much heavier than a keV could also form the dark matter but,
in this case, the temperature $T_{max}$ defined in sect.~\ref{cosmology}
should be just
right to give the correct gravitino abundance. This may be regarded
as an unappealing scenario,
since it requires a correlation between two quantities, $F$ and 
$T_{max}$, which {\it a priori} originate from different physics.
A bleak feature of dark matter gravitinos is that terrestrial experiments
looking for galactic halo particles have no hope to detect any signal.

It is interesting to observe that, in contrast with 
the case of gravity-mediated
theories, the gravitino could form the dark matter even in the presence of
considerable $R$-parity breaking. Indeed, if the $R$-breaking interaction
violates baryon (but not lepton) number, then the gravitino has no
kinematically accessible channels as long as $\sqrt{F}<\sqrt{k}~2\times 10^9$
GeV. 
However, certain baryon-number violating couplings are, in this case,
tightly constrained
by proton decay into gravitinos~\cite{choi97}. On the other hand,
if the $R$-breaking interaction violates lepton number, then the
gravitino can decay ${\tilde G}\to \nu \gamma$. For instance with a
superpotential interaction $\lambda_{ijk}L_L^iL_L^jE_R^k$, the gravitino
decay occurs via a one-loop diagram involving lepton--slepton exchange.
The largest contribution comes from the coupling $\lambda_{i33}$ which,
in the limit of degenerate ${\tilde \tau}_L$ and ${\tilde \tau}_R$ 
gives~\cite{borgani96}
\beq
\Gamma ({\tilde G}\to \nu_i \gamma )=\frac{\alpha \lambda_{i33}^2}{128\pi^4}
\frac{m_{3/2}m_\tau^2}{M_P^2}\ln^2\frac{m_{\tilde \tau}}{m_\tau}~.
\eeq
The strongest constraint on this decay comes from limits on the diffuse
photon background; for a keV gravitino mass, this implies~\cite{ressel90}
that the lifetime should exceed $10^{26}$ sec, a time much longer than
the age of the Universe. In turn, this requires
$\lambda_{i33}\lappeq 10^{-2}$. This bound for $i=2,3$
can be more stringent than present 
bounds~\cite{barger89,dreiner97} 
{}from low-energy processes, but still allows for significant
$R$-parity violating effects in high-energy experiments.

It has also been suggested~\cite{borgani96} 
that there could be two components of
relic gravitinos playing a cosmological r\^ole. One is given by the
gravitinos, which were thermalized in the early Universe, the other is
given by the gravitinos from the NLSP decay, which act as a 
hot component in a mixed
warm/hot dark matter scenario. However, in 
order to have a significant component
of gravitinos from NLSP decay, one would need a rather extreme mass hierarchy,
with sleptons heavier than 7 TeV for a (purely $B$-ino) neutralino of 30 GeV.

Another option for a dark matter candidate is given by the lightest messenger
particle~\cite{dimopoulos96c,han97}. 
Indeed, the superpotential in eq.~(\ref{phix}) satisfies 
a ``messenger number" invariance and therefore justifies the presence of
a stable messenger particle. In the minimal case of messengers belonging
to a {\bf 5} and 
${\bf \bar 5}$ of $SU(5)$, the lightest messenger
particle has the same quantum numbers
as the scalar left-handed neutrino. In order not to overclose the Universe,
this particle should be lighter than about 3 TeV. Direct 
observations~\cite{beck94,garcia95}
have already ruled out at 90\% CL a 3 TeV weakly interacting scalar particle
accounting for more than 30\% of a galactic halo with local density
of 0.3 GeV/cm$^3$. However, the authors of ref.~\cite{han97} have
suggested a model in which the
messenger dark matter particle would evade detection.

The presence of a ``messenger number" invariance is indeed in most cases
quite disturbing. The relic abundance of the lightest messengers overcloses
the Universe for typical values of the messenger mass. Breaking
this invariance with renormalizable interactions between the messenger and
observable sectors generally reintroduces the flavour problem.
Therefore one
should maybe rely on dimension-five Planck-mass suppressed interactions, which
allow the messenger to decay without inducing significant flavour violation
in the observable sector~\cite{dine96,dimopoulos96c}.

Another conceivable option for a dark matter particle is to have a global
symmetry in the strongly interacting supersymmetry-breaking
sector, which forces the
stability of the lightest ``secluded baryon" $B_\varphi$~\cite{dimopoulos96c}. 
In this case a lower bound on
the $B_\varphi$ 
relic abundance can be derived from unitarity of the annihilation
cross section~\cite{griest90}:
\beq
\Omega_{B_\varphi}h^2>(m_{B_\varphi}/300~{\rm TeV})^2~.
\eeq
Therefore a stable particle with mass in the 100 TeV range,
strongly interacting under a new gauge force
of the secluded sector, could be a good dark matter candidate.

It has also been suggested that, in theories with low-energy
supersymmetry breaking, flat directions along some field configurations
can support stable non-topological solitons. These states can be
abundantly produced in the early Universe and form the dark 
matter~\cite{kusenko97}.

\section{Basic Tools for Studying Dynamical Supersymmetry Breaking}
\label{tools}
\setcounter{equation}{0}

Before describing the attempts to construct realistic models with
dynamical supersymmetry breaking (DSB), 
we want to review in this section, for the ease of the reader, 
the basic tools necessary to
analyse the non-perturbative dynamical properties of supersymmetric theories.
We will just recall here the basic ingredients we need for our analysis,
and refer the reader to refs.~\cite{intriligator96,shifman97,peskin97} for
recent reviews on non-perturbative aspects of supersymmetric theories. For
reviews on DSB, see also refs.~\cite{skiba97,thomas98,poppitz98}.

\subsection{The Witten Index}
\label{witindex}

The Hamiltonian $H$ of a supersymmetric
system equals the square of the supercharge $Q$ ($H=Q^2$), and
$Q$ transforms a bosonic state ($b$) into a fermionic state ($f$):
\beq
Q|b\rangle =\sqrt{E} |f\rangle ~~~~~
Q|f\rangle =\sqrt{E} |b\rangle ~.
\eeq
Together, these properties imply that
bosonic and fermionic states of non-zero energy $E$ always come
in pairs. On the other hand, states with vanishing energy 
are annihilated by $Q$, so that they are not necessarily paired.
The Witten index~\cite{witten82} is defined as
\beq
\tr (-1)^F\equiv \sum_{E}n_B(E)-n_F(E)=n_B(0)-n_F(0)~,
\label{index}
\eeq
where $n_B(E)$ and $n_F(E)$ are respectively the number of bosonic and 
fermionic states with energy $E$. As $n_B(E)=n_F(E)$ for $E\not=0$,
only the supersymmetry-preserving vacua ($E=0$)  can contribute to a
non-zero index. Thus, a sufficient
criterion for unbroken supersymmetry is $\tr(-)^F\not=0$. 
Conversely, a necessary requirement for broken supersymmetry is that
$\tr (-1)^F$ be zero (or ill-defined).
The crucial property of the index is that it is a discrete
quantity, and is thus constant under ``continuous'' deformations
of the parameters of the theory. If the index is found to be non-zero
at weak coupling, supersymmetry is certainly unbroken, even at strong
coupling. We should add that by ``continuous'' deformations 
we mean those
that do not modify the asymptotic behaviour of the action
in field space. In particular, a change in the large field behaviour
of the superpotential can affect the index. The picture in that case is 
 that vacua  ``come in from'' (or ``go to'') infinity. 
In sect.~\ref{mech} 
we will discuss a specific example in which this pathology is
explicitly realized.
Therefore when $\tr(-1)^F\not = 0$ over a set of non-zero
measure of parameters, it will stay so almost everywhere, and
supersymmetry may be broken only at special points. 

On the other
hand, when $\tr(-1)^F=0$, we can only conclude that supersymmetry
is {\it probably} broken, but counter-examples with
$n_B(0)=n_F(0)\ne 0$ are known~\cite{witten82}.
In this case, more information may be obtained with the use of
holomorphy~\cite{seiberg93}. If the vacuum energy $E_{vac}$
is described by a holomorphic function $W$ of fields and couplings
via $E_{vac}=K_{ij}^{-1}
\partial_i W\partial_j^*W^*$ then, barring singularities in the
K\"ahler metric $K_{ij}$,
phase transitions are precluded. 
In other words, if $E_{vac}$ is non-zero
over a finite range of parameters, it will remain non-zero
for all values, except at special isolated
points~\cite{hsu97}.  
In particular if $E_{vac}\not = 0$
at weak coupling, supersymmetry will also be broken at strong coupling.

In ref.~\cite{witten82} the index of supersymmetric pure gauge
theories was calculated
in a finite volume and found to be non-zero. For example,
in $SU(N)$ gauge theories, it is $\tr(-1)^F=N$. 
The Witten index does not vary if we add
massive vector matter as, with a continuous change of parameters,
it can be calculated in the limit
of very large mass, where the effective theory is just pure Yang--Mills.
Thus $E_{vac}=0$ for any finite value of the mass $m$. In particular,
the theory with massless vector matter, when it exists, preserves supersymmetry.
This conclusion provides a useful criterion to search for dynamical
supersymmetry breaking.
Gauge theories that break supersymmetry dynamically must
either be chiral or have massless fermions at every point
of their parameter space (we will discuss an example of
such a case in sect.~\ref{mech}).

\subsection{Global Symmetries, $R$ Symmetry, and Supersymmetry Breaking}
\label{rsymmetry}

A {\it sufficient} condition for the occurrence of DSB has been
suggested by Affleck, Dine, and Seiberg (ADS) in 
refs.~\cite{affleck84a,affleck84c,affleck85}. 
It relies on two basic requirements.
The 
first one is that there be no non-compact flat directions in
the classical scalar potential; the second one is that there exist a 
spontaneously broken global
symmetry.
Under these circumstances, supersymmetry must be broken.
Indeed, a spontaneously broken global symmetry implies the existence of
a Goldstone boson, and unbroken supersymmetry leads to an additional
massless scalar 
to complete the supermultiplet.
This extra massless mode corresponds to 
a non-compact flat direction\footnote{In special cases, the
extra massless scalar could be a Goldstone boson itself, thus evading the
conclusion of non-compact flat directions and, ultimately, of supersymmetry
breaking.}, in
contradiction with the first requirement. Whence 
the conclusion that supersymmetry
is broken. We will refer to this condition for DSB as to the ADS criterion.

The ADS criterion is useful in strongly coupled theories, where the
breakdown of a global symmetry
can be established by the complexity (or absence) of solutions to 't Hooft's 
anomaly matching~\cite{affleck84a}, or directly by use of instanton 
calculations~\cite{meurice84}.
Even though this argument applies to any global symmetry, in all known models
satisfying these criteria, the spontaneously broken symmetry is actually
an $R$ symmetry. As pointed out in ref.~\cite{nelson94}, there is indeed
a deep connection between $R$ symmetry and supersymmetry breaking.
Consider theories that, after the strong gauge dynamics have been
integrated out, have a low-energy description in terms of a 
Wess--Zumino model, although typically with a complicated K\"ahler
potential. For these theories,
the presence of an $R$ symmetry is a {\it necessary} condition
for supersymmetry breaking, if the superpotential is a {\it generic}
function of the fields,
{\it i.e.} if it contains enough terms.
Theories with non-generic superpotentials, which typically have
some interaction or mass terms set to zero or fine-tuned, are unstable
under small variations in the couplings.

Let us consider the effective theory described by a Wess--Zumino
model with $n$ chiral superfields $\phi_i$, $i=1,\dots,n$ and no gauge fields.
The condition for unbroken supersymmetry is given by the $n$ equations
$\partial_{\phi_i} 
W=0$. These are just $n$ complex equations in $n$ complex unknowns, 
so that for a generic superpotential, with no symmetry requirements, there is
always a solution. This simple fact reflects all the difficulty in 
achieving DSB.
Consider, on the other hand, a theory with an $R$ symmetry which is 
spontaneously broken,
say, by the VEV of the field $\phi_1$~\cite{nelson94}. One can parametrize the 
fields by $\phi_1$ and by the 
$R$-invariants $\psi_i=\phi_i^{1/r_i}/\phi_1^{1/r_1}$, for
$i=2,\dots,n$, where $r_i$ is the charge of $\phi_i$. Since $W$ has charge 2,
it will be of the form $W=\phi_1^{2/r_1}\wi(\psi)$. By using $\phi_1\not = 0$
 the zero energy condition is
\bea
\partial_{\phi_1}W=&{2\over r_1}\phi_1^{-1+2/r_1}\wi(\psi)=0& \Rightarrow 
\wi =0
\label{rsym0}
\\
\partial_{\psi_i}W=&\phi_1^{2/r_1}\partial_{\psi_i}\wi(\psi)=0&\Rightarrow 
\partial_{\psi_i}\wi=0 \quad i=2,\dots,n.
\label{rsym}
\eea
Equations~(\ref{rsym0}) and (\ref{rsym}) correspond to $n$ equations in 
$n-1$ variables $\psi_i$, which cannot
be  solved for a {\it generic} $\wi$. Apart from singular points in the K\"ahler
metric, which need a specific study, we conclude that supersymmetry is broken.
Notice that the case of a non-$R$ spontaneously broken global
symmetry
does not lead to the same conclusion. Now $W$ is just a function of the
invariants $I_A$, $A=1,\dots,k$, where $k<n$. So, when written in terms of $I_A$,
the stationary conditions for $W$ are just $k$ equations in $k$ variables.
In general these equations admit 
a solution, precisely as in the absence of a symmetry. Conversely, this proves
that, for a generic superpotential with no $R$ symmetries, a 
spontaneously broken global symmetry leads to a flat direction
with zero energy. 

These arguments  show that the presence of an $R$ symmetry is a {\it
necessary} condition for supersymmetry breaking, while a spontaneously broken
$R$ symmetry provides a {\it sufficient} condiion~\cite{nelson94}. These
conclusions hold only for theories that allow a calculable low-energy
description in terms of a Wess--Zumino model with
a generic superpotential. There are interesting cases in which
these conditions are not satisfied.
For instance,
in some cases, the effective superpotential generated by the
non-perturbative dynamics is non-generic and DSB can occur without an
$R$ symmetry.

Notice that the familiar perturbative O'Raifertaigh model~\cite{orafertaigh}
corresponds to the case of a generic superpotential with an
$R$ symmetry and with a non-compact
pseudoflat direction, {\it i.e.} with constant positive energy. 
Everywhere along the pseudoflat direction, away from the origin,
the $R$ symmetry is spontaneously broken, and the {\it sufficient} condition
for supersymmetry breaking is satisfied. At the origin, the $R$ symmetry
is preserved, but supersymmetry remains broken. The {\it necessary}
condition for supersymmetry breaking is verified.

The connection between spontaneously broken $R$ symmetry and DSB may be
the cause of a phenomenological problem.
Since the gluino mass has $R$ charge
2, the $R$ Goldstone boson, usually referred to as the $R$-axion, 
couples to the QCD anomaly.
In the absence of explicit $R$ breaking, the $R$-axion decay
constant is very
constrained by astrophysical considerations~\cite{turner90,raffelt90}.
However, as we will discuss at the end of sect.~\ref{early},
there are several ways to overcome this problem.

\subsection{Flat Directions and Supersymmetric QCD}
\label{flat}

At tree level, in the absence of a superpotential, a supersymmetric gauge theory 
typically has a large set of vacua. These 
are the points with vanishing $D$-terms:
\beq
D_A\equiv \sum_i\phi_i^\dagger T_A\phi_i=0 ~.
\label{daflat}
\eeq
Here $T_A$ are the gauge generators in the representation under which the
chiral superfields $\phi_i$ transform.
 Understanding the space of flat directions
(usually referred to as classical moduli space) 
is crucial to study a model. It is often a non-trivial
problem to find explicitly all the solutions to eq.~(\ref{daflat}). 
In some cases
the techniques of refs.~\cite{affleck84,affleck84b,affleck85} 
can be useful. Fortunately there
is a general theorem~\cite{buccella82,procesi85,gatto85,gatto87,luty96} 
stating that the space of solutions
to eq.~(\ref{daflat}) is in a one-to-one correspondence with the VEVs
of the complete set of gauge-invariant functions
of the chiral fields $\phi_i$. In other words, the moduli space is the space
of independent chiral invariants. 
The coordinates on the moduli space correspond to massless chiral
supermultiplets.
In general, the global description
of this space is given in terms of a set of invariants satisfying  certain 
constraints. These constraints, which select the independent invariants,
 are determined by Fierz identities
in the gauge contractions and Bose symmetry of the scalar fields.
Although finding all the invariants and the constraints can sometimes be
difficult, this theorem greatly simplifies the search for the
solutions of eq.~(\ref{daflat}), and in practice it 
is very useful. 

After adding a superpotential $W$, some flat directions are 
lifted, {\it i.e.} the $F$ terms are non-vanishing along the
$D$-flat direction. In particular, if one
can show that every invariant is fixed by the condition $F_i=-\partial_{\phi_i}
W=0$, then all flat directions have 
been lifted, and one of the requirements of the ADS criterion
discussed in sect.~\ref{rsymmetry}
is satisfied.
In terms of the invariants the vacua are described by the zeros of
holomorphic functions, and this simplifies things considerably.

As an explicit example consider an $SU(3)\times SU(2)$ model~\cite{affleck85}
with
matter content $Q({\bf 3},{\bf 2})$, $\bar U({\bf \bar 3},{\bf 1})$, 
$\bar D({\bf \bar 3},{\bf 1})$
$L({\bf 1},{\bf 2})$. A complete set of gauge invariants is
\beq
X=Q\bar U L,\quad \quad Y=Q\bar DL,\quad\quad Z=Q\bar U Q\bar D.
\eeq
As we will show in sect.~\ref{mech}, by adding a tree superpotential
\beq
W = Q\bar UL,
\label{tredue}
\eeq
this model breaks supersymmetry dynamically. It is easy to see that
the superpotential in eq.~(\ref{tredue}) lifts 
all $D$-flat directions. 
Multiplying the equation $0=\partial_{\bar U}
W$ by $\bar U$ and $\bar D$, we respectively get $X=0$ and $Y=0$.
In the same way, by contracting $\partial_LW=0$ with $Q\bar D$
we get $Z=0$. 

This example actually illustrates a general technique which is very useful
to verify if a certain flat direction is lifted by the superpotential.
One should first construct gauge invariants by contracting in all possible 
ways the equations $\partial_{\phi_i}W=0$ with chiral superfields. If this
procedure determines all
independent chiral invariants, then all $D$-flat directions have been
lifted by the superpotential.

Another important example is given by
supersymmetric quantum chromodynamics 
(SQCD) \cite{taylor83,affleck83,affleck84b} with gauge group
$SU(N_c)$ and $N_f$ flavours of chiral multiplets
in the fundamental and antifundamental representations 
$Q_i,\Qb^i$, $i=1,\dots,N_f$. The moduli space is parametrized by
the mesons $M^i_j=\Qb^iQ_j$, the baryons $B_{i_1\dots i_{N_c}}=
\epsilon_{\alpha_1\dots\alpha_{N_c}}Q^{\alpha_1}_{i_1}\dots 
Q^{\alpha_{N_c}}_{i_{N_c}}$ (where $\alpha_i$ are gauge indices)
and the antibaryons $\bar B^{i_1\dots\i_{N_c}}$. Notice that the
baryons exist only for $N_f\geq N_c$ (since $Q_i$ are bosonic operators)
and that for $N_f<N_c$ the
mesons provide a complete, non-redundant parametrization of the
moduli space. On the contrary, for $N_f\geq N_c$
mesons and baryons are redundant and satisfy certain classical constraints.
For instance for $N_f=N_c$ there are just one baryon $B$ and one
antibaryon $\bar B$, and they satisfy the constraint
$\det M-B\bar B=0$.

A last ingredient of great use in determining the low-energy
dynamics of supersymmetric gauge theories is holomorphy~\cite{shifman86,
amati87,shifman91,seiberg93}. In these theories the
gauge kinetic term and the superpotential are determined
by chiral operators integrated over $\int d^2\theta$. The corresponding
coupling constants can therefore be treated themselves as 
spurionic chiral superfields, with 
non-vanishing VEVs only in their scalar components. The way these couplings
appear in the effective action must respect the background
supersymmetry under which also the couplings are treated
as chiral superfields. For instance if
a term $\lambda \phi^3$ appears in the tree-level $W$, the appearance
of $\lambda^*$ in the effective superpotential is forbidden.

As far as the gauge coupling constant is concerned, various different
definitions (schemes) can be given. Not for all of them can we 
extend the coupling to a chiral superfield, and the ``physical'' coupling
defined by the behaviour of scattering amplitudes is not one of these.
As discussed in 
refs.~\cite{shifman86,shifman91} (for additional discussions, see also
refs.~\cite{kaplunovsky94,arkani97a}), the holomorphic quantity is 
the coefficient of the gauge
kinetic term in the Wilsonian effective action,
the so-called Wilsonian coupling $g_W$. This and the topological angle 
$\Theta$ 
define a chiral superfield $S_W$, whose scalar component is just
$1/g_W^2-i\Theta/8\pi^2$. As physical quantities should not change
when $\Theta\to \Theta +2 \pi$, it is easily proved that holomorphy
implies that $S_W$
is renormalized only at one loop, to all orders 
in perturbation theory\footnote{When there is
an anomalous symmetry under which $\Theta\to \Theta+\alpha$,
with $\alpha$ a real number, one can also exclude non-perturbative
renormalizations of $S_W$ (see {\it e.g.} ref.~\cite{arkani97a}).}.
This property of $S_W$ is nothing but a special case of the 
non-renormalization theorem that applies to chiral operators.
The Wilsonian coupling is very useful: since it runs only at one loop,
the RG-invariant strong interaction scale is just
\beq
\Lambda^b=\mu^b e^{-{8\pi^2\over g_W^2}+i\Theta}=\mu^b e^{-8\pi^2S_W}~.
\label{wilson}
\eeq
Here $b$ is the one-loop $\beta$-function coefficient, equal to $3N_c-N_f$
in massless SQCD, and $\mu$ is the renormalization scale.
Notice that $\Lambda$ is also a chiral superfield, with
definite quantum numbers under all anomalous symmetries, under
which $\Theta$ is shifted. The quantum numbers of $\Lambda$ constrain
the form of non-perturbative renormalizations of the 
superpotential~\cite{seiberg93}. 

Since $S_W$ runs only at one loop, its matching at thresholds at which
heavy
states are integrated out is simply done 
at tree level, for instance by requiring continuity of $S_W$ through the
threshold\footnote{For a discussion of the various scheme choices for
$g_W^2$, see ref.~\cite{finnell95}.}. This makes $S_W$ very useful to describe effective theories.
As an example, if we integrate out one flavour of quark superfields
with mass $m$
in SQCD, the effective scale of the low-energy $N_f-1$ theory is 
simply 
\beq
\Lambda_{eff}^{3N_c-N_f+1}=m~\Lambda^{3N_c-N_f}~.
\label{matchm}
\eeq

As another example,
consider the theory along a $D$-flat direction in which some
flavours get VEVs. As previously discussed, this flat direction can be
described by the VEVs of some gauge invariants, in this case the mesons
or the baryons. For instance,
by giving a VEV to just one meson $M_{N_f}^{N_f}$ in massless SQCD,
the gauge group is broken down to $SU(N_c-1)$ with $N_f-1$ flavours, as
one flavour disappears
because it is eaten by the Higgs mechanism. In this case
the low-energy scale is $\Lambda_{eff}^{3N_c-N_f-2}=\Lambda^{3N_c-N_f}/
M_{N_f}^{N_f}$. For $N_f<N_c$, when all mesons acquire VEVs, the 
low-energy theory is pure $SU(N_c-N_f)$ with a scale 
\beq
\Lambda_{eff}^{3
(N_c-N_f)}=\frac{\Lambda^{3N_c-N_f}}{\det M}~.
\label{matchv}
\eeq

Let us now briefly describe the properties of SQCD with different numbers
of flavours $N_f$.
Consider first the case $N_f=N_c-1$. The classical moduli space is
described by the $(N_c-1)\times(N_c-1)$ meson matrix $M$. At
a generic point $\det M\not = 0$, all flavours get VEVs,
and the gauge group is completely broken. The global
symmetries of the system and holomorphy~\cite{affleck84b,affleck85,seiberg93}
constrain $W_{eff}$ to have the form
\beq
W_{eff}=c~{\Lambda^{2N_c+1}\over \det M},
\label{weff}
\eeq
where $c$ is a constant. Notice that $\Lambda^{2N_c+1}\propto
{\rm exp}(-8\pi^2/g_W^2)$, which is precisely the suppression
factor of a one-instanton amplitude. The calculation of instanton
effects in the broken phase is reliable, as large instantons
are suppressed by the gauge-boson mass, and this calculation explicitly 
shows that
$c\not = 0$~\cite{affleck84b,shifman88,finnell95}.

The result in eq.~(\ref{weff}) can be extended to the
massive case where $W_{tree}=\tr ~mM$, and $\det (m)\not =0$.
Using the
arguments of ref.~\cite{seiberg93}, based on holomorphy,
one can conclude that 
the effective superpotential
is just the sum of $W_{tree}$ and eq.~(\ref{weff}),
for any $m$ and $\Lambda$.
This is easily seen in the simplest case of $SU(2)$ with one flavour,
where there is just one meson $M=\Qb Q$ and a mass term $W_{tree}=mM$.
By using the global symmetries, the general form of the effective
superpotential is $W_{eff}=mM\, f(\Lambda^5/mM^2)$. The function $f(z)$
is holomorphic (analytic with no singularities at finite $z$), 
and its asymptotic behaviour is known. The limit $z\to 0$ corresponds to 
the free
field theory, so that $f(0)=1$. The limit $z\to \infty$ corresponds to
the massless theory, where, by eq.~(\ref{weff}), $W_{eff}=c\Lambda^5/M$,
so that 
$\lim_{z\to \infty}f(z)=cz$. With the help of holomorphy, we can
reconstruct $f(z)=1+cz$.

In the case of SQCD with $N_f<N_c-1$ massless flavours, the global symmetries 
and holomorphy constrain the superpotential to be
\beq
W_{eff}=c^\prime ~
\left(\Lambda_{N_c,N_f}^{3N_c-N_f}\over \det M\right)^{1/(N_c-N_f)} ~,
\label{weffgen}
\eeq
where $\Lambda_{N_c,N_f}$ is the corresponding dynamical scale.
Unlike the case $N_f=N_c-1$, the power of $\Lambda_{N_c,N_f}$
does not coincide with the one-instanton effect, so we
cannot perform a direct calculation of the constant $c^\prime$.
Indeed, 
at a generic  point on the  classical moduli
space, there is now an unbroken $SU(N_c-N_f)$ gauge group. So one
expects additional non-perturbative effects other than instantons.
Equation~(\ref{weffgen}) can however be obtained from the
theory with $N_c-1$ flavours by adding a mass $m$ to $N_c-1-N_f$
flavours. We have argued above that the exact superpotential
for this theory is just eq.~(\ref{weff}) plus the mass term~\cite{seiberg93}.
By integrating out the mesons containing a massive quark, we
obtain $W_{eff}$ for the theory with $N_f$ flavours. This has
the form of eq.~(\ref{weffgen}) with $\Lambda_{N_c,N_f}^{3N_c-N_f}=m^{N_c-N_f-1}
\Lambda^{2N_c+1}$, which is precisely the scale determined by matching
the theory with $N_f$  to the theory with $N_c-1$ flavours, see 
eq.~(\ref{matchm}). The constant $c^\prime$ can then be computed in terms
of the constant $c$, which appears in eq.~(\ref{weff}), and one concludes
that $c^\prime \ne 0$~\cite{cordes86,shifman88,finnell95}.

The superpotential in eq.~(\ref{weffgen}) 
has no supersymmetric minima at finite $M$,
but vanishes at infinity as an inverse power of $M$. Thus
massless SQCD with $0<N_f\leq N_c-1$ is not a well-defined
theory, and it has no stable vacuum. On the other hand by adding
a mass term $m\tr M$ to eq.~(\ref{weffgen}), one finds $N_c$
supersymmetric vacua characterized by 
\beq
\langle M_i^j\rangle =\delta_i^j
\Lambda_{N_c,N_f}^2\left ({m\over \Lambda_{N_c,N_f}}\right )^{-(N_c-N_f)/N_c} 
e^{2\pi i k/N_c},\quad k=1,\dots,N_c.
\label{vacua}
\eeq
Notice that $N_c$ is precisely the number of vacua suggested by the Witten 
index
$\tr(-1)^F=N_c$, calculated in supersymmetric pure $SU(N_c)$ theories
at finite volume (see sect.~\ref{witindex}).

As we said, for $N_f=N_c-1$, the effective superpotential in
 eq.~(\ref{weffgen}) is generated by
the one-instanton amplitude. What is then its origin for smaller $N_f$?
In order to answer this question we must consider the effective gauge theory
at a generic point where $\det M\not =0$. Here the VEVs
of the $N_f$ flavours break $SU(N_c)$ to pure $SU(N_c-N_f)$ with no 
charged matter.
Recalling the expression of the dynamical scale for the effective theory, 
see eq.~(\ref{matchv}),
we observe that the superpotential in eq.~(\ref{weffgen}) is
$W_{eff}=\Lambda_{eff}^3$. This is precisely the term that would be
generated by the gauge kinetic term $\int d^2\theta W^\alpha W_\alpha
$ in $SU(N_c-N_f)$, if the glueball field $W^\alpha W_\alpha$ were to receive
a VEV $\sim \Lambda_{eff}^3$. Therefore the 
interpretation of eq.~(\ref{weffgen}) is just that
gauginos $\lambda^\alpha\lambda_\alpha=W^\alpha W_\alpha |_{\theta=\bar 
\theta =0}$
condense in the vacuum of the low-energy pure $SU(N_c-N_f)$ 
theory\footnote{The same conclusion can be reached~\cite{amati87}, 
more rigorously, by using
eq.~(\ref{vacua}) and the Konishi anomaly~\cite{konishi84,konishi85}.}.
This result confirms other approaches where $\langle \lambda\lambda\rangle
=\Lambda_{eff}^3$
was derived by direct instanton 
calculus~\cite{novikov83,novikov85,novikov86,amati87} 
or by an effective 
Lagrangian for the glueball field~\cite{veneziano82}.
Notice that the $SU(N_c)$ theory has a discrete $Z_{2N_c}$ $R$ symmetry
under which $\lambda\to e^{2\pi i k/2N_c}\lambda$, $k=1,\dots, 2N_c$,
broken down to $Z_2$ by $\langle\lambda\lambda\rangle\not = 0$. 
Again, the resulting $N_c$-equivalent 
vacua are in agreement with the index $\tr(-1)^F=N_c$.

For $N_f= N_c$,  SQCD confines~\cite{seiberg94} with the light
bound states given by the meson matrix $M_i^j$ and the baryons
$B$, $\bar B$. Quantum effects (instantons) modify the 
classical constraint $\det M-B\bar B=0$
 to 
\beq
\det M-B\bar B=\Lambda^{2N_c}.
\label{smooth}
\eeq
This field equation, defining the so-called quantum moduli space (QMS), 
can be obtained by
introducing a Lagrange-multiplier superfield $A$ with superpotential
\beq
W_{quantum}=A\left (\det M-B\bar B-\Lambda^{2N_c}\right ).
\label{qms}
\eeq
Notice that  on any point of
the QMS the field $A$ pairs up with a linear combination of
$M,B,\bar B$ and becomes massive. The above picture was derived
inductively in ref.~\cite{seiberg94}, as it satisfies a series
of non-trivial consistency checks. In particular the massless
spectrum from eq.~(\ref{qms}) satisfies at any point 't Hooft's
anomaly matching conditions, for flavour and $R$ symmetries. Thus
for $N_f=N_c$ massless SQCD not only exists but has an
infinite degeneracy of vacua.

For $N_f=N_c+1$ there is also confinement, but the classical moduli space
is not modified~\cite{seiberg94}\footnote{For a general classification of
theories with this property, see also ref.~\cite{csaki97}.}. 
For $N_f>N_c+1$ the low-energy
description can be done in terms of a dual gauge theory whose
gauge bosons and matter fields are composites of the original 
ones~\cite{seiberg95}.
These last cases, although of great theoretical interest, do not
enter directly the model-building discussion of sect.~\ref{models}.
We refer the interested reader to the original papers and to the 
reviews in refs.~\cite{intriligator96,peskin96,shifman97}.

\subsection{Mechanisms for Dynamical Supersymmetry Breaking}
\label{mech}

In this section we review a few simple models that illustrate
the known mechanisms by which supersymmetry is dynamically broken.

The minimal model with calculable DSB is the 3-2 model~\cite{affleck85},
based on gauge group $SU(3)\times SU(2)$. This model has been
introduced in the previous section, where we proved that the superpotential
in eq.~(\ref{tredue}) lifts all flat directions. 
Let us consider 
the limit in which $SU(3)$ 
is much stronger than $SU(2)$ ($\Lambda_3\gg \Lambda_2$) so that,
in first 
approximation,
the $SU(2)$
non-perturbative dynamics can be neglected. 
The $SU(3)$ factor has two flavours, so according to
the discussion in sect.~\ref{flat} a dynamical superpotential is
generated by instantons. With the help of holomorphy, we obtain that
the exact superpotential
is just
\beq
W_{eff}={\Lambda_3^7\over (Q\bar D)(Q\bar U)} +\lambda (Q\bar U)L ~,
\label{break32}
\eeq
where the fields inside the brackets form a $({\bf 1},{\bf 2})$ 
under $SU(3)\times SU(2)$.
The $L$ equation of motion, $(Q\bar U)=0$, is inconsistent
with the equation $
0=\bar D\partial_{\bar D}W=\Lambda_3^7/[(Q\bar D)(Q\bar U)]$, because
we know that there are no flat directions
and $(Q\bar D)$ cannot escape
to infinity.  There can thus be
 no supersymmetric vacuum.
Notice the special r\^ole played by the fields $L$ and $\bar D$, one of
which appears
{\it only} in the tree-level operator and the other one {\it only} in the
denominator of the dynamical term. Their equations of motion
are mutually inconsistent, because they want the same combination
of fields (in this case $Q\bar U$) to be zero in one case and infinity
in the other. This is indeed a general feature of a large class of
models in which DSB is triggered by a non-perturbative
superpotential.

Supersymmetry breaking in the 3-2 model can also be established using the
ADS criterion. The superpotential in
eq.~(\ref{tredue}), besides lifting all flat directions,
also preserves an anomaly-free global $U(1)\times U(1)_R$ (where the
second factor is an $R$ symmetry) with charges $Q(1,0)$, $\bar U(2,2)$,
$\bar D(-4,-4)$, $L(-3,0)$. 
The non-perturbative 
superpotential forces the fields away from the origin (in jargon
``lifts the origin''),
thereby spontaneously breaking the $R$ symmetry. 
The conditions of the ADS criterion are met and supersymmetry
is broken. 

Some additional comments on the 3-2 model are in order.
By simple scaling arguments the VEVs of all scalar fields
scale like $v\sim\Lambda_3/\lambda^{1/7}$, see eq.~(\ref{break32}).
For $\lambda^{1/7}\ll 1$ the VEVs are much larger than $\Lambda_3$, so
that the group $SU(3)$ is broken in the weak regime. Here
the vacuum of the model can be studied in perturbation theory.
More precisely, non-perturbative corrections to the K\"ahler
potential are negligible and typically the tree-level
approximation is sufficient to characterize the spectrum around the
minimum of the potential. Indeed, in the limit $\lambda^{1/7}\ll 1$,
the three original moduli $X,Y,Z$ describe the light degrees of freedom.
The vacuum can be studied by considering the non-linear $\sigma$-model
for the light states
$X,Y,Z$, obtained by integrating out the heavy modes
in a theory with an originally flat tree-level K\"ahler 
metric~\cite{affleck85,poppitz94}.

Let us assume the addition of the operator $\delta W= (Q\bar U)(Q\bar D)/M$ to
the tree-level superpotential in 
eq.~(\ref{tredue}), where $M$ is some heavy-mass scale
\cite{seiberg93,nelson94}.
By holomorphy, the full non-perturbative superpotential now is 
\beq
W_{eff}={\Lambda_3^7\over (Q\bar D)(Q\bar U)} +\lambda (Q\bar U)L+
{(Q\bar U)(Q\bar D)\over M}~.
\label{nongen32}
\eeq
It is easily checked that the new superpotential still lifts 
all flat directions.
The perturbation $\delta W$, however, breaks the $R$ symmetry explicitly.
However, supersymmetry
remains broken since the equation of motion for $L$ 
still requires $(Q\bar U)=0$,
while the origin is lifted by  the instanton term. 
The necessary condition for DSB discussed in sect.~\ref{rsymmetry}
({\it i.e.} presence of an $R$ symmetry) is not respected, because the
superpotential in 
eq.~(\ref{nongen32}) is a {\it non-generic}
function of the fields. The lack of genericity is a direct consequence
of holomorphy: not every possible $R$-breaking term appears in 
eq.~(\ref{nongen32}). 
This illustrates an example of a model with DSB, but no harmful $R$-axion.
For instance, assuming $M=M_P$ and $v> \sqrt F\gsim 10 ^5$ GeV, we find
that the $R$-axion is heavier than 10 MeV, large enough to suppress
fast emission in stellar cooling.

In ref. \cite{intriligator96a} the exact superpotential for the 3-2
model was derived for any value of $\Lambda_3/\Lambda_2$. As expected from
holomorphy, supersymmetry was found to be always broken. It
is interesting to consider the limit $\Lambda_2\gg\Lambda_3$, in which
only the $SU(2)$ dynamics is important. Now DSB is achieved by a different
mechanism. The $SU(2)$ theory has 4 doublets (2 flavours),
and confines with a QMS. All global symmetries are preserved only at the
origin, but the origin is removed from the moduli space by the
quantum deformation.

To give a simple characterization of the DSB
mechanism on the QMS, 
consider an $SU(2)$ gauge theory with 4 fundamentals $Q_i$, $i=1,\dots, 4$,
coupled to 6 singlets $R^{ij}=-R^{ji}$~\cite{izawa96a,intriligator96a}
\beq
W_{tree}= \lambda R^{ij}Q_iQ_j.
\label{ittree}
\eeq
This model has an $SU(4)$ flavour symmetry under which $R$ is a {\bf 6}
antisymmetric tensor. 
The superpotential lifts all mesonic $D$-flat directions $M_{ij}=Q_iQ_j$,
while $R^{ij}$ remain flat. The $SU(2)$ dynamics is a special case
of the $N_f=N_c$ theories discussed in the previous section.
For $N_c=N_f=2$ the baryons and the mesons fit into the same 
antisymmetric tensor $M_{ij}$ with classical constraint $\pf M=0$.
The quarks $Q_i$ are confined in the mesons $M_{ij}$ and at
the quantum level the exact effective superpotential is given by
\beq
W_{eff}= \lambda R^{ij}M_{ij} +A(\pf M-\Lambda^4),
\label{ittot}
\eeq
where $A$ is a Lagrange-multiplier superfield and
 $\Lambda$ is the scale of the $SU(2)$ dynamics.
Equation~(\ref{ittot}) just describes the usual O'Raifertaigh
model. 
Supersymmetry  is broken as $F_{R^{ij}}$ and $F_A$ cannot
both vanish at the same time. Notice also that eq.~(\ref{ittot})
satisfies an $R$ symmetry, and that the r\^oles played by $R_{ij}$ and $A$
are similar to those of $L$ and $\bar 
D$ in the 3-2 model. It is now the quantum deformation that
lifts the origin, the only point satisfying $\partial_{R_{ij}}W=0$.

As $R_{ij}$ describe flat directions at tree level, this model does
not satisfy the ADS criterion. However, at the quantum level, these flat
directions are lifted with a vacuum energy $E_{vac}\sim \Lambda^4$.
The quantum lifting of classical flat directions,
which occurs in other DSB models~\cite{shirman96}, shows how
the first requirement of the ADS criterion is not strictly necessary.

The model under consideration is a vector-like gauge theory, so one may wonder
how DSB is reconciled with the Witten index.
Let us assume that we give mass to all quarks by adding 
$W_{mass}=m^{ij}Q_iQ_j$ to eq.~(\ref{ittree}). Evidently, as $R^{ij}$ 
are free to slide, there is always a point in $R$ space where the
effective
quark mass $R^{ij}+m^{ij}$ is zero. Actually by the field redefinition
$R^{ij}+m^{ij}\to R^{ij}$, the renormalizable Lagrangian is the same
as that of the original theory. 
 So quarks do not decouple in the limit
$m\to \infty$, and the index is not that of a pure $SU(2)$ Yang--Mills theory.
On the other hand, a mass term $m_R\pf R$ added to eq.~(\ref{ittot})
explicitly breaks the $R$ symmetry  
and restores supersymmetry. 
For small  $m_R$ the VEV of $R_{ij}$ is proportional to $\Lambda^2/m_R$. Thus
in the limit $m_R\to 0$ the supersymmetric vacuum escapes to infinity,
and the index has a discontinuous change at $m_R=0$~\cite{intriligator96a}.

There are also chiral gauge theories that are believed to break
supersymmetry in the strongly coupled regime, {\it i.e.} for values of
the VEVs smaller than $\Lambda$, where the K\"ahler potential in general
receives uncontrollable corrections. One such simple case
is an $SU(5)$ gauge theory with matter in a single $\bar F={\bf \bar 5}$ and
$A={\bf 10}$. This theory is easily analysed classically. 
Since we cannot construct
chiral gauge invariants involving only $\bar F$ and $A$, 
there are
no $D$-flat directions and, also, no superpotential
can be added. At the tree level the only ground state is at the origin,
where the gauge group is unbroken. At the origin the theory is strongly coupled,
and we cannot make a direct and reliable calculation of the vacuum.
There is, however, plentiful indirect evidence that supersymmetry is 
broken \cite{affleck84a,meurice84}. The theory has a non-anomalous
$U(1)\times U(1)_R$ global symmetry with charges $\bar F(3,0)$
and $A(-1,-2)$. In ref. \cite{affleck84a} the solutions to 't Hooft
anomaly matching conditions for the low-energy effective theory
were studied under the assumption of unbroken
supersymmetry. The solutions were found to be fairly complicated, as
the ``simplest'' solution found in ref.~\cite{affleck84a}  involves 
five light fermions. Therefore it seems more plausible
that the global Abelian symmetry is broken. If that is the case,
the ADS conditions are satisfied and supersymmetry is expected to
be broken. 

A different argument for evidence of supersymmetry breaking in the
$SU(5)$ model is 
based~\cite{meurice84,affleck84} on
the Konishi anomaly equation~\cite{konishi84,konishi85}:
\beq
{1\over 2{\sqrt{2}}}\left \{\bar Q_{\dot \alpha},\psi_a^{\dot\alpha}\varphi_a\right\}=
\varphi_a{\partial W\over \partial \varphi_a}+C_a{g^2\over 
32\pi^2}\lambda\lambda ~.
\label{kanomaly}
\eeq
Here $\psi_a$ and $\varphi_a$ are respectively the 
fermion and scalar components
of a chiral superfield with Casimir $C_a$ and the index $a$ in
eq.~(\ref{kanomaly}) is not summed. In the 
present model $W=0$, so that the first term on the r.h.s. is absent.
Then by taking the VEV of eq.~(\ref{kanomaly}), one finds that
$\langle \lambda\lambda\rangle$ is an order parameter for supersymmetry
breaking. Notice that $\langle\lambda\lambda\rangle\not = 0$
breaks the $R$ symmetry, so that the Konishi anomaly establishes
a rigorous connection between the spontaneous breaking of $R$ and
that of supersymmetry. By applying
the techniques of refs. \cite{novikov83,rossi84,amati85,amati87}, the authors of
 ref.~\cite{meurice84}
concluded that the product  $\langle\lambda\lambda\rangle^2\langle
\lambda\lambda
\bar FAAA\rangle$ is non-zero. Here $\bar 
F$ and $A$ indicate the scalar components
of the corresponding fields and an $SU(5)$ invariant contraction inside the 
VEVs is understood. As there are no flat directions, the solution
 $\langle\lambda\lambda\rangle= 0$, $ \langle\lambda\lambda \bar FAAA\rangle 
=\infty$ should be discarded. Then $\langle\lambda\lambda\rangle\not = 0$
and supersymmetry is broken\footnote{A possible loophole
of this argument is that other non-perturbative effects cancel
the one-instanton contribution to the relevant
Green's functions. While this cancellation
seems very unlikely it is impossible to exclude it rigorously.}.

More recently further circumstantial evidence for broken
supersymmetry in this model (and in similar ones)
has been gathered.
One interesting approach~\cite{murayama95,poppitz96} is to add one flavour
$\phi ={\bf 5}$, $\bar \phi={\bf \bar 5}$ to make it a calculable theory.
With the added matter fields there are $D$-flat directions
along which the theory becomes weak and can be studied. At a generic
point on the classical moduli space, $SU(5)$ is broken to pure
$SU(2)$, with no
charged matter. Gaugino condensation in $SU(2)$
generates a dynamical superpotential $W_{dyn}=\Lambda_5^6/[(AA\phi)(A\bar F
\bar \phi)]^{1/2}$. The runaway behaviour of $W_{dyn}$
is avoided by adding a tree-level mass term $W_{tree}=m\phi\bar \phi$, which
removes all flat directions\footnote{All flat directions are removed since,
after integrating
out $\phi$ and $\bar\phi$, 
we get the original $SU(5)$ model back.}. The total superpotential
$W_{eff}=W_{dyn}+W_{tree}$ is studied with the same technique as we 
described for the 3-2 model. Supersymmetry is found to be
broken for small $m$, where a perturbative control of the K\"ahler
potential around the minimum is available. One can, however, conclude 
 by continuity that the Witten index is zero for any finite $m$ and
also\footnote{The possibility $\lim_{m\to \infty} 
E_{vac}\to 0$ cannot be rigorously 
excluded \cite{hsu97}, although it seems unlikely.}
for $m=\infty$, which corresponds to the original theory. Holomorphy 
also implies a non-zero vacuum energy $E_{vac}$ for any finite $m$.
Finally we should add that another ``proof'' of broken supersymmetry
for this model has been given in ref. \cite{pouliot96}, by
adding enough flavours for a weakly coupled dual description to 
exist. By giving mass to the additional flavours,
supersymmetry
is broken in the dual theory
just by an O'Raifertaigh-type tree-level superpotential.

The model just discussed is the simplest of a class of models
based on gauge
group $SU(N)$, $N$ odd, with matter consisting of $N-4$ antifundamentals
$\bar F_i$ and one antisymmetric tensor $A$ \cite{meurice84,affleck85}.
These theories, with the
inclusion of the most general tree-level cubic superpotential,
have no flat directions and preserve a non-anomalous  $R$ symmetry.
In all models,
supersymmetry is expected to be broken in the strongly coupled regime.
Indeed, without tree-level superpotential, these models have flat
directions described by the invariants $ A\bar F_i \bar F_j$. On a 
generic point along a flat direction, the gauge symmetry is broken and
the effective theory is given exactly by the $SU(5)$ model previously
discussed, which has been shown to break supersymmetry.

More models can be constructed from this class,  as 
suggested in ref. \cite{dine96}. The strategy is to take the
$SU(N)$ group and remove some ``off-diagonal'' generators to reduce it to
$SU(k)\times SU(N-k)\times U(1)$, while keeping the original
matter content. The smaller gauge symmetry allows more $D$-flat
directions, but also
more  superpotential terms. Indeed a generic cubic superpotential
lifts all flat directions in these ``daughter'' theories \cite{leigh97}.
Notice that the 3-2 model is obtained through this decomposition from
the $SU(5)$ with ${\bf \bar 5}\oplus {\bf 10}$. In ref. \cite{dine96}
new models in this class, such as $SU(4)\times U(1)$ or $SU(6)\times U(1)$,
were shown to break supersymmetry and to lead to
interesting applications (see sect.~\ref{u1m}).
In refs. \cite{csaki96a,csaki96b}
also more complicated cases such as $SU(2n)\times SU(3)\times U(1)$,
$SU(2n+1)\times SU(4)\times U(1)$ and $SU(2n)\times SU(5)\times U(1)$ 
were shown to break supersymmetry dynamically. In the latter cases
one of the group factors has a dual description, in which 
supersymmetry breaking is manifest. In particular, in the dual theory, some
Yukawa couplings flow to mass terms, giving mass to a number of
flavours. By integrating them out the resulting effective theory 
generates a superpotential, and supersymmetry is broken in the standard way.
Finally, in ref. \cite{leigh97}, an argument for
broken supersymmetry was given 
in the general case $SU(k)\times SU(N-k)\times U(1)$.    
The approach was to consider the original $SU(N)$ theory
with the addition of an adjoint representation $\Sigma$ and a superpotential
containing the terms 
\beq
W_\Sigma=M\Sigma^2 +\lambda\Sigma^3.
\eeq
The classical vacua determined by $W_\Sigma$ have
gauge group $SU(k)\times SU(N-k)\times U(1)$ and
correspond to the ``daughter'' theories of interest.
By considering a dual \cite{kutasov95,kutasov96} of the full 
theory with $\Sigma$ one finds \cite{leigh97} that it has no
supersymmetric minima. Thus one expects no such state in any
of the ``daughter'' theories.  

To conclude this review of DSB mechanisms we mention
an example \cite{intriligator95} where confinement triggers supersymmetry
breaking.
Consider an $SU(2)$ gauge theory with matter $\psi$ in
an isospin-3/2 representation.
At the classical level there is one $D$-flat direction parametrized
by $u=\psi^4$. 
In ref. \cite{intriligator95} it is argued
that the theory confines with one massless
chiral multiplet parametrized by the modulus $u$. This hypothesis is consistent
with  't Hooft's matching of the $R$-symmetry anomalies $\tr R$ and $\tr R^3$.
Consider
now the addition of a perturbation
\beq
W_{tree}= {\psi^4\over M}~,
\eeq
where $M$ is some large mass. Classically, there
is a unique supersymmetric minimum at the origin $\psi=0$.
Notice that, when written in terms of $u=\psi^4$, the superpotential
is linear. With this parametrization, however, the classical
K\"ahler potential $K^c=(uu^\dagger)^{1/4}$ is singular at 
the origin. Therefore the classical potential $V(u)=|\partial_u W|^2/
K^c_{uu^\dagger}$ is zero at $u=0$. On the other hand,
if the theory confines, $u$ is the correct degree of freedom to describe
the low-energy effective theory, so the quantum K\"ahler metric should be
smooth (and positive) when expressed in terms of $u$. 
For instance at the origin we expect $K_{uu^\dagger}={\rm const}+
uu^\dagger+\dots$ Thus 
\beq
V=K_{uu^\dagger}^{-1}|\partial_u W|^2 >0
\eeq
at every point and supersymmetry is broken.
This phenomenon, DSB from smoothing a singularity in the K\"ahler metric,
takes place in other models too. One of these is the 4-3-1 model of 
refs. \cite{csaki96a,leigh97}, where the confined field
is a cubic  (rather then quartic) monomial of the elementary
fields. Therefore supersymmetry can be broken with a renormalizable
superpotential.

\section{Models for Dynamical Supersymmetry Breaking and Gauge Mediation} 
\label{models}
\setcounter{equation}{0}

\subsection{Early Attempts}
\label{early}

When the mass splittings inside supermultiplets arise at tree level,
the supertrace sum rule ${\rm STr {\cal M}^2=0}$ holds~\cite{ferrara79}.
This theorem prevents the construction of simple and  realistic
models in which supersymmetry is broken at tree level in the SM
sector. A generic implication is the existence
of sparticles below the mass range of quarks and leptons~\cite{dimopoulos81b}. 
The supertrace mass formula, however, does not hold
beyond tree level. It was soon realized that when the splittings
inside supermultiplets arise from radiative corrections,
the sparticles can all be made consistently heavier than
the SM particles. This was a motivation of the first gauge-mediated
supersymmetry-breaking models. Indeed the aim of 
refs.~\cite{dimopoulos81a,witten81,dine81,dine82b} 
was to build a ``supersymmetric technicolour''
theory, in which the breakdown of supersymmetry is due to some strong
gauge dynamics, while its mediation to the SM particles is just due 
to the usual SM gauge interactions. Those models came early
in the understanding of strongly coupled supersymmetric
theories, and some of the dynamical assumptions on which they were based
turned out to be false in the light of deeper tools
of analysis, such as the Witten index~\cite{witten82}. The idea
of gauge mediation emerged, however, as an independent, and general,
aspect of those models. Indeed the paradigm discussed
in refs.~\cite{dine82a,dine82c,alvarez82,nappi82,dimopoulos83},
consists of an O'Raifertaigh~\cite{orafertaigh} 
sector coupled to the messengers 
at the tree level. 

As an example   let us consider the messenger
sector described by~\cite{dine82a}
\beq
W=X(\lambda_1 \Phi_1^d\bar\Phi_1^d +\lambda_2 \Phi_1^t\bar\Phi_1^t -\mu^2) +
m_1\Phi_1^d\bar \Phi_1^d+m_2\Phi_1^t\bar\Phi_1^t+
M_3\Phi_1\bar \Phi_2+M_4\Phi_2\bar\Phi_1 ~,
\label{Oraferty}
\eeq
where $\Phi_{1,2}\oplus \bar\Phi_{1,2}$ collectively denote 
${\bf 5}\oplus {\bf \bar 5}$ of messenger fields, while $\Phi^{d}$ and $\Phi^t$
respectively denote their doublet and triplet components. The field
$X$ is a singlet. In a certain range of parameters the
absolute minimum at tree level is at $\Phi_{1,2}=\bar \Phi_{1,2}=0$,
with $F_X=\mu^2$ and $X$ undetermined. For $\lambda_1/\lambda_2\not = 
m_1/m_2$, the one-loop effective potential
generically fixes the VEV of $X$ such that $m_{1,2}'=m_{1,2}+X \lambda_{1,2}$
are both non-zero. In such a way the messenger supermultiplets are split
and non-zero soft masses arise
for both gauginos and sfermions, as discussed in sect.~\ref{structure}.
Notice that the above superpotential is not the most
general one that is consistent with the symmetries of the model, and that in
particular $m_{1,2}$ explicitly break the $R$ symmetry, which one could have
used to ``naturally'' enforce linearity in $X$. The weak principle of
naturalness can however be invoked in supersymmetric model
building, as the superpotential is not renormalized.
Notice also that it is crucial to have  $\lambda_1/\lambda_2 \not = 
m_1/m_2$, or else the masses $m_{1,2}$ can be eliminated by a shift in $X$,
after which the mimimum is at $X=0$. This preserves an $R$ symmetry and 
gauginos remain massless. 

In theories with tree-level supersymmetry breaking, like the one above,
the fundamental mass scales are just  inputs, whose origin is left 
unspecified. This implies, in particular, that the small size of the weak 
scale
itself, albeit stable against large quantum corrections, is left unexplained.
Undoubtedly, a more fundamental theory would be one
where the relevant mass scales arise dynamically.
The scale of supersymmetry breaking {\it can} indeed have this nice
property~\cite{witten81a}. This is because when supersymmetry is unbroken at tree
level, it will stay so to all orders in perturbation theory. The  reason
for that is the non-renormalization theorem, whose range of applicability
 however does not extend beyond perturbation theory.
Indeed there are theories where supersymmetry does break
 non-perturbatively.
Consequently the ratio between the 
scale of breaking and the fundamental mass scale
goes like $e^{-1/g^2}$, and it is exponentially
small at weak coupling.
This can explain in a very natural way the smallness of the supersymmetry
breaking 
(or weak) scale with respect to the Planck mass.  
Notice, indeed, the analogy with the case of technicolour theories, in which
a {\it chiral} symmetry,
unbroken in perturbation theory, is used to generate the weak-scale hierarchy.
A great advantage of supersymmetry, however, is that in many cases the 
non-perturbative effects that determine its breaking can be studied exactly.

The explanation of the gauge hierarchy is  one general motivation for being 
interested in field theories with DSB. 
As a
 matter of fact, in gauge-mediated
models the motivation is even stronger that in the gravity-mediated case.
This is because in gauge-mediated models, the whole dynamics of breaking
and mediation takes place at very low scales, where we should be
able to describe it field-theoretically. In principle this may not
be the case in the gravity-mediated scenario, 
where string theoretic effects may 
play a crucial r\^ole\footnote{Of course we must always plead ignorance of what 
makes the cosmological constant vanish, in gauge and in gravity mediation.}.

The first attempts to build realistic models of DSB
with gauge mediation were discussed in ref.~\cite{affleck85}.
Many of the gauge theories that are known to break supersymmetry
have flavour symmetries, which can remain unbroken in the vacuum.
A simple strategy~\cite{affleck85} is then to consider models with 
an anomaly-free flavour 
group $G_F \supset SU(3)\times SU(2)\times U(1)$ (or better $G_F\supset SU(5)$,
and consider only complete $SU(5)$ multiplets in order to 
preserve the unification of gauge couplings)
and weakly gauge it.  The resulting model still breaks supersymmetry as long as
the SM gauge force is weak, since the vacuum energy must be continuous in
the gauge coupling constant. 
Moreover the SM sector
``knows'' about supersymmetry breaking just by its gauge interactions.
In principle this idea seems  very promising. In practice, though,
there are problems to implement it, some of which are largely present today.
The main problem is the loss of asymptotic freedom in the SM gauge factors.
Models that both break supersymmetry and have $G_F\supset 
 SU(3)\times SU(2)\times U(1)$ have a large 
gauge group $G_S$.
{}From the point of view of the SM group (or $G_F$) the different ``colours''
of  $G_S$ are just different flavours. This means
that, $G_S$ being large, there are in general many flavours of 
SM matter in the supersymmetry-breaking 
(messenger) sector. In practice this makes
the gauge couplings of the SM blow up a few decades above the 
messenger scale. This is often well below the GUT scale. 
For instance, consider the class of DSB models
of refs.~\cite{affleck84a,meurice84},
based on $G_S=SU(N)$ ($N$ odd) with an antisymmetric tensor $A$ and
$N-4$ antifundamentals $\bar F_i$ in the matter sector. The smallest
model where an anomaly-free $SU(5)$ gauge group can be embedded is 
based on $SU(15)$, for which there are 15 families of messengers
above the supersymmetry-breaking scale. A way out could be to embed only
$SU(2)_W\times U(1)$ among the SM group factors in the messenger sector. 
But this would be problematic:
on one side gauge unification would be typically lost and on the
other the gluino mass would be tiny as it arises at a high
order in perturbation theory.

Another problem of the early attempts was the spontaneous
breakdown of the $R$ symmetry, which is rather generic in models
with spontaneous supersymmetry breaking, as discussed
in sect.~\ref{rsymmetry}. The resulting axion 
couples to the QCD anomaly and is phenomenologically problematic
as its scale $f_a$ is typically low $\sim 10$--$100$ TeV.
This can however be considered a smaller problem. Indeed there may exist
different sources of explicit $R$ breaking, which are
small enough not to restore supersymmetry (at least in a nearby vacuum) and
large enough to give the axion a mass that renders it phenomenologically 
harmless. In ref.~\cite{nelson94} it was shown that in some models
this r\^ole can be played by $1/M_P$ suppressed dimension-five operators in $W$.
Moreover, in ref.~\cite{bagger94} it was pointed out that, when 
$R$ is broken at the same scale as supersymmetry, the cancellation of the
cosmological constant, by adding a constant term to $W$ in supergravity,
implies an amount of explicit $R$ breaking that gives the axion an
acceptably large mass.
Another way out is to consider
special  $R$-breaking deformations of the DSB 
model~\cite{nelson94,seiberg93,poppitz96aa,leigh97}.
In some cases (see the example of the 3-2 model in sect.~\ref{mech}) the
stationary equations are not affected and supersymmetry remains broken,
although $R$-breaking effects give mass to the $R$-axion,
sometimes just from the K\"ahler potential 
(see the 4-3-1 model of ref.~\cite{leigh97}).
Alternatively, one can introduce a new strongly coupled gauge group under
which the $R$ symmetry carries an anomaly. Non-perturbative effects
give a large mass to the axion and, in most cases, do not restore
supersymmetry~\cite{nelson94}. 
Finally, there are now models where $R$ is broken at a  higher scale
than supersymmetry. This scale can consistently be in the astrophysically
allowed window $10^{9}$--$10^{12}$ GeV~\cite{turner90,raffelt90}, 
so that the $R$-axion could indeed 
solve the strong-CP problem.

\subsection{Messenger $U(1)$ Models}
\label{u1m}

\begin{figure}
\hspace{2cm}
\epsfig{figure=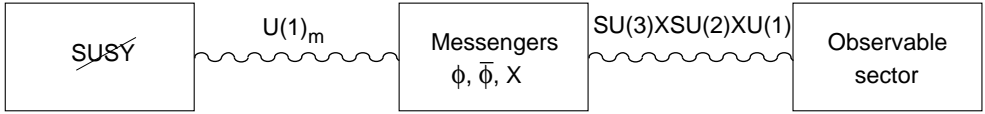,width=13cm}
\caption[]{\it
The modular structure of the messenger $U(1)$ models. 
\label{gmfig11}}
\end{figure}

The revival of DSB and gauge mediation, which was started by Dine, Nelson and 
collaborators~\cite{dine93,dine95,dine96} 
in 1993, had the main goal of tackling the
difficulties due to the loss of asymptotic freedom. In  order
to build realistic models, the approach was to assume that the DSB sector
is completely neutral under the gauge forces of the SM. The basic
structure of this class of models is shown in fig.~\ref{gmfig11}.
 There are basically
three sectors, one to break supersymmetry, 
one to mediate it and one containing the ordinary particles.
These sectors only communicate with one another by weak gauge forces.
Supersymmetry is first broken in sector I, and the news of it is transferred
to sector II (the messengers) via an Abelian gauge group $U(1)_m$,
denoted as messenger hypercharge. Sector III, the observable sector, is
neutral under $U(1)_m$, so that the breakdown of supersymmetry is felt
there only at the next step, through the usual gauge interactions. 

This structure may seem complicated, but
models in this class provided the first examples of a realistic theory
with calculable, dynamical, and universal  soft terms. Moreover, this
structure is quite generic and adaptive, as there is a vast class of
DSB models that can play the r\^ole of sector I.
Let us briefly outline a prototypical example discussed in ref.~\cite{dine96}.
The DSB sector is based on an $SU(6)\times U(1)\times U(1)_m$
gauge group, which comes from the reduction procedure illustrated in 
sect.~\ref{mech}, with the following matter content
\beq 
A({\bf 15},+2,0)\quad F({\bf 6},-5,0)\quad\bar F^{\pm}(\bar{\bf 6},-1,\pm 1)
\quad\bar F^0(\bar{\bf 6}, -1,0)\quad S^\pm({\bf 1},+6,\pm 1)\quad S^0({\bf 1},
+6,0).
\label{sixone}
\eeq
The tree-level superpotential is
\beq
W_I=\la{1} A\bar F^+\bar F^-+\la{2} F(\bar F^+S^-+\bar F^-S^+) 
+\la{3} F\bar F^0S^0.
\label{wsixone}
\eeq
This superpotential lifts all flat directions and respects a non-anomalous 
$R$ symmetry.
Moreover gaugino condensation generates a superpotential
\beq
W_{dyn}={\Lambda^7\over \sqrt{A^4\bar F^+\bar F^-\bar F^0\ F}} ~.
\label{lambda7}
\eeq
In eq.~(\ref{lambda7}) gauge indices have been neglected, and
there is only one gauge-invariant contraction, which 
respects the flavour $SU(3)$ symmetry of the $SU(6)$ strong interactions
under which $(\bar F^\pm ,\bar F^0)$ form a triplet.
Equation~(\ref{lambda7}) lifts the
origin, forcing the   $R$ symmetry to  be spontaneously broken. 
 Then, according to the arguments discussed in sect.~\ref{tools},
supersymmetry is also broken by 
non-zero $F$-terms.  Balancing the terms in
$W_{I}+W_{dyn}$ leads to a typical VEV for the scalars
$v\sim \Lambda/{\lambda}^{1/7}$, where $\lambda$ generically  denotes the
size of the Yukawa couplings in eq.~(\ref{wsixone}). For small 
enough Yukawas, $v\gg \Lambda$ and the theory can be analysed at weak 
coupling along the $D$-flat directions. Here the K\"ahler potential
can be reliably approximated by its tree-level value and the vacuum can be
studied in perturbation theory.  We refer for the details
to ref.~\cite{dine96}. The main points are the following. {\it i)} 
At the vacuum a $U(1)'_m\subset SU(6)\times U(1)\times U(1)_m$  is unbroken. 
{\it ii)} There are two light chiral superfields 
$\chi^\pm\subset \bar F^0$ with charge $\pm 1$ under $U(1)'_m$.
The scalar components of $\chi^\pm$ have a positive supersymmetry-breaking
 mass $m_\chi^2$.
{\it iii)} No $D$ term for $U(1)_m'$ is generated,
as there is an unbroken charge parity under which the $U(1)_m$ vector is odd.
This last property is very useful for practical model 
building~\cite{dine93,dine95,dine96}.

In the end, {\it i)--iii)} are the only properties of sector I, which are
 needed to build
a realistic messenger sector II.  This sector is described by
the following superpotential
\beq
W_{II}=k_1 X\varphi^+\varphi^-+k_2 X^3+ k_3 X\Phi\bar\Phi ~,
\label{wII}
\eeq
where $X$ is a singlet, $\Phi,\,\bar\Phi$ are the messengers,
and $\varphi^\pm$ have
charge $\pm 1$ under $U(1)_m$ and are singlets under the rest of the
gauge group. The fields $\varphi^\pm$
communicate with $\chi^\pm $ in module 
I via the $U(1)_m'$ gauge bosons. At two loops
the scalar components of
$\varphi^\pm$ receive soft masses 
$m^2_{\varphi}=-m_\chi^2(\alpha_m/\pi)^2\ln
(v/m_\chi)$, where $v$ is a scalar VEV in sector I, determining
the scale where $m_\chi$ is effectively generated.
In the calculable regime (see above) $v\gg\Lambda>  m_\chi$ holds,
and one obtains $m_\varphi^2<0$. 
This negative mass is crucial to obtain the desired vacuum.
The scalar potential in sector
II is the sum of the $F^2$-terms generated from $W_{II}$,
the $D_m^2\propto (|\varphi^+|^2-|\varphi^-|^2)^2$ term, and the 
negative soft masses for $\varphi^\pm$. In a certain range of parameters,
the minimum of the potential has a non-vanishing VEV for $\varphi^\pm$, 
$X$ and $F_X$, while the messenger $\Phi,\,\bar\Phi$ VEVs are zero. 
This determines 
a one-scale messenger sector with $X\sim \sqrt{F_X}\sim m_\varphi
\sim 10$--$100$ TeV.

We stress that models in this class were the first explicit examples of
calculable DSB with gauge mediation. The price that is being payed
is the modular structure: the three boxes in fig.~\ref{gmfig11}. The
supersymmetry-breaking sector is far removed from the observable
sector. A measure of
this fact is that the superfield $X$ has only a small
overlap with the goldstino: $F_X\sim (\alpha_m/\pi)^2 m_\chi^2\ll
m_\chi^2\lsim F_0$, where $F_0$ is the goldstino
decay constant. In these models, for $\alpha_m\sim 10^{-2}$ one
has $\sqrt{F_0}\gsim 10^7$ GeV. This scale is indeed in the most
dangerous range for the gravitino problem, as it requires a reheating
temperature below the TeV scale,
see sect.~\ref{cosmology}.

Moreover, the desired vacuum in sector II
turns out to be only  local.
This is easily seen by considering first the $m_\varphi^2=0$ limit.
Now we find a flat direction, see eq.~(\ref{wII}),  with $X=0$
and $F_X=k_1\varphi^+\varphi^-+k_3\Phi\bar \Phi=0$, along which
both $U(1)_m$ and $SU(3)\times SU(2)\times U(1)$ are broken. When 
the soft mass $m_\varphi^2<0$ is turned on, the potential along this
direction becomes increasingly negative until $\varphi^\pm \sim m_\chi$,
above which $m_\varphi^2$ goes to zero. At $\varphi^\pm \sim m_\chi$,
however, the potential is $V\sim -|m_\varphi^2 m_\chi^2|$ which is
much deeper than in the desired vacuum $V\sim -
 |m_\varphi^4|$. Notice that in the deeper minimum colour is broken.

The question of whether the local minimum at $F_X\not = 0$ is cosmologically
acceptable was discussed in detail in
ref.~\cite{dasgupta97}. There are two aspects
to this problem. One concerns the ``likelihood'' of being placed
in such a vacuum in the early Universe without being pushed out by thermal
fluctuations. The answer to this question is both difficult and dependent
on details of the early history of the Universe, but it is not unreasonable
to assume that the answer is positive. The other issue is 
the lifetime of the false vacuum in the cold Universe. For the 
model to be acceptable, 
this lifetime should necessarily be larger than the  age of the Universe. 
This requirement puts restrictions on
the parameter space. By studying a numerical approximation
to the bounce action, one finds~\cite{dasgupta97} that
the coupling $k_1$, see eq.~(\ref{wII}), must be smaller than $0.1$.
This bound is necessary but probably not sufficient. However,
one can conclude that with some adjustment of parameters this model can be
easily  made acceptable.
Another possibility is to enlarge the spectrum of sector II in order to make
the desired vacuum the global one.
By adding~\cite{dasgupta97,arkani97} another singlet $X'$ with the general
couplings
\beq
\Delta W= k_1' X'\varphi^+\varphi^-+k_2' {X'}^3+k_3' X'\Phi\bar\Phi
+(X,X'~ {\rm terms})
\label{sprime}
\eeq
the dangerous flat direction is lifted and the desired
vacuum is global for a range of parameters. 

\subsection{$SU(N)\times SU(N-k)$ Models with Direct Gauge Mediation}
\label{sun}

A first interesting attempt to build models of direct gauge mediation,
in which the messengers themselves belong to the DSB sector,
has been performed by Poppitz and Trivedi~\cite{poppitz97}
and by Arkani-Hamed, March-Russell, and Murayama~\cite{arkani97}.
The goal is still to find a DSB model with flavour group $G_F\supset SU(5)$,
but now the problem of Landau poles is circumvented by considering 
theories for which $X\gg \sqrt {F_X}$ in a natural way. The larger
value of the messenger mass $X$ can displace the Landau pole above the
GUT scale $M_G$, even for a ``large'' number of messenger
families $N_f\sim 5$--$10$. The idea is to consider DSB models like
those introduced in refs.~\cite{poppitz96aa,poppitz96a}, 
which have classically flat
directions that are only lifted by non-renormalizable operators 
$W_{nr}= \phi^n/M_P^{n-3}$. In the models of interest a non-perturbative
superpotential $W_{dyn}\sim \Lambda^{3-p}\phi^p$ is generated
by the gauge dynamics, where we collectively
denote all the fields by $\phi$. Thus for $p<1$, in the
limit $M_P\to \infty$, the potential slopes to zero at infinity
and the theory has no vacuum. For finite $M_P$, the vacuum is stabilized
at a geometric average $\phi^{n-p}\sim M_P^{n-3}\Lambda^{3-p}$ and
supersymmetry can be broken with generic $F$ terms $F_\phi\sim \phi^{n-1}/
M_P^{n-3}$. With the messenger masses and splittings given by
$\phi$ and $F_\phi$, one naturally obtains a two-scale model
with $F_\phi/\phi^2$ suppressed by a positive power of $\Lambda/M_P$.

 Notice
that $\phi$ is fixed by $\phi^{n-2}/M_P^{n-3}=F_\phi/\phi\sim 10^4$ 
GeV (we consider
large $N_f$). On the other hand, an acceptable suppression
of the gravity-mediated soft masses requires $\phi\lappeq 10^{15}$ GeV,
see eq.~(\ref{flapro}),
which translates\footnote{A weaker bound on $n$ can be obtained if we
assume that the mass scale of the non-renormalizable interaction is
lower than $M_P$~\cite{poppitz97}.} 
into $n\lappeq 7$. In the $SU(N)\times SU(N-1)$ and
$SU(N)\times SU(N-2)$ models of refs.~\cite{poppitz96aa,poppitz96a}, 
the flat direction
is lifted by a baryon, so that $n=N-1$ and $N-2$ respectively,
which puts a generic upper bound $N\lappeq 9$. This is already a powerful 
constraint.
The first attempt in ref.~\cite{poppitz97} 
focused on $SU(N)\times SU(N-2)$. In these
theories the flat directions can be lifted~\cite{poppitz96a} by 
preserving a flavour group $G_F=Sp(N-3)$,
which is also unbroken at the vacuum. The minimal $N$ for which an anomaly-free
$SU(5)\subset Sp(N-3)$ can be gauged is $N=13$.
Thus the gravity contribution to soft terms  is not suppressed but at best
comparable to the one induced by gauge effects. Models like these
are denoted as ``hybrid''. 

In ref.~\cite{arkani97}, the problem of large $N$ was tackled by allowing
an anomalous embedding of $SU(5)$ and by adding the suitable spectator fields
to cancel anomalies. 
On one side, this allows smaller $N$ but, on the other, the final
theory may not break supersymmetry, as 
the spectators can enter the superpotential.
A working example was, however, found in ref.~\cite{arkani97}, based on
a known~\cite{poppitz96aa,poppitz96a} $SU(7)\times SU(6)$ model.
The particle content is given by $Q({\bf 7},{\bf 6})$, $L^I({\bf \bar 7},{\bf
1})$, $I=1,\dots,6$,
$R_I({\bf 1},{\bf \bar 6})$, $I=1,\dots,7$. A group $SU(5)_W$ acting on the
indices $I=1,\dots,5$ is weakly gauged and we denote the corresponding
elements in $R_I$ and $L^I$ just by $\R$ and $\L$, dropping the indices.
Then a spectator field $\phi({\bf 1},{\bf 1},{\bf 5})$ 
under $SU(7)\times SU(6)\times 
SU(5)_W$ is added to cancel the $SU(5)^3_W$ anomaly. 
Imposing an $R$ symmetry and a
global $U(1)$ the most general superpotential is
\beq
W=\la{1} \L Q\R+ \la{2} L^6 Q R_6 +{\la{3}\over M_P}\L \phi Q R_6 +
{\la{4}\over M_P^3}B_7+{\la{5}\over M_P^4}\B \phi ~,
\label{wlqr}
\eeq
where the last two terms involve the $SU(6)$ baryons $B_7=\R^5R_7$ and
$\B_I=(\R^4)_IR_6R_7$ ($I=1,\dots,5$), 
with the obvious contractions. Equation~(\ref{wlqr})
 lifts
all flat directions involving the fields in this sector. According
to the discussion in ref.~\cite{arkani97}, the vacuum lies along 
a two-dimensional
$D$-flat direction where the diagonal $SU(5)\subset SU(6)\times 
SU(5)_W$ is unbroken. This $SU(5)$ is identified with the SM gauge interactions.
The VEVs are $\langle \R \rangle
=v_{\R} \times \identity_{5\times 5}$ and $\langle R_{6,7}^{j}\rangle =
\delta^{6 j}v_{6,7} $, where $j$ is the SU(6) gauge index, with
$|v_{\R}|^2=|v_6|^2+|v_7|^2$, as required by
$SU(6)$ flatness.
This direction is parametrized by the two baryons $B_{6,7}=\R^5R_{6,7}$.
The $R$ VEVs,
through $\la{1,2}$ in eq.~(\ref{wlqr}), give mass to all $SU(7)$ matter.
Moreover, under the unbroken
$SU(7)\times SU(5)_W$, the massive $Q\oplus \L\oplus L^6$ decompose as 
$({\bf 7},{\bf 5})\oplus ({\bf \bar 7},{\bf 5})
\oplus ({\bf 7},{\bf 1})\oplus ({\bf \bar 7},{\bf 1})$. 
Then the  supersymmetry-breaking dynamics
is very simple.  The low-energy $SU(7)$ pure gauge
theory generates a superpotential via gaugino condensation $W_{dyn}\propto
\Lambda_7^{15\over 7}B_6^{1\over 7}$. The competition of this term
with $B_7$ in eq.~(\ref{wlqr}) gives rise to a stable vacuum,
where supersymmetry is broken: $F_{B_{6,7}}\not = 0$.

It is useful to work in components and define $R_{{\tilde 6},{\tilde 7}}$
as the orthogonal
combinations of ${R}_{6,7}$ such that $R_{\tilde 7}$ has no VEV in
the scalar component. Then, with an appropriate gauge choice, we can
parametrize $\R =X\times \identity_{5\times 5}$,
$R_{\tilde 6}^j=\delta^{6j}X$, $R_{\tilde 7}^6=Y$,
$R_{\tilde 7}^j={\bar \Phi}^j$ ($j\le 5$), where $X,Y,\bar \Phi$ are
light fields and $\langle X \rangle =v_{\R}$, $\langle Y \rangle =
\langle \bar \Phi \rangle =0$. Notice that all $SU(5)_W$ charged fields
in $\R$ and $R_{\tilde 6}$ are eaten by the super-Higgs mechanism.

Parametrically the scalar and auxiliary field VEVs are given by
$\R \sim X\sim \Lambda_7^{5\over 12}M_P^{7\over 12}$ and
$F_\R \sim F_X \sim \Lambda_7^{25\over 12}M_P^{-{1\over 12}}$.
Thus there are 7 flavours of $SU(5)_W$
that get both supersymmetric  masses and splittings from $\R, F_\R\not = 0$
and act just as conventional messengers. Notice, in addition, that
they are a crucial
part of the DSB sector: the $SU(7)$ gaugino condensate that drives
DSB is proportional to $B_6$, {\it i.e.}  the messenger mass determinant.

This particular model suffers from a
problem: the dominant contribution to the ordinary sfermion mass squared is 
negative. This a generic problem of the existing models
of direct gauge mediation. We will discuss here one source of negative
contributions to
the mass squared, while another contribution 
will be discussed in sect.~\ref{plateau}.

The most important effect is related to $\phi({\bf 1},{\bf 5})$. 
This field is  needed to 
cancel the $SU(5)_W^3$ anomaly, {\it i.e.} to match the composite
$\B^j=(\R^4R_6R_7)^j=X^5\bar \Phi^j=({\bf 1},{\bf \bar 5})$, 
which is massless in the theory without $\phi$.
Its mass, see eq.~(\ref{wlqr}), is roughly $ \la{5}X^5/M_P^4$, which is
below the weak scale. The field $\B$ is however directly involved
in DSB. Its scalars get positive soft masses ${\cal O}(F_X/X)^2\sim 
(100~ {\rm TeV})^2$, leading to a positive supertrace in the $\phi\oplus \B$
sector. The supertrace
contributes, via two-loop RG evolution, to the masses of squarks
and sleptons 
\beq
\delta m_{\tilde Q, \tilde L}^2 \sim -\left(
{\alpha\over 4\pi} {F_X \over X}\right)^2\ln {X\over
m_\phi}<0 ~.
\label{negmass}
\eeq
This negative contribution
is parametrically $\ln(X/m_\phi)\sim 30$ times bigger that the standard
messenger result in gauge mediation. 

\begin{figure}
\hspace{3cm}
\epsfig{figure=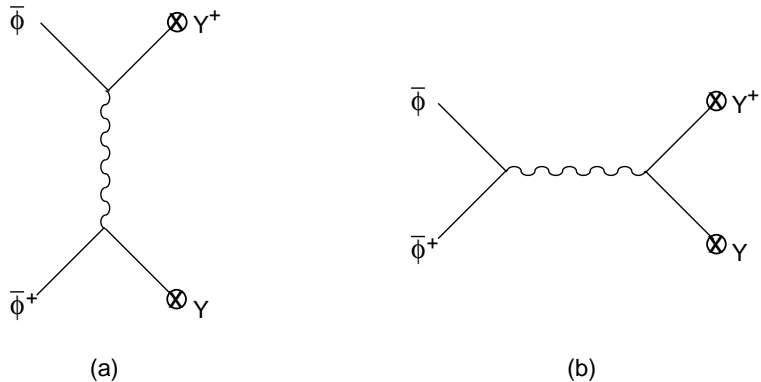,width=10cm}
\caption[]{\it
Tree-level contributions to the $\bar \phi$ soft masses from the $Y$ VEV.
\label{gmfig12}}
\end{figure}

This problem~\cite{arkani97,poppitz97a}
is common to models based on $SU(N)\times SU(N-k)$,
which have $k$ such light states. A positive mass supertrace
in the messenger sector
arises at tree level by decoupling the heavy vector superfields. It is easily
illustrated by using our parametrization in components, where
$R_{\tilde 7}$ contains a light $SU(5)_W$ antifundamental $\bar \phi$
(which we earlier parametrized by the baryon $\B$) and the singlet $Y$.
At the vacuum, $Y$ has zero scalar VEV, but $F_Y\sim F_X\not =0$.
Then, see fig.~\ref{gmfig12},
exchange of massive vectors in the coset $SU(6)\times 
SU(5)_W/SU(5)$, gives rise to positive soft  masses for $\bar\phi$
\beq
m_{\bar \phi}^2={|F_Y|^2\over 2 M_V^2}\left (1 - {1\over N}\right )>0~,
\eeq
where $M_V$ is the vector mass, and 
where $N=6$ corresponds to the present model.
The negative $-1/N$ contribution, which 
arises from the diagram of fig.~\ref{gmfig12}b is always
subdominant.

\subsection{Models with Quantum Moduli Spaces and Plateau Models}
\label{plateau}

Another pathway for constructing simpler DSB theories with gauge mediation
was opened by Izawa and Yanagida~\cite{izawa96a} and by Intriligator and 
Thomas~\cite{intriligator96a}. These authors considered gauge theories with
quantum moduli spaces (QMS)~\cite{seiberg93}, 
which do not break supersymmetry, and coupled 
them to singlets to obtain effective O'Raifertaigh models.
Now all field configurations where supersymmetry is unbroken belong only
to the classical moduli space, and are removed by quantum deformation.
The prototype example has been discussed in sect.~\ref{mech}, and it is 
based on
an $SU(2)$ gauge theory with 4 fundamentals $Q_i$, $i=1,\dots,
4$ coupled to 6 singlets $R^{ij}=-R^{ji}$, with the
tree-level superpotential given
in eq.~(\ref{ittree}). At the quantum level, the superpotential is given
by eq.~(\ref{ittot}) and supersymmetry is broken
as $F_{R^{ij}}$ and $F_A$ cannot
both vanish at the same time. 
However, the determination of the vacuum state
requires control of the K\"ahler potential. For the model in question,
the only range where this can be done perturbatively is 
$|\lambda \sqrt {\pf R}|\gg \Lambda$.
The dynamics in this region will be discussed  in the next section.
The bottom line is that, at large $X\equiv \sqrt{\pf R}$, the effective
potential is well approximated by $V(X)\simeq {|\lambda \Lambda^2|^2/ 
Z_R(XX^\dagger)}$, where $Z_R$ is the $R$ wave function.
By solving the one-loop RG equation for $Z_R$, one finds that $V$ increases
monotonically with $R$ in the perturbative region\footnote{Heuristically, 
$V$ behaves like the running 
coupling of a pure Yukawa theory since $Z_R$, at one loop, is only affected by 
$\lambda$. As $Z_R$ removes the flatness of the potential, the first
condition of the ADS criterion is verified only after including
one-loop effects.}. On the one hand this is
good, as the potential is stabilized at large $R$, on the other it is
problematic, as $R$ is driven to the region where we cannot control
 the K\"ahler potential. 

Thus it is natural to expect that the vacuum is either at
$R={\cal O}(\Lambda)$ or at $R=0$. In the former case, the field $R$
could be directly used to give mass to the messengers by adding
a term $R\Phi\bar \Phi$ to eq.~(\ref{ittree}). However,
the modified model has supersymmetry-preserving vacua where $R=0$
and $F_R=\lambda \Lambda^2+\Phi\bar \Phi=0$. These are similar  to those
existing in sector II of the messenger $U(1)$ models, discussed 
in sect.~\ref{u1m}. 
This is a problem, since there is no obvious parametric suppression
of the rate of decay of the false vacuum and the theory 
is probably cosmologically
unacceptable. 

A more realistic model requires an intermediate step in
the coupling between $R$ and the 
messengers~\cite{hotta97,izawa97,izawa97a,nomura97,nomura97a}. 
In particular, this was done in the model of 
ref.~\cite{hotta97} by adding a singlet $Y$ with superpotential
\beq
W_Y=\lambda_Y Y^3+\lambda_\Phi Y\Phi\bar\Phi.
\label{hoho}
\eeq
In this model, the superpotential in eq.~(\ref{ittree}) preserves only
an $SO(5)$ subgroup of the $SU(4)$ flavour group, under which the fields
$R$ transform as a ${\bf 1}+{\bf \bar 5}$: $R=(R_0,R_a)$, $a=1,\dots ,5$.
In a range of parameters, the vacuum has $F_{R_0}\not = 0$ and $F_{R_a}=0$.
Furthermore the field $Y$ is assumed to mix with 
$R_0$ in the K\"ahler potential
$K\supset \beta R_0Y^\dagger+ {\rm h.c.}$, where $\beta$
is a constant of order unity\footnote{Notice that we use  a 
field parametrization different from that in ref.~\cite{hotta97}. 
There,
$Y$ is chosen to couple to a combination of the mesons $M_{ij}$ in
the superpotential, while the kinetic terms are diagonal. The two
formulations are equivalent.}.
The basic dynamical assumption of ref.~\cite{hotta97} is that the vacuum of the
original model ($\beta=0$) is at  $R_0=R_a=0$.
At $\beta\not = 0$ the interactions with $Y$ and $\Phi,\bar\Phi$,
modify the vacuum in a perturbative manner. In a range 
of parameters one has a minimum with
 $\langle Y^2\rangle \sim \beta \Lambda^2$,
$F_Y\sim \Lambda^2/(4 \pi)^2$ and $\Phi=\bar \Phi=0$. Notice that $F_Y$
is generated by radiative corrections to the K\"ahler potential from
the interaction in eq.~(\ref{hoho}).
Then one has $F_Y/Y\sim \Lambda/(4\pi)^2$, so that the
usual gauge-mediated contributions to gauginos and sfermion masses squared
correspond respectively to two- and four-loop effects.
In this model, however, there are additional
and generically larger contributions to the sfermions. 
This is because the second term
in eq.~(\ref{hoho}) induces, at one loop, a non-vanishing supertrace in the 
messenger sector. Inserting this supertrace in the usual two-loop diagrams
for sfermion masses, we realize that there is an effective three-loop 
contribution to these masses. In order to avoid
sfermions that are a factor ${\cal O}(4\pi)$ heavier than gauginos,
the choice $\lambda_\Phi \lappeq \lambda_Y / (4\pi)$ must be made.
In spite of this adjustment of couplings, we still consider
the above model as an instructive example towards a realistic
gauge-mediated model with $X$ not much larger than $\sqrt{F_X}$.

A characterization of a class of realistic and calculable models
based on QMS was given in 
refs.~\cite{murayama97,dimopoulos97b}.
The basic  idea is to gauge part of the flavour group
of models like those based on the superpotential in eq.~(\ref{ittot}), so that
the fields $R$, which couple to the
mesons on the QMS, are no longer singlets. 
This allows the creation of 
a stable minimum far away from the origin $R\gg \sqrt 
{F_R}$. The theories are based on a gauge group 
$G=SU(5)_W\times G_B \times G_S$,
where $SU(5)_W$ again  stands for $\gws$, and $G_S$ is the strong
factor with a QMS. The messenger-sector superpotential is
\beq
W=\lambda_1R \Phi\bar \Phi +\lambda_2 R Q\Qb ~,
\label{simple}
\eeq
where $R$ transforms non-trivially only under $G_B$,
$\Phi\oplus \bar \Phi$ are  $SU(5)_W$ fundamentals and singlets of $G_S$. 
Finally, $Q\oplus \Qb$ form
a vectorial representation of $G_S$, but are singlets under $SU(5)_W$.
All the fields transform non-trivially under $G_B$. 
The following requirements must be satisfied.
{\it i)}~$G_S$ has a QMS, {\it i.e.} the Dynkin index $\mu_Q$ of  $Q\oplus \Qb$ 
equals  that of the adjoint $\mu_{G_S}$;
$G_S=SU(2)$ with four fundamentals is an example.
{\it ii)} $R$ contains just one $D$-flat direction parametrized
  by one invariant $u=u(R)\sim R^k$. 
{\it iii)} Along $u$, all the $\Phi,\bar \Phi$ and $Q,\Qb$  get
 masses $\sim \lambda_1 R$ and $ \sim\lambda_2 R$, respectively.

Far along $u\not = 0$, the low-energy theory
has gauge group $G=SU(5)_W\times H_B\times G_S$  (with $H_B\subset G_B$).
The only massless matter  
is given by the observable-sector fields plus the gauge singlet $u$. The
additional group factors are just pure gauge. It is also assumed that
$H_B$ is weak, so that only the strong dynamics of $G_S$ is relevant.

The effective theory picture is  simple.
Calling $\Lambda$ the strong scale of the original $G_S$ ($X\ll \Lambda$),
the scale $\Lambda_{eff}$ of the effective theory below $X$
is determined by one-loop matching, $\Lambda_{eff}=\Lambda
(\lambda_2 X/\Lambda )^{\mu_Q/3\mu_{G_S}}$.
Above the $G_S$ scale $\Lambda_{eff}$, the effective theory has
a vanishing superpotential,
$W=0$. Below the scale $\Lambda_{eff}$, $G_S$ confines and generates
a superpotential for $X$ via gaugino 
condensation.
The symmetries, holomorphy, and the QMS relation 
$\mu_Q=\mu_{G_S}$, constrain $W_{eff}$ to be linear in $X=u^{1/k}$
(where in the previous example $X=\sqrt {\pf R}$)
\beq
W_{eff}= \Lambda_{eff}^3=\Lambda^{3-\mu_Q/\mu_{G_S}}(
\lambda_2 X)^{\mu_Q/\mu_{G_S}}=\lambda_2 X\Lambda^2~.
\label{linearw}
\eeq
Now supersymmetry is broken at any point on the 
$X$ complex line. For $\mu_Q\not =\mu_{G_S}$ the scalar
potential $|\partial _XW_{eff}|^2$ would push $X$ either to the origin
or to infinity, where supersymmetry is in general  restored.
Models with the simple microscopic  superpotential in 
eq.~(\ref{simple})
have in general other flat directions at $X=0$
(involving the massless $\Phi$ and $Q$) where supersymmetry can be restored.
Unlike the models discussed at the beginning of this section, 
however, it is now possible to stabilize
$X\sim R$ far away from these points, {\it i.e.}  
$\langle X\rangle\gg \Lambda$.

In the region
$\lambda_2 X\gg \Lambda$, the K\"ahler potential is perturbatively calculable.
It is given by loops involving the heavy superfields $Q,\Qb$,
the messengers  $\F,\Fb$ 
and the heavy vector superfields in the coset $G_B/H_B$. All
these particles  get a mass from the VEV of $X$. At lowest order
the result corresponds to the one-loop anomalous dimension of $R$
\beq
K(X,X^\dagger)= XX^\dagger \left [ 1 +\fourpisq\bigl ( C_Bg_B^2
-C_1 \lambda_1^2-C_2\lambda_2^2\bigr )\ln(XX^\dagger/M_P^2)+\dots \right ] ~,
\label{oneloop}
\eeq
where the $C$'s are positive coefficients and $M_P$ is the cut-off scale.
By resumming the logs via the RG,
one gets
$ K(X,X^\dagger)= XX^\dagger Z_R\left (XX^\dagger/M_P^2\right )$,
where $Z_R$ is the running $R$ wave function.
The effective potential  is then
\beq
V_{eff}= {|\partial_X W|^2\over \partial_X\partial_{X^\dagger} K}\simeq
{\lambda_2^2 \Lambda^4\over  Z_R\left (XX^\dagger/M_P^2\right )}.
\label{effpot}
\eeq
 The logarithmic evolution of $Z_R$ with
$|X|$ can generate local minima in the effective potential.
Unlike the pure Yukawa case, $d \ln Z_R^{-1}/d \ln X$ is not necessarily
negative at every $X$, as the gauge and Yukawa contributions can
balance each other at some scale. 
It is crucial that $X$ be part of a field $R$, which is charged
under $G_B$. This provides
the gauge contribution $g_B^2$ without which $X$ is always  pushed 
towards the origin, where we lose control of our approximation and typically
restore supersymmetry. 
The stationary points arise via the Coleman--Weinberg mechanism and
are determined by the zeros of the anomalous dimension for $R$
\beq
8 \pi^2 {d \ln Z_R\over d\ln|X|}=C_B {\bar g}_B^2-C_1{\bar \lambda_1^2}
-C_2{\bar \lambda_2^2}=0~.
\label{stationary}
\eeq
Here ${\bar g}_B, \bar \lambda_1, \bar \lambda_2$ are the running couplings
evaluated at the scale $|X|$.  Moreover, 
in order to ensure that the stationary
point be a minimum, one should verify that $d^2 Z_R/d (\ln X)^2$ is negative.
This mechanism for stabilizing a tree-level flat potential
is just Witten's inverse hierarchy~\cite{witten81a}.  It is now natural
to expect $X\gg {\sqrt {F_X}}\sim \Lambda $. This hierarchy ensures
that the minimum on the plateau, even if it is not the true ground 
state, is nonetheless stable on cosmological time scales.
In ref.~\cite{dimopoulos97b} the rate of tunnelling into the 
true supersymmetric
vacuum at $X=0$ was estimated by using the semiclassical 
approximation~\cite{coleman}, and was found to be larger than the lifetime
of the Universe as long as $X\gappeq 10~\Lambda$.  Again, here there is the open
question of why this particular local minimum is chosen by the
cosmological evolution.

A simple way to implement this idea is to consider $G_S=SU(2)$ and
$G_B =SU(2)\times SU(2)\sim SO(4)\subset SU(4)$, where $SU(4)$ is the flavour
group of the prototype model based on the superpotential in eq.~(\ref{ittot}).
The spectrum under $SU(2)_S\times SU(2)_{B1}\times  
SU(2)_{B2}\times SU(5)$ is
\bea
Q = ({\bf 2},{\bf 2},{\bf 1},{\bf 1}),
\quad\bar Q = ({\bf 2},{\bf 1},{\bf 2},{\bf 1}),
\quad R = ({\bf 1},{\bf 2},{\bf 2},{\bf 1}),\nonumber\\
  \F_3\oplus \F_2\oplus \F_1= 
({\bf 1},{\bf 1},{\bf 2}, {\bf 3}\oplus {\bf 2}\oplus {\bf 1})\quad 
\Fb_3\oplus
\Fb_2\oplus \Fb_1 = ({\bf 1},{\bf 2},{\bf 1},{\bf \bar 3}
\oplus {\bf \bar 2}\oplus {\bf \bar 1})~,
\label{2cubematter}
\eea
where the $\F_i$'s have been split into their $\gws$ irreps. Notice that
a pair of SM singlets $\F_1,\Fb_1$ is added in order
to cancel the $SU(2)_{B1}\times SU(2)_{B2}$ global anomaly.
The tree-level superpotential is given by
\beq
      W =  R\left (\lambda Q\Qb +\lambda_3 \F_3 \Fb_3+ \lambda_2 \F_2 \Fb_2
+\lambda_1 \F_1 \Fb_1\right ).
\label{microw}
\eeq
The classically flat direction of interest is
$X = (\det R)^{{1 \over 2}}$. Along $X$, we obtain $\langle R\rangle
\propto \identity_2$,
$H_B=SU(2)$ and conditions {\it i)}--{\it iii)}
are satisfied.   By studying the RG equations one finds~\cite{dimopoulos97b}
a  significant region
of parameter space where this model has a local mimimum on the 
plateau at large $X$. There it behaves as
a conventional gauge-mediated supersymmetry-breaking model with two 
messenger families. The messenger mass $X$ can be anywhere between $10^9$ and
$10^{14}$ GeV, where the lower bound is determined by asking perturbative
control over the K\"ahler potential. Notice also that at the upper edge
$X\sim 10^{13}$--$10^{14}$ nucleosynthesis is problematic, see 
sect.~\ref{cosmology}. Around the minimum
the scalar part of the goldstino superfield $X$ gets a mass
$m_X^2\sim (\alpha/4\pi)^2 (F_X/ X)^2$,
which is of the order of the sfermion masses, $\sim 10^2$--$10^3$ GeV.
At this stage,
the pseudoscalar part is the $R$-axion. Notice that $\langle X\rangle $, 
the axion scale,  as far as gauge mediation
is concerned, can consistently be in the $10^9$--$10^{12}$ GeV axion 
window. So ${\rm Im} X$ provides an interesting QCD axion candidate.

A special and more ambitious possibility is to try to
identify $Q,\Qb$ with $\F,\Fb$, so that
the messengers themselves trigger supersymmetry breaking.
This was done in ref.~\cite{murayama97}, in a theory based on
$Sp(4)\times SU(5)\times SU(5)$, and in ref.~\cite{dimopoulos97b},
in a theory based on $\fivecube$. Let us consider, for illustrative
purposes, the $\fivecube$ model, which is indeed quite simple.
The matter content is given by
$R(\one,\five,\fivebar)$, $\F(\fivebar,\one,\five)$ and  
$\Fb(\five,\fivebar,\one)$. Again $SU(5)_W$ is
already broken to $\gws$ and the standard matter transforms as
${\bf 5}\oplus \bar{\bf 10}$ of $SU(5)_W$. The most general
renormalizable superpotential is
\beq
W=\lambda R\F\Fb ~,
\label{tree5}
\eeq
which lifts most flat directions. 
Along  $u\equiv \det R=X^5$ this model satisfies conditions {\it i)--iii)} and 
$W_{eff}=\lambda X \Lambda_1^2$. Notice that $\langle R\rangle =X~\identity_5$
breaks $SU(5)_2\times SU(3)\times SU(2)\times U(1)\to  SU(3)\times SU(2)\times 
U(1)$, the SM gauge group. A detailed numerical 
study~\cite{dimopoulos97b} shows that a stable minimum can be consistently
found for any $X\gsim 10^{14}$ GeV. This is low enough to suppress gravity
contributions to soft terms. Minima at smaller $X$ turn out
to imply a Landau pole in $\lambda$ below the Planck scale. 
The fields $\F$, $\Fb$ with mass $\lambda X$ and splittings $\lambda F_X$,
correspond to five flavours of conventional messengers. Notice that this
model is a special case of the $SU(N)\times SU(N-k)$ 
models discussed in sect.~\ref{sun} ($k=0$), in which we do not need
non-renormalizable interactions to stabilize the potential.
The problem of the positive supertrace is solved since
there are no light messengers. 

The actual problem is that, in both models 
in which $Q,\bar Q$ are identified with the messengers,
there are fatal negative
contributions to the sfermion masses
coming from the massive vector superfields. Indeed the SM
gauge bosons live in the diagonal subgroup of $SU(5)\times SU(5)$.
The vector multiplets corresponding to the coset space interact 
with the SM matter and feel supersymmetry breaking, 
since their masses are determined by the superfield $X$.
Their contribution to the sfermion masses has been calculated in 
ref.~\cite{giudice97}.
It turns out that although
squark squared masses are stabilized at low
energy by the positive gluino contribution,  slepton squared masses
are negative.
It is not clear at the moment if there exist simple and viable models with 
gauge messengers. Notice that this problem is fatal for the models in 
which $Q,\bar Q$ are identified with the messengers, but of course
it is not present in models with no gauge messengers, as in
the example presented above, based on $G_S=SU(2)$ and 
$G_B=SU(2)\times SU(2)$. 
The negative contributions from  heavy vector loops
are present also in the $SU(N)\times SU(N-k)$ models. But in that case 
there is an even larger negative contribution from the light messengers,
as discussed in sect.~\ref{sun}.

An interesting and viable $SU(5)^3$ theory has been discussed in 
ref.~\cite{shirman97}. The embedding 
of the SM gauge group is changed in order to
avoid gauge messengers. Under $SU(5)_1\times SU(5)_2\times SU(5)_W$
the spectrum contains $R(\fivebar,\five,\one)$, $\F(\one,\fivebar,\five)$  
$\Fb(\five,\one,\fivebar)$, $A({\bf 10},\one,\one)$ and $\bar F(\fivebar,\one,
\one)$. Now, along the flat direction $\det R$, the weak factor $SU(5)_W$ is 
unbroken, unlike the previous case, since the quantum number assignment
is now different. The analysis in ref.~\cite{shirman97} shows that
the superpotential in eq.~(\ref{tree5}) and the strong dynamics lift 
all flat  directions but $\det R$.
Along $\det R$, $ SU(5)_1\times SU(5)_2$ is broken to the diagonal $SU(5)_D$,
whose matter content is just $A\oplus \bar F= {\bf 10}\oplus {\bf \bar 5}$.
Thus the low-energy $SU(5)_D$ dynamics break supersymmetry in the strong
coupling regime~\cite{affleck84a,meurice84}. Although this happens in an
uncalculable way, one can estimate the vacuum energy to be~\cite{shirman97}
\beq
V\sim \Lambda_D^4\sim \left ({\Lambda_1^8\Lambda_2^{10}
\over \det R}\right )^{4/13} ~,
\label{runaway}
\eeq
where $\Lambda_D$ and
$\Lambda_{1,2}$ are
the effective scales for $SU(5)_D$ and $SU(5)_{1,2}$.
Equation~(\ref{runaway}) 
leads to a runaway behaviour of
the field $R$. In general this slope can be compensated at large $R$
by Planck-suppressed corrections to the tree-level superpotential. The 
lowest-dimension operator of this kind 
is $W_1=\det R/M_P^2$. By adding $W_1$ to eq.~(\ref{tree5}), 
the vacuum relaxes
at $\langle R\rangle \sim (M_P^{13}\Lambda_1^8\Lambda_2^{10})^{1/31}$.
Around this minimum, $\F$ and $\Fb$ act as 5 flavours of messengers, while the
heavy gauge fields are neutral under $SU(5)_W$. 
By 
fixing $F_R/R\sim 10^4$ GeV the messenger scale $\langle R\rangle $ is 
then $\sim 10^{13}$ GeV. This scale could be problematic for
nucleosynthesis, see sect.~\ref{cosmology}.
  
Planck-suppressed effects in the K\"ahler potential play a central r\^ole
in the mechanism proposed in ref.~\cite{NiPo}. 
This scenario requires a simple
messenger sector superpotential $W=kX^3+\lambda X\Phi\bar \Phi$. The coupling
of $X$ to a bilinear in the supersymmetry-breaking sector,
\beq
\int d^4\theta {QQ^\dagger X\over M_P},
\label{qqx}
\eeq
generates a tadpole
for $X$ when $(QQ^\dagger)\to |F_Q|^2\theta^2\bar\theta^2$. Over a wide 
range of $k$ and $\lambda$ this tadpole leads to an absolute minimum in
the messenger sector at $\Phi,\bar \Phi=0$ and at
\beq
X\sim m_{3/2}^{2\over 3}M_P^{1\over 3}\quad\quad {F_X\over X}\sim X,
\label{vacnilles}
\eeq
where $m_{3/2}\sim F_Q/M_P$ has been used. Notice that, because of the 
well-known ``destabilizing'' effects of 
singlets~\cite{polchinski82,nilsred,lahanas},
the visible scale $X$ of supersymmetry breaking is parametrically
larger that $m_{3/2}$, a property crucial to solve the flavour problem. 
In the vacuum defined by eq.~\ref{vacnilles}
gauge mediation proceeds in the usual way. This scenario gives
automatically $F_X\sim X^2$. In order to have $F_X/X\sim 10^5$ GeV it must
be $F_Q\sim 10^8$ GeV so that $m_{3/2}$ is in the MeV range. This scenario
shares an obvious similarity with that discussed below eq.~\ref{hoho},
as the VEV of $X$ is triggered by the K\"ahler potential. It also shares the
overall structure of messenger $U(1)$ models, but now the 
connection between the first two boxes in Fig. 11 is determined by $1/M_P$
effects. An advantage of the scenario of ref.~\cite{NiPo} is that an absolute
minimum is obtained rather naturally.

A different class of models, based on a superpotential such as 
eq.~(\ref{simple}),
has been discussed in ref.~\cite{luty97b}. The field $R$ is now a gauge
singlet. The corresponding flat direction is lifted by two competing
quantum effects: gaugino condensate in one group factor lifts the origin
$R=0$ and DSB in another factor
stabilizes the potential. A subclass is based on the gauge group
$SU(N)_S\times SU(5)_B\times SU(5)_W$. The fields in eq.~(\ref{simple})
tranform as
$\F(\one,\fivebar,\five)$, $\Fb(\one,\five,\fivebar)$, $Q^i=({\bf N},\one,
\one)$ and $\bar {Q_i}({\bf \bar N}, \one ,\one )$, where $i$ is a 
flavour index $i=1,\dots, N_f$  and $Q\bar Q$ in
eq.~(\ref{simple}) is identified with $\sum_i Q^i\bar{Q_i}$.
The matter content is completed by $A(\one,{\bf 10},\one)$
and $\bar F(\one,\fivebar,\one)$, which do not appear in the superpotential.
Along $R\not = 0$, $SU(N)_S$ is a pure supersymmetric Yang--Mills theory, 
and gaugino condensation generates
a superpotential $W_S\propto R^{N_f/N}$. Then for  $N_f<N$, the potential
$V_S=|\partial_R W_S|^2$ will ``push'' $R$ away from the origin. 
However, at $R\not = 0$, the gauge group $SU(5)_B$, whose only 
matter is contained in
${\bf 10}\oplus \fivebar$, breaks supersymmetry with a vacuum
energy $V_B\sim \Lambda_B^{32/13}R^{20/13}$, which stabilizes $R$.
The bottom line is that, for $\Lambda_B\ll \Lambda_S$, $R$ is stabilized
in a supersymmetry-breaking ($F_R\not=0$) minimum far away from the origin. 
This is typically a false
vacuum but, as was the case for plateau models, it is cosmologically 
acceptable, since the true vacuum is ``far'' away in field space. 

An interesting mechanism to stabilize a flat direction, different than the one
presented in this section, can be obtained from confinement, as
discussed in ref.~\cite{luty97c}.
This, however, can only happen in the region
of field values of the order of the confinement scale $\Lambda$.
The resulting phenomenological models have interesting features, such
as composite messengers, but require assumptions
about uncalculable terms in the K\"ahler potential.

\subsection{Models with Composite $X$}

In refs.~\cite{randall97,haba97,haba97a,shadmi97,csaki98} it 
was assumed that the
field $X$ corresponds to a gauge-invariant composite of the 
supersymmetry-breaking sector. This assumption was made to avoid intermediate
sectors involving apparently {\it ad hoc} singlet fields.
Indicating by $K$ the invariant playing the r\^ole of $X$, the
coupling to the messenger will have the form 
\beq
W={K\F\Fb\over M^{d-1}} ~,
\label{highd}
\eeq
where $d\geq 2$ is the dimension of $K$, and $M$ is a fundamental
mass scale. The basic feature
of eq.~(\ref{highd}) is that while $X^2$ scales like $1/M^{2(d-1)}$,
$F_X$ scales only like $1/M^{d-1}$. By treating $1/M$ as a small parameter,
one expects a qualitative difficulty in getting $F_X<X^2$ to
avoid a tachyon. In ref.~\cite{shadmi97} it was found that this
generically requires a new mass scale $M\ll M_P$ to suppress these operators,
since
$M=M_P$ is too large. Alternatively~\cite{haba97,shadmi97}, one must
assume some small (less than $10^{-6}$) coupling in the supersymmetry-breaking
sector, or try to stabilize the superpotential by some higher-dimensional
operator~\cite{haba97a}. 
A possibility is that this new mass scale arises from some
confining gauge dynamics involving fields in both the observable and
supersymmetry-breaking sectors. But this seems difficult to realize
without creating new and dangerous terms in the effective Lagrangian.

The scenarios characterized in ref.~\cite{randall97}, and constructed 
in ref.~\cite{csaki98}, realize the new mass scale as the mass
of two composite fields $S$ and $\bar S$. These are coupled to both 
sectors and by integrating them out, they mediate the interaction
in eq.~(\ref{highd}). This intermediate step in the communication of
supersymmetry breaking avoids the appearance of dangerous interactions.
Thus the effective scale in eq.~(\ref{highd})
is $M=M_P^k\Lambda_c^{1-k}$, with $0<k<1$, where $\Lambda_c$
is a combination of strong-dynamics scales smaller than $M_P$. 
One of the examples of ref.~\cite{csaki98} is based on a gauge group
$SU(3)\times SU(2)\times [SU(2)\times SU(4)]_c\times SU(5)_W$, where the
factor in brackets provides the confining dynamics. It is rather non-trivial 
that the superpotential from confinement, together with a limited set
of tree-level operators, gives rise to the needed interactions. It is also
conceivable that realizations simpler  than those presented
in ref.~\cite{csaki98} exist.
Nevertheless, albeit without gauge singlets, the overall modular structure
of fig.~\ref{gmfig11} seems to be present here, with the confining
sector playing the r\^ole of module II. More generally one could measure
the ``remoteness'' of the supersymmetry-breaking sector by observing
that $F_X$ is much smaller than $F_0$,
the gravitino decay constant, as is always the case  when $X$ is
composite. Nonetheless,  the composite scenario, together with the ones
discussed in the previous sections, illustrates the many different ways
gauge mediation can be realized in the context of dynamical supersymmetry
breaking.
It is noteworthy that, in essentially all these models, the phenomenologically
relevant parameters correspond to the quantities $M$, $F$, and $N$ introduced
in sect.~\ref{structure}.

\section{The Origin of $\mu$ and $B\mu$}
\label{muprob}
\setcounter{equation}{0}

\subsection{The $\mu$ Problem}

Usually, the $\mu$-problem~\cite{kim84} is referred to as 
the difficulty in generating the correct mass scale for the Higgs
bilinear term in the superpotential
\beq
W=\mu H_1H_2~,
\label{mu}
\eeq
which, for phenomenological reasons, has to be of the order of the weak
scale. If this term were present in the limit of exact supersymmetry,
then it would have to be of the order of the Planck scale or some
other fundamental large scale (as the GUT scale). In the gravity-mediated
scenario, however, if this
term is absent in the supersymmetric limit and the K\"ahler potential is 
general enough, the correct 
$\mu$ is generated after supersymmetry is broken~\cite{giudice88}.
In particular the two terms
\beq
K=H_1 H_2 \left ({X^\dagger \over M_P}+{XX^\dagger \over M_P^2}+\dots\right )
~,
\label{muplanck}
\eeq
where $X$ is the supersymmetry-breaking chiral spurion, generate respectively
$\mu \sim F_X/M_P$ and $B\mu \sim (F_X/M_P)^2$. 
Therefore the sizes of both $\mu$ and $B\mu$ are of the order of the weak
scale.
Here $B\mu$ is the 
supersymmetry-breaking counterpart of $\mu$, defined by the following term
in the scalar potential
\beq
V= B \mu H_1 H_2 + {\rm h.c.}
\eeq

In gauge mediation the problem is somehow more acute. Now $\mu$ must be related
to the scales $X$ and $F_X$ in the messenger sector, and the natural
value is $\mu \sim(1/16 \pi^2) F_X/X$. Indeed it is not
hard to accomplish that, as we will show below. 
The real difficulty is to generate {\it both}
$\mu$ and $B$ of the same order.
Consider the simplest possibility of a direct coupling $W=\lambda X H_1  H_2$.
Even putting $\lambda \sim 1/16 \pi^2$ by hand does not accomplish
the job since $\mu = \lambda X$ while $B\mu = \lambda F_X$, so that $B= F_X/X
\sim 10$--$100 $ TeV. The source of this problem is not the tree-level origin
of $\mu$, but the fact that $\mu$ and $B\mu$ are generated at the same order
in the ``small'' parameter $\lambda$. 

To illustrate this, let us consider 
the more realistic situation of a radiatively generated $\mu$. A simple option
to do that is to couple $H_1$ and $H_2$ to a pair of messengers~\cite{dvali96}
\beq
W=\lambda H_1\Phi_1\Phi_2+\bar\lambda H_2\bar\Phi_1\bar\Phi_2\, .
\label{couplinga}
\eeq
Assuming that a single  $X$ determines the messenger masses,
$W=X(\lambda_1 \Phi_1\bar\Phi_1+\lambda_2 \Phi_2\bar\Phi_2)$, then
the one-loop correction to the K\"ahler potential is 
\beq
\int d^4 \theta\left[ {\lambda\bar\lambda\over 16 \pi^2}f(\la{1}/\la{2})
  H_1H_2 {X^\dagger \over X}+{\rm h.c.}\right] ~.
\label{Kahlermu}
\eeq
Here $f(x)=(x\ln x^2)/(1-x^2)$ and we have neglected higher-derivative
terms, which only give ${\cal O}(F_X^2/X^4)$ corrections. Both $\mu$ and
$B \mu$ are generated from eq.~(\ref{Kahlermu}) and we again obtain
\beq
B = {B\mu\over \mu}={F_X\over X}\, .\label{problematic}
\eeq
A similar result is expected also in the general case
where $X^\dagger/X$ in eq.~(\ref{Kahlermu}) is replaced by some
function $G(X,X^\dagger)$. Only for specific, {\it tuned}, 
functions\footnote{For instance,
$G=\ln(X/X^\dagger)$ gives $B=0$, but
we do not know of any explicit example giving this result.}
$G$ can one get a small $B$.

Equation~(\ref{problematic}) is the expression of the $\mu$-problem in
gauge mediation. 
The electroweak-symmetry breaking
condition can be satisfied 
only at the price of a quite unreasonable fine-tuning.
In order to solve this problem and avoid the relation in 
eq.~(\ref{problematic}), $\mu$ should
neither be generated from a direct coupling to $X$ at tree level 
nor via loop corrections to the 
K\"ahler potential. Let us now discuss possible solutions that have
been proposed in the literature. 

\subsection{Dynamical Relaxation Mechanism}

The basic point of the solution outlined in ref.~\cite{dvali96} 
is that $\mu$ arises only from higher-derivative operators of the form
\beq
\int d^4\theta H_1 H_2 D^2 G(X^\dagger, X)\, .
\label{qqq}
\eeq
Here $D_\alpha$ is the supersymmetric covariant derivative and $G$
a function determined by loop corrections.
The crucial point is  that a 
$D^2$ acting on any function of $X$ and $X^\dagger$ always produces
an antichiral superfield. Then, independently of the form of $G$, no
$B$ term is generated along with $\mu$ from eq.~(\ref{qqq}). 
Different operators coming from higher-order
corrections can be responsible of generating $B$ of the correct size.

A possible implementation of this mechanism is given by the
superpotential
\beq
W=S\left( \lambda_1 H_1 H_2 +\frac{\lambda_2}{2}N^2+\lambda \Phi \bar \Phi
-M_N^2\right) +X\Phi\bar \Phi~,
\label{spot}
\eeq
where $\Phi \oplus \bar \Phi$ are the messengers, ${\bf 5}\oplus
{\bf \bar 5}$ of $SU(5)$,
and $S,N$ are gauge singlets.
It is assumed that the DSB dynamics provides a VEV for $X$, $F_X$
and also the mass $M_N^2\sim F_X$. For a given model of DSB, one should
also check that the interactions in eq.~(\ref{spot}) do not 
destabilize the original DSB vacuum.
We will later illustrate an 
explict DSB model, which generates all the appropriate mass terms.

The dynamics of this model is easily described. In the limit in which the 
source $F_X$ of supersymmetry breaking is turned off, there is a minimum of the
energy at $S,H_1,H_2 =0$ with $F_S=0$ saturated by a non-zero VEV
$\la{2}\langle N\rangle^2=2M_N^2$. Around this vacuum $S$ and $N$ pair up and
get a mass $\sqrt{ 2\la{2}}M_N$. At this stage $\mu\propto \langle S
\rangle =0$.
Notice now that messenger exchange in one-loop diagrams produces 
a wave-function mixing between
$S$ and $X$ in the K\"ahler potential
\beq
  \int d^4\theta \frac{5\lambda}{16 \pi^2}SX^\dagger \ln (XX^\dagger) 
+{\rm h.c.}~
\eeq
Thus, when $F_X$ is turned on, a one-loop tadpole $\delta V \sim SF_X^2/X$
appears in the scalar potential and a non-zero $\mu$ is generated
\beq
\mu =\lambda_1\langle S\rangle 
=-\frac{5}{32\pi^2}\frac{\lambda\lambda_1}{\lambda_2}\frac{F_X^2}{XM_N^2}
 ~.
\label{zip}
\eeq
At the same order no $B\mu$ is generated or, equivalently, $ F_S=0$.
This is understood by noticing that {\it i)} $F_S=F_S(N)$ is a function
of $N$; {\it ii)} at one loop the effective potential depends on $N$
only via $|F_S(N)|^2$, as $N$ does not interact directly with $\F, \Fb$.
Then, at one loop, the minimization of the energy term $F_S^2$ ensures that
$B\propto F_S$ relaxes to zero. 
At two loops, soft masses for $N$ are generated and $F_S$ becomes
non-vanishing:
\beq
B \mu = \lambda_1\langle F_S \rangle =-\frac{10\lambda^2\lambda_1
\lambda_2}{(16\pi^2)^2}
\left( 1+\frac{5F^2}{8\lambda_2^2M_N^4}\right){F_X^2\over X^2} ~.
\label{zap}
\eeq
In addition to these terms, new
contributions to the soft masses for $H_1$, $ H_2$ arise at two loops
\beq
\delta m_{H_1}^2=\delta m_{H_2}^2=10
\left( \frac{\lambda \lambda_1}{16\pi^2}\right)^2{F_X^2\over X^2} ~.
\label{hb}
\eeq
In an operator language, it is easy to verify that $\mu$ in this
model is generated by eq.~(\ref{qqq}) after decoupling $N$ and $S$.
Notice that the fields $N$ and $S$ have large masses of order $\sqrt{F_X}$,
and therefore belong to the messenger sector and  no 
singlet fields are contained in the low-energy theory. 

This mechanism can be implemented in a DSB theory, where $M_N^2$
in eq.~(\ref{spot}) is not just an input. 
Models with QMS are well suited for this:
 the meson condensate $\langle Q\bar Q\rangle = \Lambda^2\sim F_X$ is the
natural source of $M_N^2$ via the microscopic coupling $S Q\bar Q$.
This was explicitly realized in ref.~\cite{dimopoulos97c} in a theory based
on the gauge group  $SU(5)_S\times SU(5)_W\times SU(2)_B$. In that model
the messengers and the confining fields coincide
$\F\oplus\Fb=({\bf 5},{\bf \bar 5},{\bf 1})\oplus ({\bf \bar 5} ,{\bf 5},{\bf
 1})$ and the superpotential is
\beq
W = S(\lambda_1 H_1 H_2 + \lambda_2 N^2/2  - \la{3} \F\Fb)
 +\la{4} Y\F\Fb + I(\la{5} Y^2 - \la{6} \psi\bar{\psi}) ,
\label{relax}
\eeq
where $I,Y$ are additional singlets, $\psi$, $\bar \psi$ are $SU(2)_B$
doublets and, finally, all $\la{i}$ are ${\cal O}(1)$ Yukawa couplings.
Equation~(\ref{relax}) has a classical flat direction where
$\la{5}Y^2=\la{6}\psi\bar \psi\equiv X^2\not = 0$ and all the other fields 
are at the origin. Along $X$ the messengers $\Phi\bar\Phi$ are massive,
so that the $SU(5)_S$ gaugino condensate generates $W_{eff}\sim X\Lambda^2$,
as in the models described in sect.~\ref{plateau}. 
Notice also that $X$ is an admixture of the singlet
$Y$ and the 
$SU(2)_B$ doublets $\psi$, $\bar \psi$, so that this second component 
can lead to a stable minimum on the $X$ plateau via Witten's
inverse hierarchy. Thus, below the scale $X$, $W_{eff}$ has the form
of eq.~(\ref{spot}), with $\lambda_3 \F\Fb\to \Lambda^2\equiv
M_N^2$. 
Moreover
messenger loops lead to the K\"ahler-potential terms needed to generate VEVs 
for $S$ and $F_S$. 

\subsection{Extensions with Light Singlets}
A possible way to dynamically generate $\mu$ and $B\mu$ is to 
add a singlet chiral superfield $S$ to the observable sector and to include
the most
general
superpotential free from mass parameters,
\beq
W=\lambda S H_1 H_2+{k\over 3}S^3.
\label{scube}
\eeq
In this case the generation of $\mu$ corresponds to the spontaneous
breakdown of a $Z_3$ symmetry under which each superfield in eq.~(\ref{scube})
is rotated by $e^{i2\pi/3}$.  If the soft terms are general 
enough, the VEVs of $S, H_1, H_2$ are all non-zero.
We stress that the singlet $S$ here has electroweak scale mass, while the 
one in the previous section had a large mass, of order $\sqrt{F_X}$, and
decoupled from the observable sector. 

We also recall that in models with low-energy supersymmetry breaking,
in contrast with the gravity-mediated case~\cite{polchinski82},
a light singlet offers less (or none at all)  danger to the stability of 
the gauge hierarchy~\cite{nemeschansky84} (see also 
refs.~\cite{dine96,ciafaloni97a}).  Consider,
for instance, a GUT embedding of the Higgs sector with a gauge singlet $S$. 
Even if 
the $Z_3$ symmetry that protects $\langle S\rangle $ is broken, for 
instance, by the Higgs triplets mass terms,
the induced tadpole is at most 
 $\delta V\sim (\alpha/4\pi)^3 S F_X^2/M_G$. Here three loops are needed 
to couple
$S$ to the messengers, and $1/M_G$ is the price paied to propagate GUT-scale
particles.  The resulting 
$\langle S\rangle$ is below the weak scale, as long 
as\footnote{This also allows, in the context of gauge mediation,
to solve the Higgs doublet--triplet splitting problem by means of the
sliding singlet~\cite{witten81a,ibanez82}.} $\sqrt {F_X}\lappeq
 10^8$ GeV.
Conversely, in DSB theories with $\sqrt {F_X}\sim 10^7$--$10^8$ GeV
the correct $\mu$ term
is provided by this tadpole; but this seems quite fortuitous.

The problem with the light-singlet approach is that in gauge mediation
the soft terms are
not general enough to give the desired vacuum. In particular
a VEV for $S$ requires either a sizeable negative soft mass $m_S^2$
or large $A_\lambda$ and $A_k$, the coefficients of the
$A$-terms associated with eq.~(\ref{scube}).
As $S$ is not directly interacting with the
messengers, the leading contributions to all these 
terms arise only through RG evolution.  
For a low messenger scale both $A$-terms are so small that they are
irrelevant in the minimization. On the other hand $m_S^2$ gets a small
contribution, which is negative for light messengers:
\beq
m_S^2\simeq -2\lambda^2\left ( 2 m_{H_1}^2  -3 \la{t}^2m_{\tilde t}^2
{\ln(X/m_{H_1})\over 8 \pi^2} \right ) {\ln (X/m_{H_1})\over 8 \pi^2}<0~.
\label{ms}
\eeq
Here $m_{H_1}$ is the soft mass of the Higgs coupled to down-type quarks.
The smallness of this term is problematic for a correct implementation
of the electroweak-breaking conditions.
Moreover one always finds~\cite{dine93} 
a light scalar Higgs boson and an almost massless pseudoscalar, and the
LEP bound on their associated production rules out this scenario.
The presence of the light pseudoscalar is caused by the spontaneously broken 
$R$ symmetry of the superpotential in eq.~(\ref{scube}), which is
explicitly violated only
by the small $A$ terms. In a detailed numerical study~\cite{gouvea97a} it
has been shown that this unsatisfactory situation persists for any value
of the messenger mass $M$. In the model with a singlet described by
the superpotential in eq.~(\ref{scube}), the soft-breaking terms induced by
gauge mediation either do not give the correct minimum or predict
unacceptably light new particles.

A more promising direction is to couple directly $S$ to some sizeable
source of supersymmetry breaking.
One possibility~\cite{dine93} is to add one light $SU(5)$
flavour $f, \bar f$ coupled to $S$ via $W=\la{f}Sf \bar f$. If 
$\la{f}\sim 1$ then $m_{S}^2$ receives a large negative contribution
from the coloured triplets soft masses, which is the analogue of the stop
contribution to the Higgs mass. Thus a VEV for $ S$ of the
order of the weak scale can be generated, together 
with a consistent mass spectrum.
However, this scenario implies the existence of exotic and stable
(or very long-lived) matter $f, \bar f$ with weak-scale mass.

Another possibility is to couple $S$ directly to the 
messengers~\cite{giudice97}.
This can be done in a theory with at least two messenger flavours
and a superpotential
\beq
W=X({\bar \Phi}_1 \Phi_1 +{\bar \Phi}_2 \Phi_2 ) +\la{S}S{\bar \Phi}_1 \Phi_2~,
\label{xs}
\eeq 
which avoids the radiative generation of the dangerous
tadpole mixing $SX^\dagger$. The form in eq.~(\ref{xs}) is fixed
by a discrete $Z_3$ symmetry under which $\bar \Phi_1 , \Phi_2$, and $S$
have charge $1/3$, $\Phi_1$ and
${\bar \Phi}_2$ have charges $-1/3$ and $X$ is neutral.
This symmetry, broken only at the weak scale, distinguishes
 between $X$ and $S$.
In the leading $F_X/X^2$ approximation, the messengers contribute
to $A_{k,\lambda}$ and $m_S^2$ at one and two loops, respectively:
\beq
{A_k\over 3}=A_\lambda=-\frac{ 5}{4\pi}\alpha_{\lambda_S}{F_X\over X}~,
\label{laaa}
\eeq
\beq
 { m}_{S}^2=\frac{5}{8\pi^2}\alpha_{\lambda_S}
\left[\frac{7}{2}\alpha_{\lambda_S}
-{8\over 5}\alpha_3-{3\over 5} \alpha_2 -{13\over 3} \alpha_1
\right] \left ({F_X\over X}\right)^2~.
\label{lamas}
\eeq
Here $\alpha_{\la{S}}=\la{S}^2/4 \pi$ and all coupling constants are
evaluated at the messenger scale $X$. Notice that the mixed gauge--Yukawa
contribution to $m_S^2$ is negative. Indeed there is a range of $\la{S}$
where $m_S^2$  is negative and of the
order of the Higgs soft term $m_{H_2}^2$ induced by stop
loops. This allows a correct electroweak-symmetry breaking. 
Moreover, the $A$ terms
are now sizeable, and there is no approximate $R$-Goldstone boson.

In ref.~\cite{ciafaloni97a} the superpotential of eq.~(\ref{scube})
was considered in the limit $k=0$. In this case, 
$S$ could indeed be the sliding singlet which explains the doublet-triplet
splitting. Let us not worry for the moment about the Peccei--Quinn
symmetry of this limiting case and about its potentially dangerous axion.
The basic point of ref.~\cite{ciafaloni97a} is that a large enough
$A_\lambda$-term for $\lambda SH_1H_2$, generated by the messenger sector,
can trigger the correct electroweak breaking. Notice that in the limit
$\lambda\ll 1$ the term $\lambda^2|H_1H_2|^2$ can be neglected and 
the potential depends on $\lambda$ and $S$ only via the combination
$\mu =\lambda S$. For a range of  parameters where $A_\lambda^2\sim
m_{H_1}^2, m_{H_2}^2\sim m_{weak}^2$ at the weak
scale, the minimum of the potential gives the correct electroweak breaking
and $\mu \sim m_{weak}$.  
This cannot be achieved within ordinary gauge mediation, but
it is not excluded that 
soft terms of the correct size may be obtained with
additional Higgs-messenger Yukawa couplings. 
In ref.~\cite{ciafaloni97a} it is also pointed out that a
Peccei--Quinn-breaking tadpole $\delta V= \rho S$ can give
the axion a sufficiently large mass without destabilizing the weak scale.
The tadpole could originate from Planck or GUT scale suppressed operators.
For instance, the term $SXX^\dagger/M_P$ in the K\"ahler potential would 
give $\rho\sim F_X^2/M_P$. Notice that the hierarchy is preserved for
$\rho/\lambda\lsim 10^6 ~{\rm GeV}^3$, that is for ${\sqrt F_X}\lsim
\lambda^{1/4} 10^6 ~{\rm GeV}$.
 Now, for $\lambda<10^{-7}$ the axion decay constant is $\langle S\rangle\gsim
10^{9}~{\rm GeV}$, so that
the cooling of red giants does not pose a problem.  For larger $\lambda$,
the axion mass $m_a^2\simeq \rho/S=\lambda \rho/\mu$ should be larger 
than 1 MeV. This bound is compatible with the stability of the weak scale
for $\lambda>10^{-5}$, and collider bounds on axion-like scalars
require $\lambda<10^{-2}$. In conclusion, there is a significant
range for $\lambda$ where the axion does not cause a problem. Notice that 
$F_X$ is constrained to be in its lower range of validity.

Another possibility, suggested in ref.~\cite{dine96}, is to invoke
a field $S$ with 
non-renormalizable 
superpotential couplings $W=S^nH_1H_2+S^{m}$, in units of $1/M_P$. Soft terms,
like those from a K\"ahler-potential interaction
$ SS^\dagger XX^\dagger/M_P^2$, induce a tree-level
$\mu= \langle S\rangle^n\sim F_X^{n/(m-2)}$, so that for $m=2n+2$ one has
$\mu \sim \sqrt{F_X}$. This requires further adjustment of the 
relevant coupling constants to get the right $\mu$.
Anyhow, an interesting 
implication of this mechanism is that $B$
is not generated at tree level along with $\mu$. As previously discussed,
this solves the 
supersymmetric CP
problem,  since the only source of $A$ and $B$ terms is the gaugino
mass via RG evolution. Moreover the size of 
the radiatively induced $B$ is small, so that $\tan \beta$
is predicted to be naturally large~\cite{babu96,dine97,rattazzi96}. 

Finally, a different mechanism for generating the $\mu$ term was suggested 
in ref.~\cite{yanagida97}. It was argued that the cancellation
of the cosmological constant in gauge mediation necessarily requires
a new mass scale. Indeed the cosmological constant receives two different
contributions
\beq
\langle V\rangle \sim \langle F\rangle^2-\frac{\langle W\rangle^2}{3M_P^2}
~.
\eeq
If supersymmetry is broken at low scales, both the auxiliary fields $F$
and the superpotential $W$ are determined by the same typical energy scale
$\Lambda$. The cancellation (of course always at the price of a fine-tuning)
between the two terms can only
be achieved if we add to the superpotential a constant term of
order $\bar \Lambda^3\sim \Lambda^2 M_P$. It is possible that the scale 
$\bar \Lambda$ is dynamically generated by some new strongly interacting
sector, which breaks the continuous $R$ symmetry. If this new sector
couples to the ordinary Higgs via non-renormalizable interactions, 
the $\mu$ term is generated by 
the condensate of a bilinear of new fields,
\beq
\mu \sim \frac{\bar \Lambda^2}{M_P} \sim \left(\frac{\Lambda}{M_P}\right)^{1/3}
~\Lambda ~.
\eeq
For interesting values of $\Lambda$, the ratio 
$(\Lambda /M_P)^{1/3}$ can mimic the loop factor and give the correct
size of the $\mu$ term.

\section{Conclusions}
\setcounter{equation}{0}

Theories with gauge-mediated supersymmetry breaking provide an interesting
alternative to more conventional theories, in which the information of
supersymmetry breaking is communicated to the observable sector by gravity.
They offer the possibility of solving the flavour problem, since the 
soft terms are generated at a low mass scale and are not sensitive to
the high-energy physics, which is presumably responsible for the breaking
of the flavour symmetry. 

Moreover, since the relevant dynamics occur at an energy scale much
smaller than the Planck mass, gauge-mediated models can be solved in
the context of field theory, with no knowledge of quantum gravity
aspects. Although this means that these models cannot describe the ultimate
unified theory, they nevertheless allow us to calculate and predict 
important aspects of the low-energy physics. In particular, because
of the recent developments in
the understanding of non-perturbative aspects of supersymmetric 
theories (for a review, see {\it e.g.} 
refs.~\cite{intriligator96,shifman97,peskin97}), 
the question of dynamical supersymmetry breaking can be addressed in
a quantitative way. This opens the possibility of constructing a theory in
which the electroweak scale can be computed in terms of a single
fundamental mass scale, like the Planck mass.

The last two years have seen a flourishing in the construction
of models with gauge-mediated dynamical supersymmetry breaking. This
activity follows the line of the seminal work in 
refs.~\cite{dine93,dine95,dine96}.
Clearly this field is still evolving; however, progress has been made. 
Models have become simpler in structure, which is to say more 
believable candidates to describe reality. For instance, some realistic
models have been found where the messengers themselves play an essential
r\^ole in breaking supersymmetry. Also, with the variety of new models, all
of the allowed values of the phenomenologically relevant parameters $M$ and
$N$ can be reproduced, with the resulting diversity of cosmological and
phenomenological implications.

Also, gauge-mediated theories have quite distinct phenomenological
features. The supersymmetric mass spectrum is determined in terms of
relatively few parameters. The most important parameter is $\Lambda
=F/M$, which sets the scale of supersymmetry breaking in the observable sector.
The supersymmetric particle masses are typically a one-loop factor smaller
than $\Lambda$, with coefficients completely determined by the particle
gauge quantum numbers. 

The second parameter is the messenger mass scale $M$. Supersymmetric particle
masses depend only logarithmically on $M$, from their RG evolution. This
scale can vary roughly between several tens of TeV and $10^{15}$ GeV. The
lower bound on $M$ is determined by present experimental limits on
supersymmetric particle masses; the upper bound from the argument that
gravity contributions should be small enough not to reintroduce the flavour
problem, see eq.~(\ref{flapro}). 

The third important parameter is the messenger index $N$. If the goldstino
is contained in a single chiral field, then $N$ is the number of messenger
flavours weighted by the corresponding Dynkin index, see eq.~(\ref{enne}).
Perturbativity of gauge couplings gives an upper bound on $N$, which
significantly depends on the messenger mass $M$, see eq.~(\ref{uppern}).
In theories in which $F/M$ is not the same for all messenger
fields, $N$ is no longer an integer, but it retains its physical meaning
of measuring the ratio between the typical gaugino and scalar squared
mass at the
scale $M$. 

Other unknown parameters are connected with the solution to the $\mu$ problem.
As discussed in sect.~\ref{muprob}, the origin of the Higgs mixing parameters
$\mu$ and $B$ still seems problematic, although definite progress has been
made towards a more direct generation of $\mu$ and $B$ from the sector
that breaks supersymmetry. From the phenomenological point of view,
a simple option is to include $\mu$ and $B$ terms by hand and impose
the constraint of electroweak breaking. We are left with a single free
parameter, usually chosen to be $\tan\beta$, apart from an ambiguity in
the phase of $\mu$. A more restrictive possibility is to assume $B=0$ at
the messenger scale $M$. In this case, there are no new free parameters and
$\tan \beta$ is determined to be very large. However, one should keep it
in mind that dynamical solutions
to the $\mu$ problem often induce new contributions to the 
supersymmetry-breaking Higgs mass terms. We then find two (or one, if the
two-Higgs-doublet sector has an isospin invariance) extra free parameters.

As we have discussed in this review, the phenomenological aspects of
gauge-mediated theories are quite diverse, depending on the parameter
choice. However, they can roughly be divided into two regimes.
Models with low values of $F_0$, and therefore comparable values of $M$,
have the characteristic that the NLSP decays promptly (for
$\sqrt{F_0}$ roughly less than $10^6$ GeV) and that the gravitino relic density
is less than the critical density (for
$\sqrt{F_0}$ roughly less than $10^7$ GeV). 
Depending on whether the NLSP is a neutralino or a stau, the characteristic
collider signals are anomalous events characterized by missing energy and
$\gamma$ or $\tau$, respectively.
For larger $\sqrt{F_0}$, the NLSP lifetime is longer and the collider
phenomenology can resemble the well-known missing-energy supersymmetric
signatures (for a neutralino NLSP) or can lead to a long-lived heavy charged
particle penetrating the detector (for a stau NLSP). In this regime, the
gravitino or the NLSP decay can cause cosmological difficulties.

Theories with gauge-mediated supersymmetry breaking have very appealing
theoretical features and quite distinctive experimental signatures.
We expect that the joint theoretical and experimental research will soon
demostrate whether these theories are relevant for the description of nature.

We thank N.~Arkani-Hamed, 
S.~Dimopoulos, G.~Dvali, S.~Hsu, K.-I.~Izawa, M.~Luty, H.~Murayama, H.P.~Nilles,
Y.~Nir,
A.~Pomarol, N.~Polonsky, E.~Poppitz,
L.~Randall, U.~Sarid, Y.~Shadmi, S.~Weinberg, T.~Yanagida,
and A.~Zaffaroni for useful discussions.

\section*{Appendix}
\bigskip

\renewcommand{\theequation}{A.\arabic{equation}}
 
In this appendix we give the complete matching conditions at the messenger
scale for the gauge-mediation mass spectrum, including next-to-leading
order radiative corrections in the gauge coupling constants and the 
top-quark Yukawa, in the limit $F\ll M^2$. 
Details of the calculation are presented in 
refs.~\cite{giudice97,arkani98}. The results are given in the so-called
$\overline{\rm DR}'$ 
scheme for soft terms, which is defined in ref.~\cite{jack94}.
The formulae given below can be implemented in numerical
codes that compute the next-to-leading order RG evolution~\cite{martin94} 
and the 
low-energy threshold corrections~\cite{pierce94,pierce97}, and then can be
used in phenomenological
studies.

We choose to define the matching scale as
\beq
M\equiv \lambda_G \langle X \rangle ,
\eeq
where $\langle X \rangle$ is the scalar component VEV of the chiral
superfield containing the Goldstino, and $\lambda_G$ is the coupling
constant in eq.~(\ref{phix}) evaluated at the GUT scale. We also
define $F\equiv \lambda_G \langle F_X \rangle$, where $F_X$ is the
auxiliary component of the superfield $X$.
Notice that the physical messenger thresholds are instead at
\beq
M_i={M\over Z_i(M)}
\eeq
where $Z_i$ is the wave-function renormalization
of the $i$-th messenger flavour.
We are also assuming that the wave functions for $\Phi_i$ and 
$\bar\Phi_i$ are equal. This is correct in most cases, and the expressions we
give here can be easily generalized to the case in which this assumption
does not hold.

The next-to-leading order expression of the gaugino mass at the scale $M$ is
\beq
{\tilde M}_{\lambda_r}(M)=k_r\frac{\alpha_r (M)}{4\pi}N\frac{F}{M}
\left[ 1+\frac{\alpha_r (M)}{2\pi}T_{G_r}\right] .
\label{gaugnn}
\eeq
Here we use the same notation as in sect.~\ref{spectrum} and $T_{G_r}
=N$ for $SU(N)$ groups and $T_{G_r}=0$ for $U(1)$ groups. Because
of the ``gaugino screening" theorem discussed in ref.~\cite{arkani98}, 
the result in eq.~(\ref{gaugnn}) is independent of new messenger interactions
and, in particular, of the assumption that the couplings $\lambda_i$ unify
at the GUT scale into a single coupling $\lambda_G$. The QCD corrections
to gluino mass are important especially at small $N$ and $M$ and can 
exceed 10{\%}~\cite{arkani98,picariello98}.

The next-to-leading expression of the supersymmetry-breaking scalar masses
at the matching scale $M$ is
\beq
m_{\tilde f}^2 (M)=2\sum_{r=1}^3 k_r \frac{\alpha_r^2(M)}{(4\pi)^2}N
\frac{F^2}{M^2} \left( C_r^{\tilde f} + \frac{\Delta_r^{\tilde f}}{4\pi}\right)
.
\label{scann}
\eeq
Here the notation is analogous to the one followed in sect.~\ref{spectrum}
and $\Delta_r^{\tilde f}$, which represents the next-to-leading correction,
is given by
\bea
\Delta_r^{\tilde f} =&&2  C_r^{\tilde f} \left[ b_r +T_{G_r} +(Nk_r-b_r)
(1-{\cal Z}_r)\right] \alpha_r (M) -4 C_r^{\tilde f} \sum_{s=1}^3 
C_s^{\tilde f} \alpha_s (M) +C_r^{\tilde f} \gamma_r \nonumber \\
&& -\left[
P_r^{\tilde f} +\frac{d_{\tilde f}}{2} a_r (1-{\cal Z}_r)\right] \alpha_t (M).
\label{deltone}
\eea
Here $\alpha_t=h_t^2/(4\pi)^2$ measures the top-quark Yukawa coupling,
$a_r=2(C_r^{{\tilde Q}_L} + C_r^{{\tilde U}_R}+C_r^{{H}_2})=
(13/9,3,16/3)$ and
the only non-vanishing coefficients $P_r^{\tilde f}$ and $d_{\tilde f}$ are
\beq
P_3^{H_2} =8,~~~~P_2^{{\tilde U}_R^3}=3,~~~~P_1^{{\tilde Q}_L^3}=\frac{2}{3}
,~~~~
P_1^{{\tilde U}_R^3}=-\frac{1}{3},~~~~P_1^{H_2}=\frac{2}{3} ,
\eeq
\beq
d_{{\tilde Q}_L^3}=1,~~~~d_{{\tilde U}_R^3}=2,~~~~d_{H_2}=3.
\eeq
We have also defined
\beq
{\cal Z}_r=\frac{1}{N}\sum_{i=1}^{N_f}n_i^r \ln Z^2_i
\eeq
\beq
\gamma_r=\frac{1}{N}\sum_{i=1}^{N_f}n_i^r \gamma_i .
\eeq
The sum is extended over the messenger fields, $n_i^r$ is twice the
corresponding Dynkin index ({\it e.g.} $n_i^r=1$ if the messengers belong
to the fundamental of the group $G_r$, and $n_i^r=(6/5)Y^2$ for hypercharge),
and the messenger index is $N=\sum_{i=1}^{N_f} n_i^r$. The 
$\gamma_i$'s (see eq.~(\ref{diman}) below)
are determined by the messengers' anomalous dimensions:
$\gamma_i=-4 \pi d\ln Z_i(M)/d\ln M$.

The ``screening theorem"~\cite{arkani98} 
does not hold for scalar masses, and therefore
eq.~(\ref{scann}) depends on the messenger wave-functions $Z_i$. Here we assume
unification of the couplings $\lambda_i$ and therefore the messenger
wave-functions can be calculated in terms of the unification scale $M_G$
and the unified coupling $\lambda_G$,
\beq
Z_i=\left[ \frac{\lambda_i^2 (M)}{\lambda_G^2}\right]^{-1/D_i}
\prod_{r=1}^3 \left[ \frac{\alpha_r(M)}{\alpha_r(M_G)}
\right]^{\frac{2C_r^i(2-D_i)}{D_i(b_r-Nk_r)}},
\eeq
\beq
\gamma_i=\frac{\lambda_i^2(M)}{2\pi} -4\sum_{r=1}^3 C_r^i \alpha_r (M) .
\label{diman}
\eeq
We have defined $D_i=2+{\rm dim}(R_{\Phi_i})$, where ${\rm dim}(R_{\Phi_i})$
is the dimension of the messenger representation ({\it e.g.} 
${\rm dim}(R_{\Phi_i})=N_G$ for a fundamental of an $SU(N_G)$ gauge group). The 
relations between the coupling constants at the scale $M$ and the GUT scale
$M_G$ are
\beq
\alpha_r^{-1}(M)= \alpha_r^{-1}(M_G) +\frac{(b_r-Nk_r)}{2\pi}\ln \frac{M_G}{M},
\eeq
\beq
\lambda_i^2(M)=\frac{\lambda_G^2E^i(M)}{1-\frac{D_i}{8\pi^2}\lambda_G^2F^i(M)}
\label{lambino}
\eeq
\beq
E^i(\mu)=\prod_{r=1}^3 \left[ \frac{\alpha_r(M_G)}{\alpha_r(\mu )}
\right]^{\frac{4C_r^i}{(b_r-Nk_r)}},~~~~F^i(M)=\int_0^{\ln M/M_G}
d\ln \mu ~E_i(\mu).
\eeq

Finally, defining the trilinear $A$ terms as in eq.~(\ref{lsodef}), their
next-to-leading order matching condition at the scale $M$ is
\beq
A_{{\tilde f}_i}(M)=-2\sum_{r=1}^3 k_r C_r^{{\tilde f}_i} 
\frac{\alpha_r^2(M)}{(4\pi)^2}N\frac{F}{M}(1-{\cal Z}_r) .
\eeq

For phenomenological applications, it is useful to rewrite the next-to-leading
corrections in the particular case in which the messengers form $N$ fundamental
and antifundamental multiplets of $SU(5)$. To further simplify 
eq.~(\ref{deltone}), we keep only QCD and top-Yukawa corrections and neglect
the weak coupling (which is justified only for $M\ll M_G$). 

For squarks different than the stop, eq.~(\ref{deltone}) becomes
\beq
\Delta_3^{\tilde q}=\frac{8}{3}\alpha_3(M) \left[ N-\frac{7}{3}
-2(N+3)\ln Z_3 \right] +\frac{2\lambda_3^2(M)}{3\pi},
\label{prim}
\eeq
\beq
Z_3=\left[\frac{\lambda_3^2(M)}{\lambda_G^2}\right]^{-\frac{1}{5}}~
\left[\frac{\alpha_3(M)}{\alpha_G}\right]^{\frac{8}{5(N+3)}}.
\eeq
For left and right sleptons, we find
\beq
\Delta_2^{{\tilde \ell}_L}=\frac{3\lambda_2^2(M)}{8\pi}
\eeq
\beq
\Delta_1^{{\tilde \ell}_L}=-\frac{8\alpha_3(M)}{15}
+\frac{3\lambda_2^2(M)}{40\pi}
+\frac{\lambda_3^2(M)}{20\pi},
\eeq
\beq
\Delta_1^{{\tilde e}_R}=4\Delta_1^{{\tilde \ell}_L}.
\eeq
Here $\lambda_2$ and $\lambda_3$ refer to the couplings which determine
the masses of the $SU(2)$ doublet and $SU(3)$ triplet messengers contained
in the fundamental representation of $SU(5)$. They are related to the
unified coupling $\lambda_G$ by eq.~(\ref{lambino}).

The corrections to the left and right stop masses are
\beq
\Delta_3^{{\tilde Q}^{(3)}_L}=\Delta_3^{\tilde q}-\frac{8}{3}
(1-\ln Z_3^2)\alpha_t(M),
\eeq
\beq
\Delta_3^{{\tilde U}^{(3)}_R}=\Delta_3^{\tilde q}-\frac{16}{3}
(1-\ln Z_3^2)\alpha_t(M).
\eeq
The correction to the soft mass of the Higgs $H_2$ is
\beq
\Delta_3^{H_2}=-8(2-\ln Z_3^2)\alpha_t(M).
\label{ultim}
\eeq

Notice that, neglecting weak next-to-leading corrections, $Z_3$ is the only
messenger wave-function appearing in the mass formulae. Indeed, in this
case, it is more convenient to use in eq.~(\ref{deltone})
the triplet messenger mass $M_3=M/Z_3$
as a matching scale, instead of $M$. The expressions
for the corrections $\Delta_r^{\tilde f}$ are then given by 
eqs.~(\ref{prim})--(\ref{ultim}), setting $\ln Z_3 =0$. This can be explicitly
verified by using the one-loop RG equations for the soft masses to relate
$m_{\tilde f}^2 (M)$ to $m_{\tilde f}^2 (M_3)$.
The result in eq.~(\ref{ultim}) plays an important r\^ole in the analysis
of electroweak breaking and it has been used in eq.~(\ref{hnlo}).

\end{document}